\documentclass[aps,onecolumn,11pt,floatfix,altaffilletter,preprintnumbers,
               tightenlines,showpacs,showkeys,notitlepage,nofootinbib,superscriptaddress]{revtex4-1}
\usepackage[colorlinks=true,citecolor=blue,linkcolor=blue]{hyperref}
\usepackage[utf8]{inputenc}
\usepackage[T1]{fontenc}            
\usepackage[normalem]{ulem}
\usepackage{amsmath,amssymb,bbm}
\usepackage{mathtools}
\usepackage{epsfig}
\usepackage{graphicx}               
\usepackage{url}
\usepackage{color}
\usepackage{multirow}
\usepackage{placeins}
\usepackage[dvipsnames]{xcolor}
\usepackage{epstopdf}
\usepackage[shortlabels]{enumitem}
\usepackage{cleveref}
\usepackage{lipsum}
\usepackage{gensymb}
\usepackage{siunitx}
\usepackage{booktabs} 
\usepackage{titlesec} 
\usepackage{subfigure}
\usepackage[subfigure,titles]{tocloft} 
 
\allowdisplaybreaks

\makeatletter

\renewcommand{\p@subsection}{}
\makeatother

\titleformat*{\section}{\centering\bfseries\scshape}
\titlelabel{\thetitle\quad}
\titleformat*{\paragraph}{\bfseries}
\titlespacing*{\paragraph}{0pt}{3.25ex plus 1ex minus .2ex}{1em}

\newcommand{\BR}{\text{BR}}

\newcommand{\iso}[2]{{\ensuremath{{}^{#2}}\ensuremath{\rm #1}}}

\usepackage{xspace}
\newcommand{\genie} {{\sf GENIE}\xspace}
\newcommand{\gibuu} {{\sf GiBUU}\xspace}
\newcommand{\nuance}{{\sf NUANCE}\xspace}
\newcommand{\nuwro} {{\sf NuWro}\xspace}

\pacs{}
\keywords{}

\newcommand{\newtext}[1]{{\bf\color{Blue} #1}}
\begin{document}

\title{An Altarelli Cocktail for the MiniBooNE Anomaly?}

\author{Vedran Brdar}
\email{vedran.brdar@northwestern.edu}
\affiliation{Fermi National Accelerator Laboratory,
             Batavia, IL 60510, USA}
\affiliation{Northwestern University, Dept.~of Physics \& Astronomy,
             Evanston, IL 60208, USA}

\author{Joachim Kopp}
\email{jkopp@cern.ch}
\affiliation{Theoretical Physics Department, CERN,
             1211 Geneva 23, Switzerland}
\affiliation{Johannes Gutenberg University Mainz,
             55099 Mainz, Germany}

\preprint{CERN-TH-2021-131, FERMILAB-PUB-21-450-T}
\preprint{MITP-21-042, NUHEP-TH/21-14}

\begin{abstract}
We critically examine a number of theoretical uncertainties affecting the
MiniBooNE short-baseline neutrino oscillation experiment in an attempt to
better understand the observed excess of electron-like events.  We re-examine
the impact of fake charged current quasi-elastic (CCQE) events, the background
due to neutral current $\pi^0$ production, and the single-photon background.
For all processes, we compare the predictions of different event generators
(\genie, \gibuu,  \nuance and \nuwro) and, for \genie, of different tunes.
Where MiniBooNE uses data-driven background predictions, we discuss the
uncertainties affecting the relation between the signal sample and the control
sample.  In the case of the single-photon background, we emphasize the
uncertainties in the radiative branching ratios of heavy hadronic resonances.
We find that not even a combination of uncertainties in different channels
adding up unfavorably (an ``Altarelli cocktail'') appears to be sufficient to
resolve the MiniBooNE anomaly.  We finally investigate how
modified background predictions affect the fit of a $3+1$ sterile neutrino
scenario.  We carefully account for full four-flavor oscillations not only in
the signal, but also in the background and control samples. We emphasize that
because of the strong correlation between MiniBooNE's $\nu_e$ and $\nu_\mu$
samples, a sterile neutrino mixing only with $\nu_\mu$ is sufficient to explain
the anomaly, even though the well-known tension with external constraints on
$\nu_\mu$ disappearance persists.
\end{abstract}

\maketitle

\newcommand{\contentsname}{}
\setlength{\cftbeforesecskip}{3pt}
\setlength{\cftsecnumwidth}{2em}
\setlength{\cftsubsecindent}{\cftsecnumwidth}
\setlength{\cftsubsecnumwidth}{3em}
{
    \cftsetpnumwidth{4em}
    \tableofcontents
}


\section{Introduction}
\label{sec:intro}

\noindent
The decision to downgrade the BooNE proposal to the MiniBooNE
experiment~\cite{Church:1997ry} has been, in retrospect, both a curse and a blessing
for neutrino physics.  On the one hand, MiniBooNE has given us one of the most
intriguing anomalies particle physics has seen in recent years: a $4.8\sigma$
\cite{Aguilar-Arevalo:2020nvw} excess of electron neutrinos ($\nu_e$) in a beam
consisting mostly of muon neutrinos ($\nu_\mu$).  This observation has led to
significant progress in our understanding of neutrino--nucleus interactions~\cite{Formaggio:2012cpf,Alvarez-Ruso2021}, progress that will be invaluable to future neutrino
experiments. The anomaly has also given rise to a tremendous amount of
theoretical and phenomenological work interpreting the excess as a hint for new
physics, for instance in the form of sterile neutrinos \cite{Fischer:2019fbw,
Gninenko:2009ks, Bertuzzo:2018itn, Dentler:2019dhz, Ballett:2018ynz,
deGouvea:2019qre, Abdallah:2020biq, Dutta:2020scq, Datta:2020auq, Abdallah:2020vgg,
Abdullahi:2020nyr, Brdar:2020tle,Abdallah:2020biq}.  On the other hand, if MiniBooNE hadn't been
stripped of its second detector, we might have known right away whether the
anomaly is due to ``new physics'' or due to imperfect modeling of Standard Model
effects.

In any case, the situation is being rectified now, with Fermilab's new
short-baseline program consisting of not one but three additional detectors:
MicroBooNE~\cite{Acciarri:2016smi}, SBND~\cite{McConkey:2017dsv}, and
ICARUS~\cite{Rubbia:2011ft, Machado:2019oxb}. These detectors are located at
different baselines, $L$, from the primary target and should
therefore be able to unambiguously determine whether MiniBooNE's $\nu_e$ excess
oscillates with $L$ or not.  Moreover, they are liquid argon time projection
chambers which, compared to MiniBooNE's mineral oil-based \v{C}erenkov
detector, offer much better event reconstruction capabilities and will
therefore be much better at distinguishing a possible neutrino oscillation
signal from various backgrounds.

Our goal in this paper is to add several novel aspects to the discussion of
background processes and theoretical uncertainties in MiniBooNE. Ultimately, we
would like to determine whether an accumulation of small deviations in
different background channels adding up in an inauspicious way -- often dubbed
an ``Altarelli cocktail''~\cite{Ellis:2017itw} -- could be sufficient to
explain the MiniBooNE anomaly.

We begin in \cref{sec:channels} by re-calculating several of MiniBooNE's most
important backgrounds.  First, in \cref{sec:CC-fp}, we address MiniBooNE's
event reconstruction: the signal process charged current quasi-elastic (CCQE)
neutrino--nucleus scattering is identified by the exclusive presence of a
single $e^\pm$ or $\mu^\pm$, and the energy of the incoming neutrino is
calculated from the energy of this charged lepton and its direction with
respect to the beam axis.  However, events may be incorrectly classified as
CCQE if additional final state particles such as pions are either reabsorbed
before they leave the target nucleus, or are missed by the detector.  The
resulting misreconstruction of neutrino energies has been discussed previously
in refs.~\cite{
  Martini:2009uj,           
  Nieves:2011pp,            
  Sobczyk:2012ms,           
  Meucci:2012yq,            
  Lalakulich:2012ac,        
  Meloni:2012fq,            
  Nieves:2012yz,            
  Lalakulich:2012hs,        
  Martini:2012uc,           
  Aguilar-Arevalo:2013pmq,  
  Coloma:2013rqa,           
  Mosel:2013fxa,            
  Megias:2014qva,           
  Ericson:2016yjn,          
  Aguilar-Arevalo:2018gpe}, 
and while it leads to distortions of neutrino energy spectra, the
effect has been found to be too small to explain the MiniBooNE anomaly.
Our novel contribution compared to previous works
will be threefold: (i) we compare predictions of different event
generators, namely \genie, \gibuu, \nuance, and \nuwro, to better estimate
how CCQE energy reconstruction depends on theory errors; (ii) we include the
impact of fake CCQE events in the $\nu_\mu$ sample, which MiniBooNE analyze
together with the $\nu_e$ sample to better constrain the neutrino flux; (iii)
we work with more up-to-date data than previous studies, in particular the
data from ref.~\cite{Aguilar-Arevalo:2018gpe}.\footnote{We do not consider
the even more recent data from the 2020 update of the MiniBooNE anomaly
\cite{Aguilar-Arevalo:2020nvw}, which corresponds to a roughly 50\% further
increase in statistics and elevates the significance of the anomaly from
4.7$\sigma$ to 4.8$\sigma$.}

Second, in \cref{sec:fp-pi0} we will study MiniBooNE's $\pi^0$ background,
comparing again the predictions of different event generators (see
ref.~\cite{Stowell:2016jfr} for previous related work in this
direction). The $\pi^0$ background arises from neutral current
(NC) interactions in which a single $\pi^0$ is produced.  To the MiniBooNE
detector, the photons from $\pi^0$ decay look the same as $e^\pm$ from a
charged current (CC) $\nu_e$ or $\bar\nu_e$ interaction.  Therefore, if one of the
two photons is missed, or if the two are so close to each other that they merge
into a single reconstructed photon, NC $\pi^0$ production can mimic CCQE
$\nu_e$ interactions, the signal MiniBooNE is looking for. 

The third background we address in this paper is the single-photon background
(\cref{sec:fp-single-photon}). Single photons can originate from radiative
decays of hadronic resonances like the $\Delta(1232)$ (``resonance-pole
terms''), from coherent production off the target nucleus, or from nucleon-pole
terms~\cite{Hill:2009ek, Zhang:2012xn, Wang:2013wva, Wang:2014nat}. As for the
$\pi^0$-induced background, a single photon can mimic the CCQE $\nu_e$ signal.
We compare predictions of different event generators and tunes to estimate the
theoretical uncertainties affecting the single-photon background. 

In the second part of the paper, \cref{sec:dd}, we shift our focus towards
data-driven estimates for the $\pi^0$s and single photons in an attempt to more
closely follow the approach the MiniBooNE collaboration is taking in predicting
backgrounds. For both the $\pi^0$ background and the single-photon background,
a suitable control sample are single $\pi^0$ events in which the two photons
from the decay are separately reconstructed. On the one hand, this control
sample constrains the rate of $\pi^0$ production. But since at MiniBooNE
energies most $\pi^0$s stem from the decay of hadronic resonances, it also
constrains the production rate of such resonances and thus rate of radiative
resonance decay events.  Even with data-driven background estimates,
theoretical uncertainties enter when translating the event rate in the control
sample into a number of expected background events in the signal region.  To
estimate these theoretical uncertainties, we develop a mock-up of MiniBooNE's
data-driven $\pi^0$ and single-photon analyses, anchoring these background
rates to the measured spectrum of $\pi^0$ events and comparing the impact of
different theoretical models and different event generators/tunes on the
translation between the control and signal samples.\footnote{MiniBooNE's
data-driven estimates of the $\pi^0$ background have also been scrutinized
recently in refs.~\cite{Ioannisian:2019kse, Giunti:2019sag}, focusing in
particular on the effect of $\pi^0$ re-absorption (an effect that has also
been included in MiniBooNE's analyses~\cite{Louis:privcomm}).}

In \cref{sec:br-scans}, we study the impact of uncertainties in the radiative
branching ratios of hadronic resonances.  We will find that these uncertainties
can affect the predictions of the single-photon background at the 10\% level.
Importantly, this uncertainty cannot be reduced even when data-driven methods
are used.

In the final part of the paper, \cref{sec:results}, we fit
a $3+1$ sterile neutrino scenario (3 standard active neutrinos and 1 additional eV-scale
sterile neutrino, $\nu_s$) to MiniBooNE data.  We first emphasize that in a full four-flavor
fit, the $\nu_e$ background can be affected by significant $\nu_e \to \nu_s$
disappearance, and the $\nu_\mu$ control sample that is used for flux
normalization can suffer from sizeable $\nu_\mu \to \nu_s$ disappearance.
In a two-flavor fit, on the other hand, disappearance effects are negligible.
We will show how this disparity affects the preferred parameter regions of the
$3+1$ scenario.
We then investigate how the fit changes depending on which event-generator
is used for the background predictions, and on whether the background
prediction is taken directly from the Monte Carlo (as in \cref{sec:channels})
or whether data-driven methods are used (as in \cref{sec:dd}).
We summarize and conclude in \cref{sec:summary}.

\section{Background estimates from Monte Carlo simulations}
\label{sec:channels}

\noindent
We begin by individually considering various background processes relevant
to MiniBooNE's sample of CCQE $\nu_e$-like
events. We focus in particular on the CC $\nu_e$ background due to the $\nu_e$
contamination in the beam (\cref{sec:CC-fp}), NC $\pi^0$ production
(\cref{sec:fp-pi0}), and NC single-photon production (\cref{sec:fp-single-photon}).
Even though the MiniBooNE collaboration is not relying on Monte Carlo
simulations alone, but rather on data-driven background estimates wherever possible,
a comparison of Monte Carlo-only predictions will give a first indication of
where large theoretical uncertainties are lurking.  We employ in particular the
following event generators: \genie v3.00.04 \cite{Andreopoulos:2015wxa}, \nuance
v3.000 \cite{Casper:2002sd}, \nuwro v19.02.2-35-g03c3382 \cite{Golan:2012wx},
and \gibuu (2019 release) \cite{Leitner:2008ue}.

While \genie, \nuwro, and \gibuu are actively used state-of-the-art tools,
\nuance is, to the best of our knowledge, not under active development any
more. Nevertheless, \nuance will be crucial for our analysis because it is the
main generator used by the MiniBooNE
collaboration~\cite{AguilarArevalo:2010zc}.  Indeed, for \nuance, we work with
flux and configuration files that were kindly provided to us by the MiniBooNE
collaboration and are dated April/May~2007.

\gibuu differs from the other
three generators in that it employs a more holistic approach: rather than
piecing together largely independent theoretical models for different kinematic
regimes (quasi-elastic scattering, resonance production, deep-inelastic
scattering, etc.) and subprocesses (primary interaction, final state
interactions, etc.), it uses the same inputs such as nuclear ground state,
nuclear potentials, and production/absorption amplitudes for all kinematic
regimes. By solving a set of quantum transport equations, one of its strengths
is the accurate simulation of final state interactions (FSI).

\nuwro has been widely used for testing new nuclear models that are yet to be 
implemented in other generators such as \genie. Neutrinos with energies between $\mathcal{O}(100)$ MeV and $\mathcal{O}(100)$ GeV can be simulated using this generator. This energy range covers quasielastic, resonant, and deep inelastic scattering. The generator also offers several options for accounting for nuclear effects such as global/local Fermi gas and spectral functions \cite{Ankowski:2005wi, BENHAR1994493}.

In the case of \genie,
we will consider six different tunes~\cite{tune-list}. The naming convention for
these tunes is {\tt G18\_XXy\_02\_11a}, where tunes with {\tt XX=01} can be considered
baseline tunes, those with {\tt XX=02} feature updated implementations of resonant and
coherent scattering, and those with {\tt XX=10} also employ updated models for CCQE and
two-particle/two-hole (2p2h) interactions~\cite{Nieves:2011pp} as well as an
improved description of the nuclear initial state in terms of a local Fermi gas
(with radius-dependent Fermi momentum, as opposed to a relativistic Fermi gas
with a Fermi momentum that is the same everywhere in the nucleus). The lower-case
letter {\tt y} indicates how FSI are treated,
with {\tt y=a} corresponding to a simple implementation of hadron--nucleus cross sections,
while {\tt y=b} stands for a more sophisticated hadronic cascade in which interactions
of hadrons with individual nucleons are recursively simulated. The code {\tt 02\_11a},
finally, describes the data sets that the models have been tuned to, which are the
same for all tunes considered here. We would like to stress that despite some differences in approaches, there is a large overlap in model choices for different tunes. For instance, the Rein-Sehgal model \cite{REIN198179} is employed across all tunes for resonance processes \cite{Avanzini:2021qlx}.

Note that \gibuu, \nuance, and \nuwro do not implement radiative decays of heavy
baryonic resonances (e.g.\ $\Delta(1232) \to N + \gamma$) by default. As these decays
are an important source of single-photon events in MiniBooNE and thus an important
background to the $\nu_e$ appearance search, we have implemented them manually
by randomly replacing the pion in $\Delta(1232) \to N + \pi^0$ events by a photon
with the same energy. We do this for 0.6\% of all $\Delta(1232) \to N + \pi^0$
events, corresponding to the branching ratio of $\Delta(1232) \to N + \gamma$
according to ref.~\cite{Zyla:2020zbs}. 

The different generators and tunes used in this work are also summarized in
\cref{tab:generators}. For more detailed description of the generators, as well 
as the comparison between them for various processes we refer the reader to \cite{Betancourt:2018bpu, Avanzini:2021qlx}.

\begin{table}
  \centering
  \begin{ruledtabular}
  \begin{tabular}{l@{\hspace{-0.2cm}}ccl}
    \bf Generator & \bf Tune          & \bf  Ref.              & \bf Comments \\
    \hline
    \nuance       &        --         &  \cite{Casper:2002sd}  & the generator used by MiniBooNE \\
    \gibuu        &        --         &  \cite{Leitner:2008ue} & theory-driven generator \\
    \nuwro        &        --         &  \cite{Golan:2012wx}  & \newtext{sandbox for other generators; several options for nuclear effects} \\ 
    \genie        & G18\_01a\_02\_11a &  \cite{Andreopoulos:2015wxa,tune-list} &
                                           \genie baseline tune; see \cite{tune-list} for
                                           naming conventions \\
                  & G18\_01b\_02\_11a &  & different FSI implementation compared
                                           to G18\_01a\_02\_11a \\
                  & G18\_02a\_02\_11a &  & updated res./coh. scattering models
                                           compared to G18\_01a\_02\_11a \\
                  & G18\_02b\_02\_11a &  & updated res./coh.~scattering models
                                           and different FSI \\
                  & G18\_10a\_02\_11a &  & theory-driven configuration; similar to G18\_02a \\
                  & G18\_10b\_02\_11a &  & theory-driven configuration; similar to G18\_02b \\
  \end{tabular}
  \end{ruledtabular}
  \caption{Event generators and tunes used in this work.}
  \label{tab:generators}
\end{table}

Our strategy is the same for each of the three considered background channels (CC
neutrino scattering, NC $\pi^0$ production, and NC single-photon production),
and can be described  as follows:
\begin{enumerate}[label=$(\roman*)$]
  \item From a Monte Carlo simulation using the \nuance generator, we predict
    the event sample under consideration. In doing so, we make our best effort
    to reproduce the cuts and implement the efficiency factors of the real
    MiniBooNE analysis (which also employed the \nuance generator).

  \item The predicted event spectrum from $(i)$ is then compared with the
    corresponding prediction obtained by the MiniBooNE collaboration
    \cite{Aguilar-Arevalo:2018gpe}; the differences, which are expected to be
    mild, are compensated by bin-by-bin tuning.

  \item We then predict the same event sample using \gibuu, \nuwro, as well as
    six different \genie tunes, using the same cuts and efficiency factors as
    for \nuance. We then apply the tuning factors determined in step
    $(ii)$ as the ratio between our \nuance prediction and MiniBooNE's.
    This final tuning greatly alleviates any residual differences between our
    simplified analysis and the one employed by the MiniBooNE collaboration,
    yielding background predictions that are accurate enough to compare to data
    in a meaningful way.
\end{enumerate}
In our analysis we only consider positive horn polarity (neutrino mode) data
which mostly drives the statistical significance of the reported excess..

\subsection{Charged Current Events}
\label{sec:CC-fp}

\noindent
We start by considering CC neutrino interactions. To MiniBooNE's $\nu_\mu \to
\nu_e$ oscillation search, such interactions are relevant not only for the
signal, but also for part of the background. This is because the beam, though
consisting mostly of muon neutrinos, unavoidably contains a small admixture of
electron neutrinos, mostly from the decays of kaons and muons.\footnote{Here,
and in the following, ``neutrino'' refers to both neutrinos and anti-neutrinos,
unless stated otherwise.} This intrinsic $\nu_e$ background accounts for
$\mathcal{O}(10\%)$ of the total background at the lowest measurable neutrino
energies $E_\nu \sim \SI{200}{MeV}$, and for almost all background events at
$E_\nu > \SI{1}{GeV}$.  On top of this, the sterile neutrino fit includes
also CC $\nu_\mu$ events, which are used as a control sample to normalize the
flux.  A change in the CC $\nu_\mu$ rate will thus indirectly affect
predictions for the intrinsic $\nu_e$ background and for the $\nu_\mu \to
\nu_e$ signal.

Following the strategy introduced in the beginning of this section, we first
compute the expected rate of CC $\nu_e$ and CC $\nu_\mu$ events using \nuance \cite{MiniBooNE:2008hfu}. Out of all simulated events, we keep those that
contain exactly one charged lepton (electron or muon) and no detectable mesons.
We define a ``detectable'' meson as a neutral pion, a charged pion above the
\v{C}erenkov threshold, or a meson heavier than a pion.  For CC $\nu_e$-like
events, we apply a 20\% detection efficiency~\cite{Aguilar-Arevalo:2018gpe},
while for CC $\nu_\mu$ events, the efficiency is assumed to be 35\%
\cite{Katori:2008zz}.

Like the MiniBooNE collaboration, we reconstruct the neutrino energy $E_\nu$ in
each event based on the assumption that the event topology is indeed
$\nu_{e,\mu} + n \to e^- / \mu^- + p$ (or the corresponding processes for
anti-neutrinos). In this case, $E_\nu$ can be calculated as
\cite{AguilarArevalo:2010zc}
\begin{align}
  E_\nu = \frac{2 m_n' E_\ell - (m_n'^2 + m_\ell^2 - m_p^2)}
               {2 \,[ m_n' - E_\ell + \sqrt{E_\ell^2 - m_\ell^2} \cos\theta_\ell \,] }  \,,
  \label{eq:Enu-qe}
\end{align}
where $E_\ell$ is the charged lepton's energy ($\ell = e,\,\mu$), $m_\ell$ is
its mass, and $\theta_\ell$ is the direction of its momentum vector relative to
the beam axis.  The proton and neutron masses are denotes as $m_n$ and $m_p$,
respectively, while $m_n' \equiv m_n - E_B$, with $E_B$ the binding energy in
the nucleus. We set $E_B = \SI{34}{MeV}$, corresponding to neutrons bound in a
\iso{C}{12} nucleus \cite{PhysRevLett.26.445}. It is important to keep in mind
that \cref{eq:Enu-qe} will yield an incorrect value for $E_\nu$ in fake CCQE
events, that is events which contain extra final state particles (for instance
pions), but in which these extra final state particles are missed, either
because they are reabsorbed by the nucleus in which they are produced, or
because they fall below the experimental thresholds \cite{Martini:2012fa,
Martini:2012uc, Ericson:2016yjn}.

\begin{figure}
  \centering
  \includegraphics[width=\textwidth]{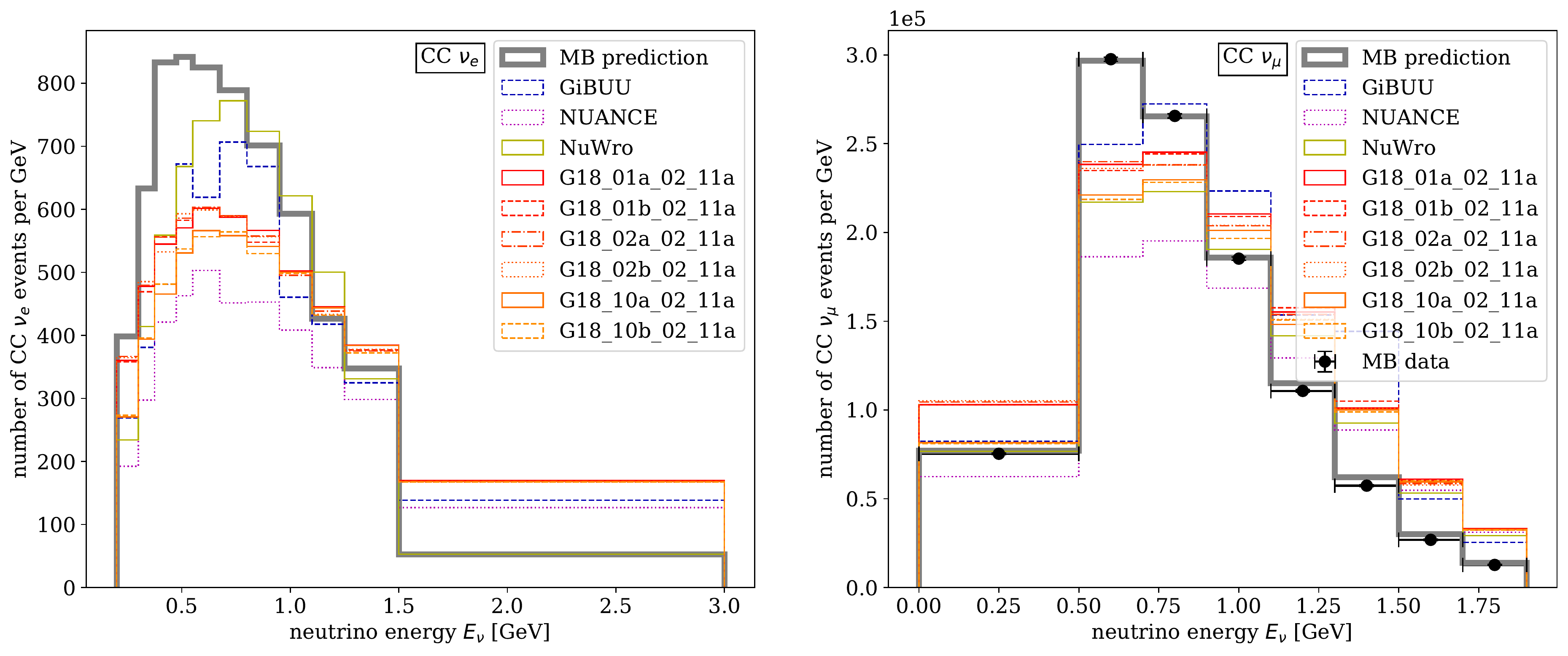}
  \caption{(a) Monte Carlo-only predictions 
    for the CC $\nu_e$ background to MiniBooNE's $\nu_e$ appearance search from
    different Monte Carlo event generators, in particular \gibuu (blue
    dashed), \nuance (purple dotted), \nuwro (yellow solid), and \genie
    (orange/red) with different tunes as explained in
    \cref{tab:generators}. The solid gray histogram corresponds to the official
    background prediction by the MiniBooNE collaboration. (b) Monte Carlo-only
    predictions for the CC $\nu_\mu$ events that are used for flux
    normalization. We show results for the same generators and tunes as in (a),
    but we also compare to MiniBooNE's data (black points with error bars).
  }
  \label{fig:ccqe-spectra}
\end{figure}

We compare in \cref{fig:ccqe-spectra} our predicted $E_\nu$ spectra (colored
histograms) to the ones used by MiniBooNE \cite{Aguilar-Arevalo:2018gpe} (gray
histograms).  For the case of $\nu_\mu$ interactions (which are observed
essentially without backgrounds), we also compare to data (black points with
error bars in \cref{fig:ccqe-spectra}).  Focusing first on the differences
between different event generators and tunes, we observe that predictions vary
by $\mathcal{O}(10\%)$.  One striking observation is that \nuwro predicts
relatively large CC $\nu_e$ rates compared to the other generators, while for
CC $\nu_\mu$ interactions, its predictions are among the lowest. 
\gibuu's predictions are overall relatively large, which is a reflection of the
well-known fact that \gibuu predicts lower pion production rates than observed
in MiniBooNE (while being consistent with the pion production rates in
MINER$\nu$A and T2K) \cite{Leitner:2008wx, Lalakulich:2012cj, Mosel:2016cwa}.
Here, this deficit means that a larger number of CC interactions will be
identified as CCQE, and fewer will be vetoed because of the presence of extra
pions.  Regarding the comparison between our predictions and MiniBooNE's there
are certain discrepancies; namely, MiniBooNE predicts somewhat higher event
rates compared to us both in the $\nu_e$ channel, and their predicted spectrum
is more peaked in the $\nu_\mu$ channel.  This indicates that our simplified
cuts do not fully capture MiniBooNE's true acceptance and efficiency.  As
discussed above, for the purpose of the sterile neutrino fits which we will
present in \cref{sec:results}, we will eliminate this bias by applying
additional energy-dependent tuning factors which are obtained as the ratio of
MiniBooNE's prediction and our prediction using \nuance.  That way, we ensure
that using the same generator as MiniBooNE, our predictions exactly match
theirs.  After this tuning, the differences between our predictions expose the
differences between event generators while being fairly robust with respect to
the simplifications of our analysis. 

\subsection{Neutral Current \boldmath{$\pi^0$} Production}
\label{sec:fp-pi0}

\noindent
Neutral pions are frequently produced in neutrino interactions. Of particular
concern to MiniBooNE's $\nu_e$ appearance search are neutral current
interactions of the form $\nu + N \to \nu + N + \pi^0$. In this case, the
$\pi^0$, or rather the two photons into which it promptly decays, are the only
visible interaction products. If one of the photons leaves the fiducial volume
before showering, or if the laboratory frame opening angle between the two
photons is small, the event will contain a single electromagnetic shower that
can be mistaken for an $e^\pm$ from a CC $\nu_e$ or $\bar\nu_e$ interaction.

To predict the contribution of NC $\pi^0$ events to the $\nu_e$ background, we
proceed as follows.  First, out of all simulated neutrino events, we select
those which have one or several $\pi^0$'s in the final states, no $e^\pm$ or
$\mu^\pm$, and no other charged particles above the \v{C}erenkov threshold.  We
then generate the photons from $\pi^0$ decay, and we apply Gaussian energy and
angular smearing to their 4-momentum vectors. We use a $10^\circ$ angular
resolution, and an energy resolution given by $(\Delta_E/E)^2 = [0.08
\sqrt{\si{GeV} / E}]^2 + [0.024 / (E/\si{GeV})]^2$ \cite{MBtalk}.

On average, photons propagate around \SI{50}{cm} before converting to an
$e^+e^-$ pair and starting an electromagnetic shower \cite{pdg}.  Therefore,
we pick the point at which this happens by randomly drawing from an exponential
distribution with a mean of \SI{50}{cm}.
Photons converting outside the detector volume
are discarded. If a photon converts in the veto region outside the fiducial
volume ($r_\text{fiducial} = \SI{574.6}{cm}$~\cite{AguilarArevalo:2008qa}),
but still inside the active volume ($r_\text{veto} = \SI{610}{cm}$), the whole
event is vetoed. The remaining events are assigned a weight factor
according to MiniBooNE's $e/\gamma$ efficiencies, published together with
ref.~\cite{Aguilar-Arevalo:2012fmn}.

\begin{figure}
  \centering
  \begin{tabular}{cc}
    \includegraphics[width=0.48\textwidth]{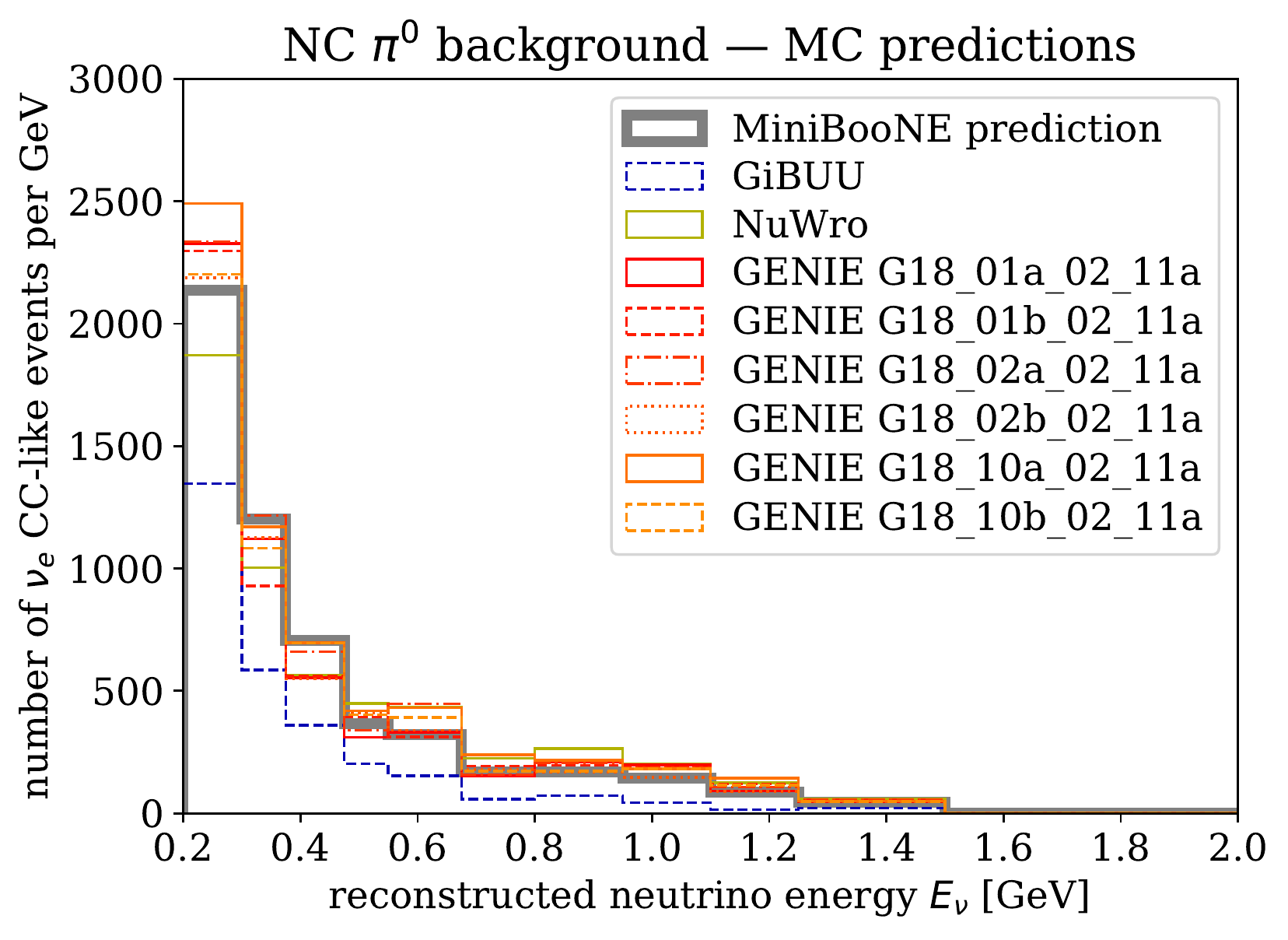} &
    \includegraphics[width=0.48\textwidth]{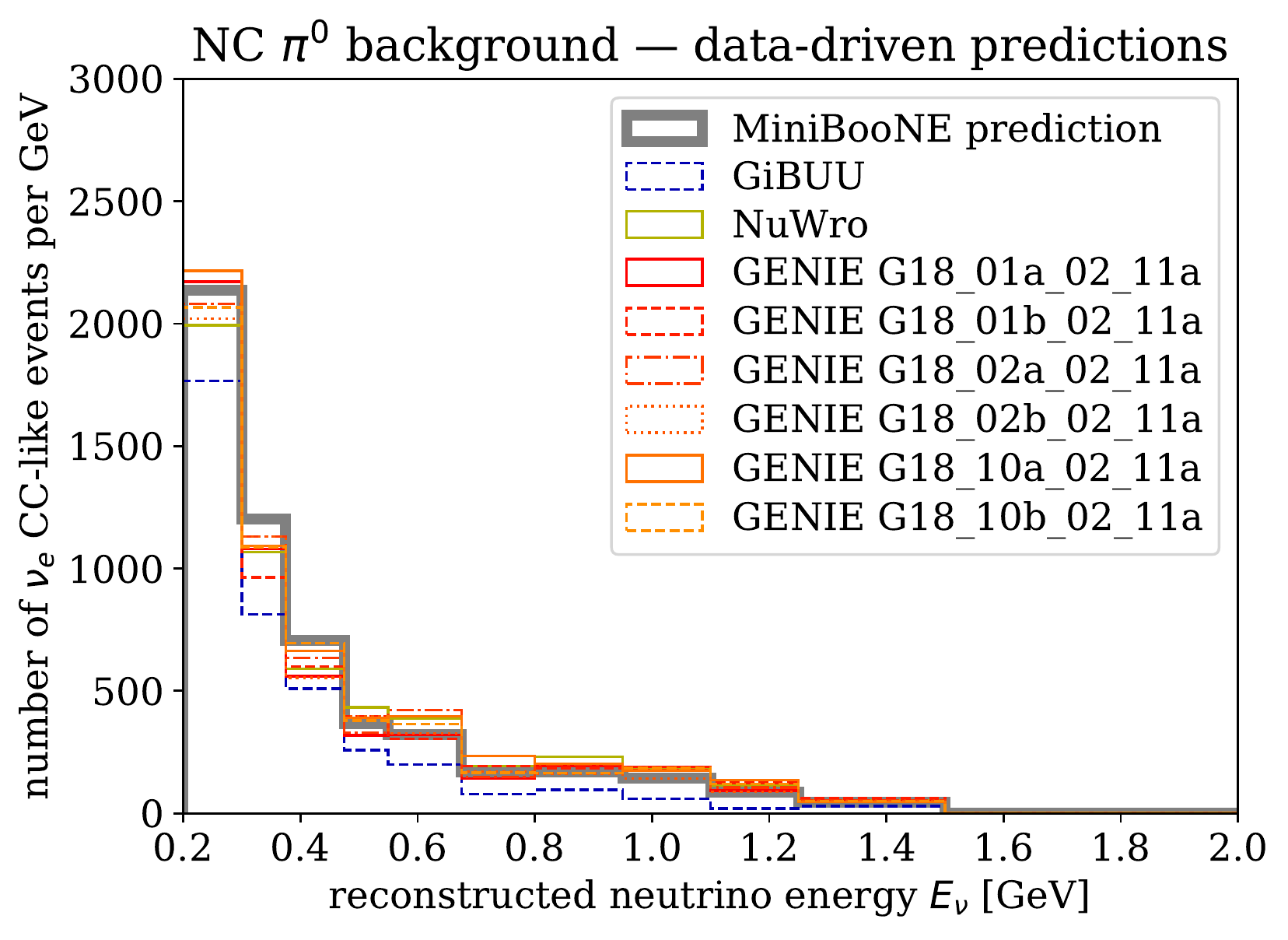} \\
    (a) & (b)
  \end{tabular}
  \caption{(a) Monte Carlo-only predictions and (b) data-driven predictions
    for the NC $\pi^0$ background to MiniBooNE's $\nu_e$ appearance search from
    different Monte Carlo event generators, in particular \gibuu (blue dashed),
    \nuwro (yellow solid), and \genie (orange/red) with different tunes as
    explained in \cref{tab:generators}. We do not show \nuance results here because
    they are used for tuning our analysis and thus would not be independent,
    The solid gray histogram corresponds to the official background prediction
    by the MiniBooNE collaboration.
  }
  \label{fig:NCpi}
\end{figure}

If there are two or more photons left in an event, we need to determine whether
they can be reconstructed separately or if they merge into one.  We do so by
applying a cut on the opening angle $\phi$ between pairs of photons. If $\phi$
is below a threshold $\phi_\text{thr}$, the two photons are merged into one. If
not, they are kept separate, and the algorithm continues to consider the next
pair of photons.  If, at the end of this procedure, exactly one photon is left,
the event is considered as a fake $\nu_e$ event, contributing to the
background in the $\nu_\mu \to \nu_e$ oscillation search. Otherwise, the event
is discarded.  We allow the threshold $\phi_\text{thr}$ to depend on the
reconstructed neutrino energy $E_\text{reco}$ that would be assigned to the
event according to \cref{eq:Enu-qe} if the photon-induced electromagnetic
shower was mis-interpreted as originating from an electron and the event was
mis-reconstructed as a CCQE $\nu_e$ interaction, with all photons merged into
one. In each $E_\text{reco}$ bin, we choose $\phi_\text{thr}(E_\text{reco})$
such that our prediction for the NC $\pi^0$ background using the \nuance
generator agrees exactly with MiniBooNE's prediction (likewise based on
\nuance) in that channel. We find that the resulting
$\phi_\text{thr}(E_\text{reco})$ decreases with energy.  With this procedure,
our \nuance prediction by construction matches exactly MiniBooNE's, while our
\genie, \nuwro, and \gibuu predictions differ from it, highlighting the
discrepancies between generators.

Our results for the NC $\pi^0$ background prediction are shown in
\cref{fig:NCpi}~(a). We see that differences between \nuance, \nuwro, and
\genie (with any tune) are small, while the \gibuu prediction is significantly
lower. As mentioned already in \cref{sec:CC-fp}, this discrepancy between
\gibuu's predictions and MiniBooNE data on single-pion production is well known
\cite{Leitner:2008wx, Lalakulich:2012cj, Mosel:2016cwa}, but given that a
similar discrepancy does not exist when comparing \gibuu to MINER$\nu$A and T2K
data, it is an open question whether it indicates a problem on the theory side
or on the experimental side.  Taking the \gibuu predictions at face value
would even increase the significance of MiniBooNE's low-E $\nu_e$ excess.

\subsection{Neutral Current Single \boldmath{$\gamma$} Production}
\label{sec:fp-single-photon}

\noindent
We next discuss single photon events. Most of these arise from radiative decays
of heavy hadronic resonances created in NC neutrino interactions, which is why
this background is referred to as the $\Delta \to N \gamma$ background in
MiniBooNE publications. Nevertheless, our event selection procedure described in the
following picks up any single-photon production channel included in the
generators' output, but not to single-photon production that happens outside
the primary target nucleus (such as the subdominant process $\pi N \to \gamma N$ scattering).
\footnote{Recently, a novel but again subdominant single
photon production channel was presented in \cite{Chanfray:2021wie}.}

We select simulated events that contain exactly one photon in the final state,
no electrons or muons, and no other charged charged
particles above the \v{C}erenkov threshold. We apply the same energy and angular
smearing as in \cref{sec:fp-pi0}. Each event is then (mis)reconstructed as
a CC $\nu_e$ interaction, misinterpreting the photon as an electron and
applying \cref{eq:Enu-qe} to determine the reconstructed neutrino energy
$E_\text{reco}$.  We finally determine an $E_\text{reco}$-dependent reconstruction
efficiency factor by demanding that, in each $E_\text{reco}$ bin, our \nuance
prediction matches MiniBooNE's prediction for the $\Delta \to N \gamma$ background.
We find efficiency factors of order 10--20\%, not too different from the
$e^\pm/\gamma$ efficiencies from the supplemental material of
ref.~\cite{Aguilar-Arevalo:2012fmn}, see \cite{Aguilar-Arevalo:2021odc}.

\begin{figure}
  \centering
  \includegraphics[width=0.6\textwidth]{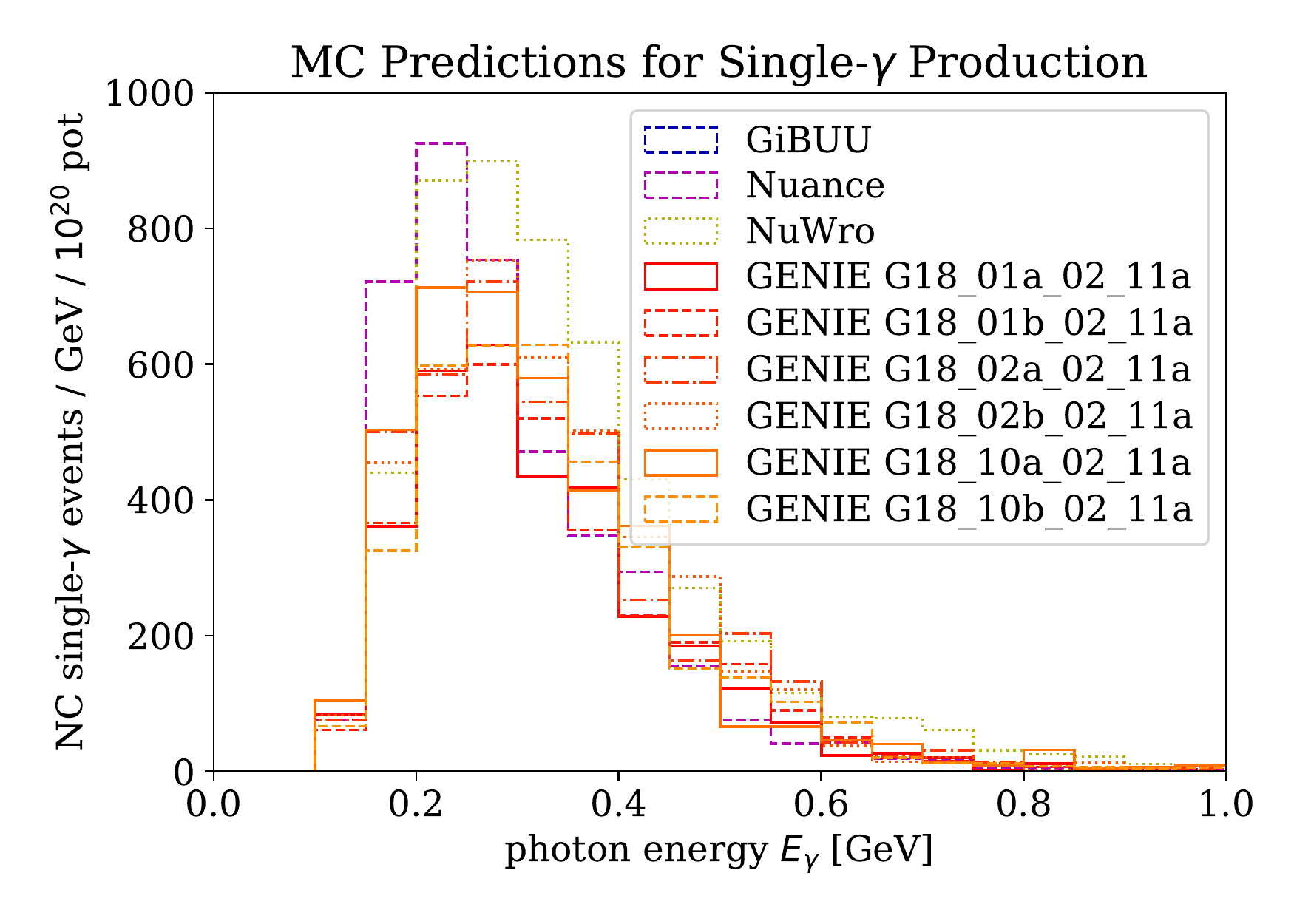}
  \caption{Predictions for the single-photon background in MiniBooNE from
    different event generators as a function of the photon energy.  We have used
    the electron/photon efficiencies published together with
    ref.~\cite{Aguilar-Arevalo:2012fmn} . The color
    scheme used here is the same as in \cref{fig:NCpi}.}
  \label{fig:single-gamma-spectrum}
\end{figure}

\begin{figure}
  \centering
  \begin{tabular}{cc}
    \includegraphics[width=0.48\textwidth]{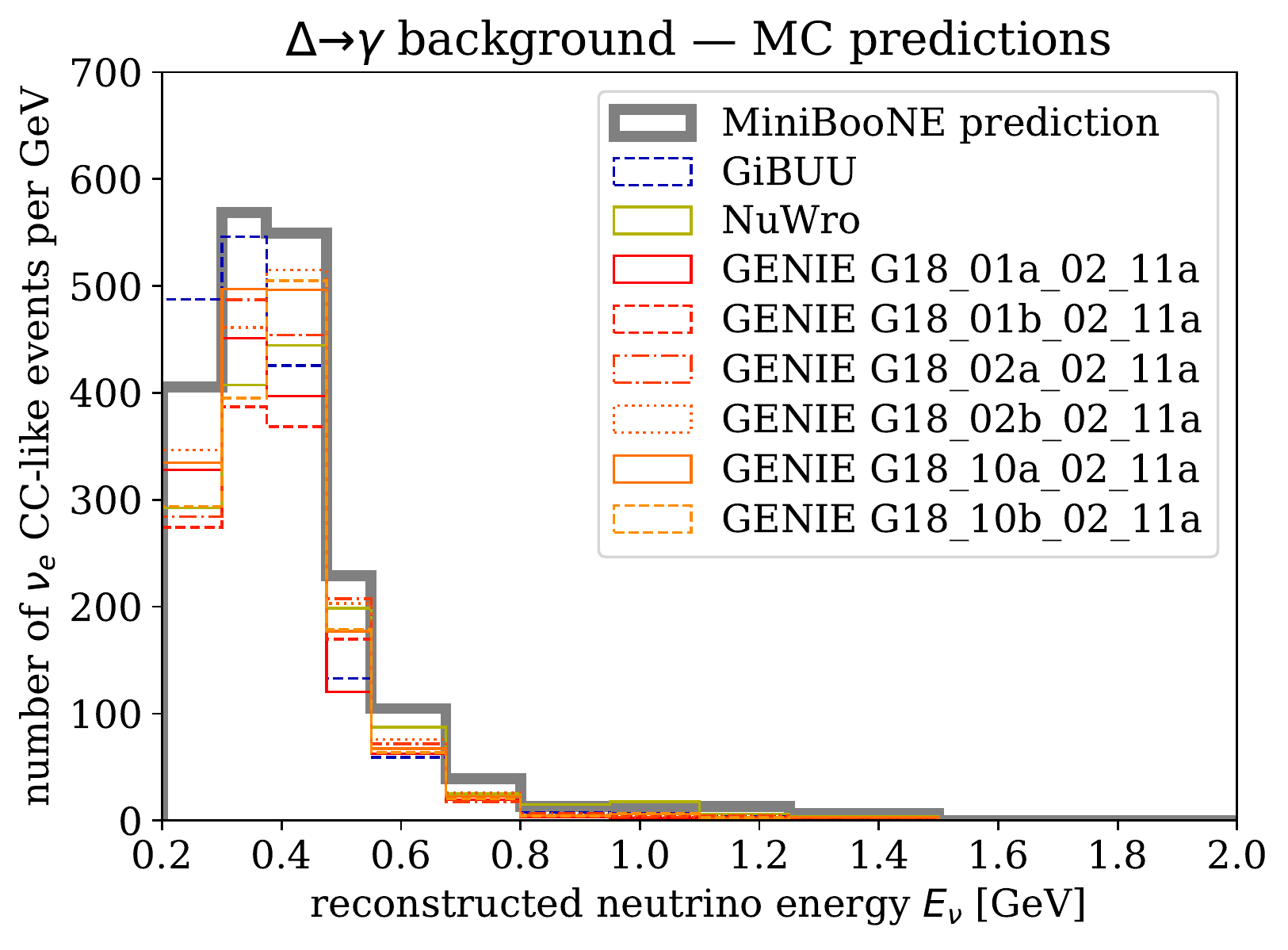} & 
    \includegraphics[width=0.48\textwidth]{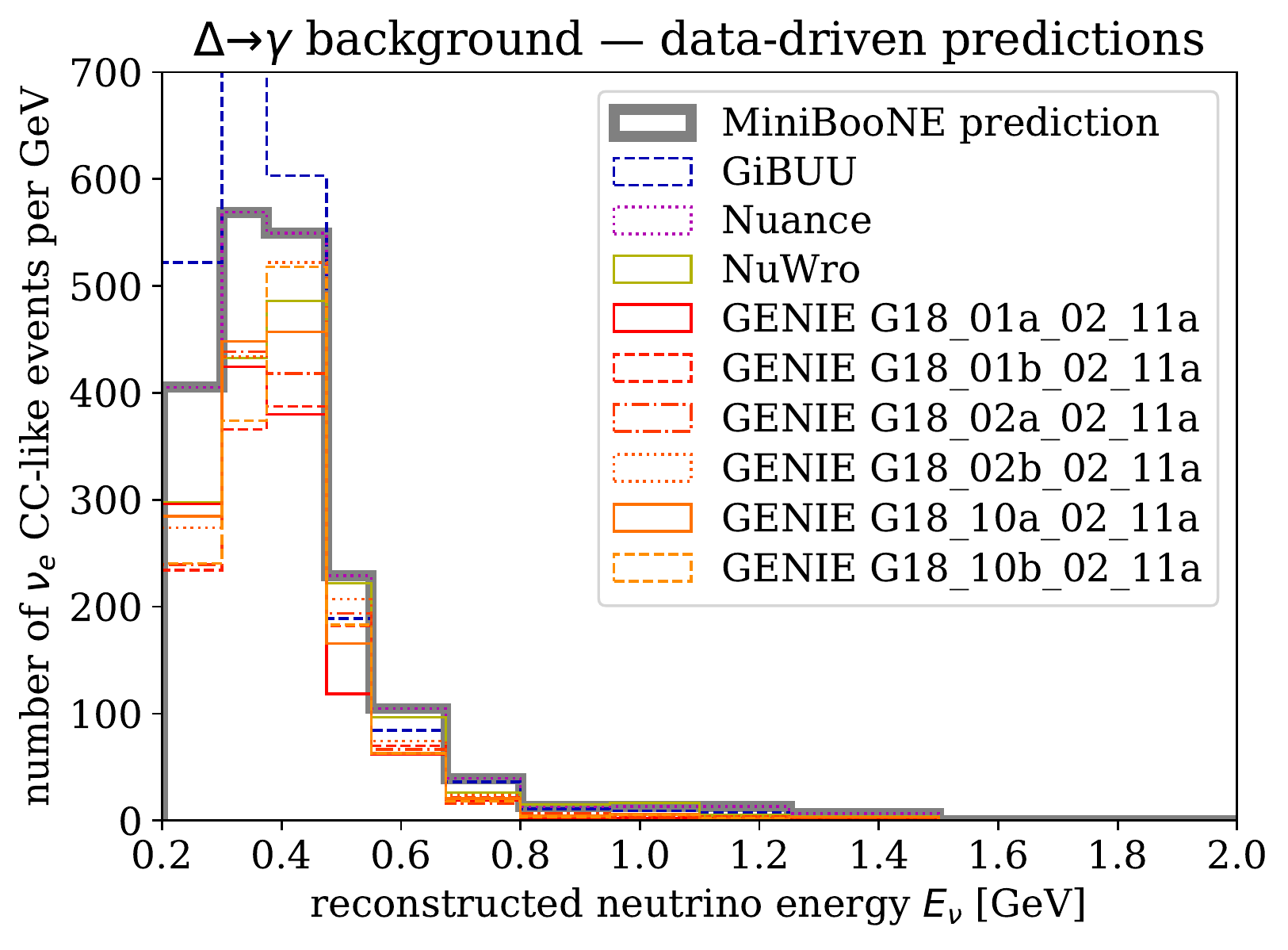} \\
    (a) & (b)
  \end{tabular}
  \caption{(a) Monte Carlo-only predictions and (b) data-driven predictions for
    the NC single-photon background to MiniBooNE's $\nu_e$ appearance search from
    different Monte Carlo event generators, in particular \gibuu (blue dashed),
    \nuance(purple dotted), \nuwro (yellow solid), and \genie (orange/red) with
    different tunes as explained in \cref{tab:generators}.  MiniBooNE's own
    prediction for this background is shown as a thick gray histogram.  We do
    not show Monte Carlo-only predictions from \nuance because, given the way
    our analysis is designed, they agree exactly with MiniBooNE's prediction.  
  }
  \label{fig:NCgamma}
\end{figure}

The comparison between event generators is shown in \cref{fig:single-gamma-spectrum}
as a function of photon energy, and in \cref{fig:NCgamma}~(a) as a function of
reconstructed neutrino energy.  We
find that \gibuu's, \nuwro's, and \genie's single-photon spectra are 10--20\%
lower than MiniBooNE's official, \nuance-based, background prediction.  The
older \genie tunes ({\tt G18\_01a\_02\_11a} and {\tt G18\_01b\_02\_11a}) seem
to give the lowest single-photon yield.

In passing, let us note that for \genie, we found an unusually large number
of events with two photons from the decays of an $\eta$ resonance. Other
generators predict far fewer such events, a discrepancy which might be due to
differences in hadronization models. Luckily, we find that the probability for
missing one of the two photons from $\eta$ decay is small. Therefore, even in
the \genie simulation, photons from $\eta$ decay account only for a handful
of background events in the $\nu_e$ appearance search. 

\section{Data-Driven Background Estimates}
\label{sec:dd}

\noindent
While our comparison of different Monte Carlo predictions for the MiniBooNE
backgrounds in \cref{sec:channels} reveals important discrepancies between the
various generators, MiniBooNE's own background prediction is to some extent
resilient to these discrepancies. This is because it is based on data-driven
techniques, meaning that certain crucial aspects, for instance
the $\pi^0$ or $\Delta(1232)$ production rate are directly measured in control
samples rather than being predicted theoretically. However, theoretical input
is still needed for translating measurements in the control regions to
background predictions for the signal region. Therefore, even data-driven
background estimation techniques are not fully immune to theoretical
uncertainties. In the following, we will investigate these uncertainties for
the $\pi^0$ and single photon backgrounds in MiniBooNE.

\subsection{Neutral Current \boldmath{$\pi^0$} Production}
\label{sec:dd-pi0}

\noindent
To emulate the data-driven estimation technique for the $\pi^0$ background, we 
select from our simulated events those containing exactly one $\pi^0$. Other than
that, events need to satisfy the same criteria as in \cref{sec:fp-pi0}: no charged
leptons ($e^\pm$ or $\mu^\pm$) and no charged particles above the \v{C}erenkov threshold
are allowed. We assume that the smearing kernel for pions is the same as for
photons, see \cref{sec:fp-pi0}.

\begin{figure}
  \centering
  \includegraphics[width=0.6\textwidth]{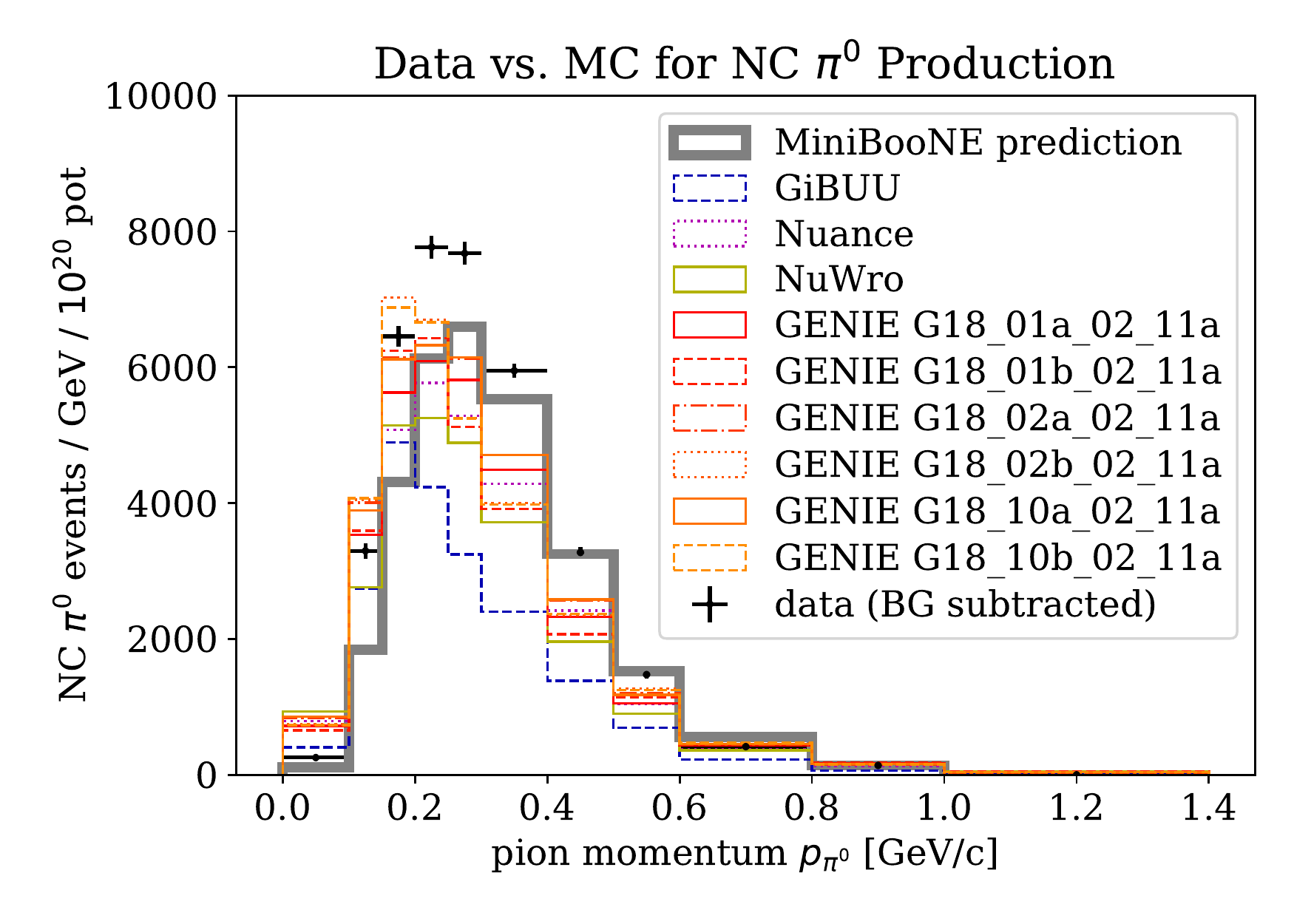}
  \caption{Comparison of MiniBooNE's data on NC $\pi^0$ production from
    ref.~\cite{AguilarArevalo:2009ww} to predictions from various
    Monte Carlo event generators as a function of the pion momentum
    $p_{\pi^0}$. The color scheme used here is the same as in
    \cref{fig:NCpi,fig:NCgamma}.  MiniBooNE's own
    prediction for the $\pi^0$ spectrum is shown as a thick gray
    histogram.
  }
  \label{fig:pi0-spectrum}
\end{figure}

We first compare the rate of $\pi^0$ production as a function of the $\pi^0$
momentum, $p_{\pi^0}$, to MiniBooNE's measurement of the $\pi^0$ spectrum
published in ref.~\cite{AguilarArevalo:2009ww}.  To do so, we apply the
efficiency factors given in fig.~5 of that reference. The result of the
comparison is shown in \cref{fig:pi0-spectrum}, where MiniBooNE's prediction
for the background (mostly mis-identified CC events, events with $\pi^\pm$, and
multi-pion events) has been subtracted from the data. We see that Monte Carlo
predictions vary by $\mathcal{O}(50\%)$, with \nuance (purple), \nuwro
(yellow), and \genie (orange/red) matching the data well, but not perfectly,
while the event rate predicted by \gibuu is too low and the spectrum is too
soft.  In other words, we observe again the well-known discrepancy mentioned in
\cref{sec:CC-fp,sec:fp-pi0} between \gibuu's predictions and MiniBooNE data on
single-pion production \cite{Leitner:2008wx, Lalakulich:2012cj, Mosel:2016cwa}.
We also note that MiniBooNE's own Monte Carlo prediction is on the low side,
showing that even with a full detector simulation, the data on NC $\pi^0$
production is very difficult to understand theoretically.

To obtain a data-driven prediction for the $\pi^0$ background to the $\nu_e$
appearance search, we extract a signal sample of $\pi^0$s faking a $\nu_e$
from the simulation using the same criteria as in \cref{sec:fp-pi0}.
The signal sample is then binned in $p_{\pi^0}$ using the same bins
as in the control sample, namely the 11 bins shown in \cref{fig:pi0-spectrum}.
Next, we reweight the signal sample in each of these bins with the ratio
of observed to simulated single-$\pi^0$ events in the control sample.  The thus
reweighted signal sample is what we call our data-driven prediction.

The result of (mis)reconstructing the data-driven background sample as CCQE $\nu_e$
interactions is shown in \cref{fig:NCpi}~(b) for the different event generators
and tunes.  We observe that the spread in our results is reduced compared to
the Monte Carlo-only predictions in \cref{fig:NCpi}~(a).  In particular, the
discrepancy between the \gibuu-based prediction and MiniBooNE's own prediction
is not quite as bad in panel (b) as in panel (a).  Among the other generators,
the spread is $\lesssim 10\%$.

\subsection{Neutral Current Single \boldmath{$\gamma$} Production}
\label{sec:dd-single-photons}

\noindent
The procedure we follow to determine the impact of different Monte Carlo
generators on the data-driven prediction for the single-photon background is
very similar to the one employed in the case of the $\pi^0$ background above.
The control sample consists once again of single-$\pi^0$ events, given that the
neutral $\Delta(1232)$ resonance, which is responsible for most of the
single-photon background, predominantly decays to pions. The signal sample is
extracted from the simulation using the same criteria as in
\cref{sec:fp-single-photon}. Both the $\pi^0$ control sample and the single-photon
signal sample are then binned according to the $\pi^0$ and photon
momentum, respectively, using the binning from \cref{fig:pi0-spectrum}.  Next, we reweight
the simulated single-photon events in each of these bins with the ratio of
observed to simulated single-$\pi^0$ events in the control sample.
Finally, we (mis-)reconstruct the reweighted single-photon events as CC $\nu_e$
interactions to obtain our data-driven background prediction to the $\nu_e$
appearance search, binned in $E_\text{reco}$.

The result is shown in \cref{fig:NCgamma}~(b). In contrast to what we observed
in \cref{sec:dd-pi0} for the $\pi^0$ background, we now find that the data-driven
technique does \emph{not} significantly decrease the spread between predictions
from different generators. This indicates that, for the single-photon background,
large theoretical uncertainties exist which are not related to the $\Delta(1232)$
production cross-section. Instead, they are due to other sources
of single-photon events, not related to the $\Delta$ resonance, such
as coherent photon production, decays of heavier resonances, etc.
Reducing these uncertainties may be possible with more sophisticated data-driven methods
separating different sources of single-photon events, and identifying suitable
control samples for each of them. However, even if this was possible, we
expect significant cross-contamination between the different control samples. For
instance, it will be well-nigh impossible to fully separate different heavy baryonic
resonances based only on their visible decay products.  Therefore, any significant
reduction in systematic uncertainties in the single photon channel will be very
challenging. Our analysis thus shows that systematic uncertainties in the
single-photon channel could play an important role in understanding the MiniBooNE
$\nu_e$ appearance anomaly.

\section{Uncertainties in the Radiative Branching Ratios of Heavy Baryonic Resonances}
\label{sec:br-scans}

\noindent
We would finally like to discuss an additional aspect that could contribute to
the MiniBooNE anomaly: the branching ratios for radiative decays of heavy
baryonic resonances ($\Delta(1232)$, $N(1440)$, etc.) are
uncertain. As an example, the Particle Data Group quotes
$\BR(\Delta(1232) \to N\gamma) = \text{0.55\%--0.65\%}$,
$\BR(p(1440) \to p\gamma)      = \text{0.035\%--0.048\%}$,
$\BR(n(1440) \to n\gamma)      = \text{0.02\%--0.04\%}$,
$\BR(p(1520) \to p\gamma)      = \text{0.31\%--0.52\%}$, and
$\BR(n(1520) \to n\gamma)      = \text{0.30\%--0.53\%}$
\cite{Zyla:2020zbs}.  These branching ratios are inferred from
baryon--photon interaction amplitudes determined in pion--nucleon and
photon--nucleon scattering, see for instance ref.~\cite{Ronchen:2015vfa}.
We will in the following investigate the potential implications of branching
ratio uncertainties.

\begin{figure}
  \centering
  \includegraphics[width=0.6\textwidth]{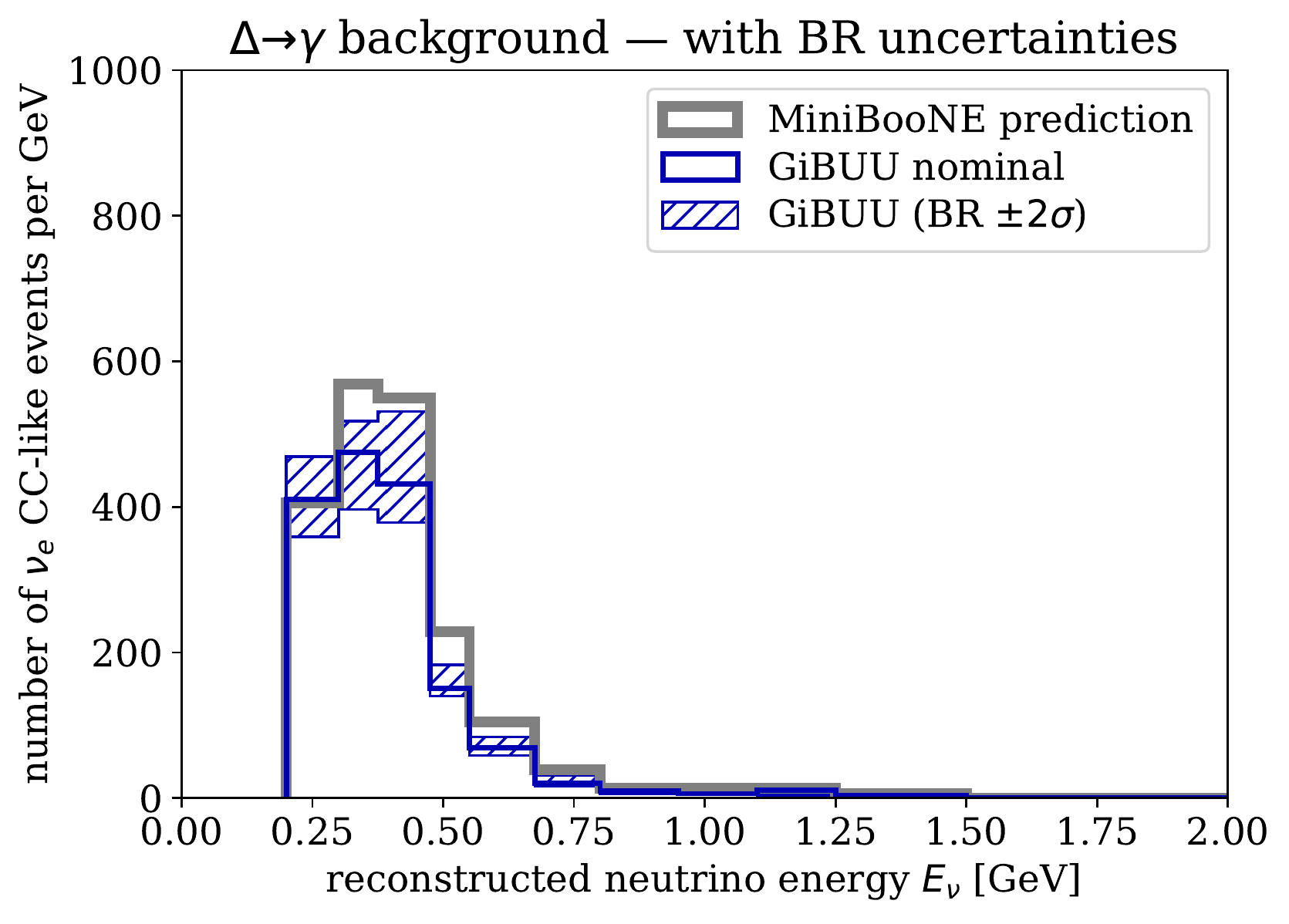}
  \caption{Potential impact of branching ratio uncertainties on \gibuu's prediction
    for the single-photon background in MiniBooNE.  We have varied the
    radiative branching ratios of the $\Delta(1232)$, $N(1440)$, and $N(1520)$ baryon
    resonances within the (probably very conservative) $2\sigma$ confidence intervals
    given in \cite{Zyla:2020zbs}, and we show the envelope of the resulting event
    spectra (blue hatched histogram).  For comparison, we also show MiniBooNE's
    \nuance-based prediction for this background \cite{AguilarArevalo:2009ww}
    (gray histogram).
  }
  \label{fig:br-scan}
\end{figure}

To begin, it is important to emphasize that state-of-the-art neutrino Monte Carlo
generators do not take uncertainties in decay
branching ratios into account, so these uncertainties need to be
carefully accounted for a posteriori.\footnote{We also note that decay data in
Monte Carlo generators is not always based on the latest version of the
Particle Data Group's compilation, which may add to the error on these data.}
We have done so by carrying out a set of 27
\gibuu runs in which the radiative branching ratios of the $\Delta(1232)$,
$N(1440)$, and $N(1520)$ resonances were varied within their $2\sigma$
uncertainty intervals from ref.~\cite{Zyla:2020zbs}.  We have analyzed the
resulting event samples using the methods described in
\cref{sec:fp-single-photon}.  In \cref{fig:br-scan}, we plot the envelope of
the 27 single-photon event spectra as a function of the (mis-)reconstructed
would-be neutrino energy (blue hatched region).  We see that the event rate
varies by $\mathcal{O}(10\%)$, especially in the low-energy bins where MiniBooNE
observes its anomaly.

We conclude that branching ratio uncertainties are non-negligible for making reliable
background predictions in a MiniBooNE-like experiment. If they are indeed as
large as given in ref.~\cite{Zyla:2020zbs}, they may make an
important contribution to the total error budget.  Whether or not they can be removed
by using data-driven techniques depends on their origin.  The helicity amplitudes
for $\Delta(1232) \to N\gamma$, for instance, are related to those for $\Delta(1232)
\to N \pi$, therefore a bias in these amplitudes will at least partially cancel between
the signal sample and a $\pi^0$ control sample.

\vspace{1ex}
To summarize our findings so far, we have identified a number of differences
between different Monte Carlo predictions of the MiniBooNE backgrounds.
Visual inspection of \cref{fig:ccqe-spectra,fig:NCpi,fig:NCgamma,fig:br-scan}
suggests that these discrepancies may alleviate the tension with the MiniBooNE
$\nu_e$ data, but will probably not be large enough to fully explain away
the observed event excess.  In the following, we will quantify this statement
by carrying out explicit fits in a $3+1$ sterile neutrino model and
determining the confidence level at which the no oscillation hypothesis
is excluded.

\section{Impact on Sterile Neutrino Fits}
\label{sec:results}

In fitting the MiniBooNE data, we will focus on the $3+1$
scenario, in which the Standard Model is extended by a single sterile neutrino
whose mass is assumed to be at the eV scale.  The only new interaction
is thus a Yukawa coupling of the form
\begin{align}
  \mathcal{L} \supset y \, (i \sigma^2 H^*) L N \,,
  \label{eq:neutrino-portal}
\end{align}
where $L$ is a Standard Model lepton doublet, $N$ is the sterile neutrino field,
$H$ is the Standard Model Higgs doublet, $\sigma^2$ is the second Pauli matrix,
and $y$ is a dimensionless coupling constant. All fermion fields are interpreted as
Weyl spinors here. In general, $y$ and $L$ carry flavor indices to allow for
different mixing between $N$ and each of the three Standard Model neutrino
flavors.  \Cref{eq:neutrino-portal} implies that the leptonic mixing matrix
$U$ is extended to a unitary $4 \times 4$ matrix, while apart from this
modification the standard expression for the neutrino oscillation probabilities
remains unchanged
\begin{align}
  P_{\alpha\beta} = \sum_{j,k} U_{\alpha j}^* U_{\beta j} U_{\alpha k} U_{\beta k}^*
                    e^{-i \Delta m_{jk}^2 L / (2 E)} \,.
  \label{eq:Pab}
\end{align}
Here, as usual $\Delta m_{jk}^2 \equiv m_j^2 - m_k^2$ is the difference between
the squared masses of neutrino mass eigenstates $j$ and $k$.

The $3+1$ scenario has been
extensively studied in the context of the MiniBooNE anomaly (and the other
short-baseline anomalies), see for instance refs.~\cite{Kopp:2011qd, Conrad:2012qt,
  Archidiacono:2013xxa, Kopp:2013vaa, Mirizzi:2013kva, Giunti:2013aea,
Gariazzo:2013gua, Collin:2016rao, Gariazzo:2017fdh, Giunti:2017yid,
Dentler:2017tkw, Dentler:2018sju, Giunti:2019sag}.  These fits have revealed
significant tension in the global data, caused mainly by the fact that
MiniBooNE (and LSND) suggest relatively large mixing between the sterile
neutrino $\nu_s$ and both $\nu_e$ and $\nu_\mu$. It is in particular the
$\nu_s$--$\nu_\mu$ mixing that is strongly constrained by $\nu_\mu$
disappearance searches such as the ones in MINOS/MINOS+~\cite{Adamson:2017uda},
IceCube \cite{TheIceCube:2016oqi, Jones:2015, Arguelles:2015}, and MiniBooNE
itself \cite{AguilarArevalo:2009yj, Cheng:2012yy}.  The scenario is also
constrained by cosmology, in particular by Big Bang Nucleosynthesis
\cite{Cyburt:2015mya} and by CMB+structure formation data \cite{Ade:2015xua},
though some or all of these constraints can be avoided in extended cosmological
scenarios \cite{Dasgupta:2013zpn, Hannestad:2013ana, Chu:2018gxk, Yaguna:2007wi,
Saviano:2013ktj, Giovannini:2002qw, Bezrukov:2017ike, Farzan:2019yvo,
Cline:2019seo, Dentler:2019dhz}. In fact adding sterile neutrinos
may even help alleviate the tension between local and cosmological determinations
of the Hubble constant \cite{Archidiacono:2020yey}.

In the following, we address two important features of fits to the MiniBooNE data:
the differences between a full four-flavor fit compared to the
two-flavor fits presented for instance in refs.~\cite{Aguilar-Arevalo:2013pmq,
Aguilar-Arevalo:2020nvw}, and the dependence of the fit results on the choice
of event generator for the background predictions.

\subsection{2-Flavor vs.\ 4-Flavor Fits to MiniBooNE Data}
\label{sec:fits-2f-vs-4f}

\begin{figure}
  \begin{tabular}{cc}
    \includegraphics[width=0.5\textwidth]{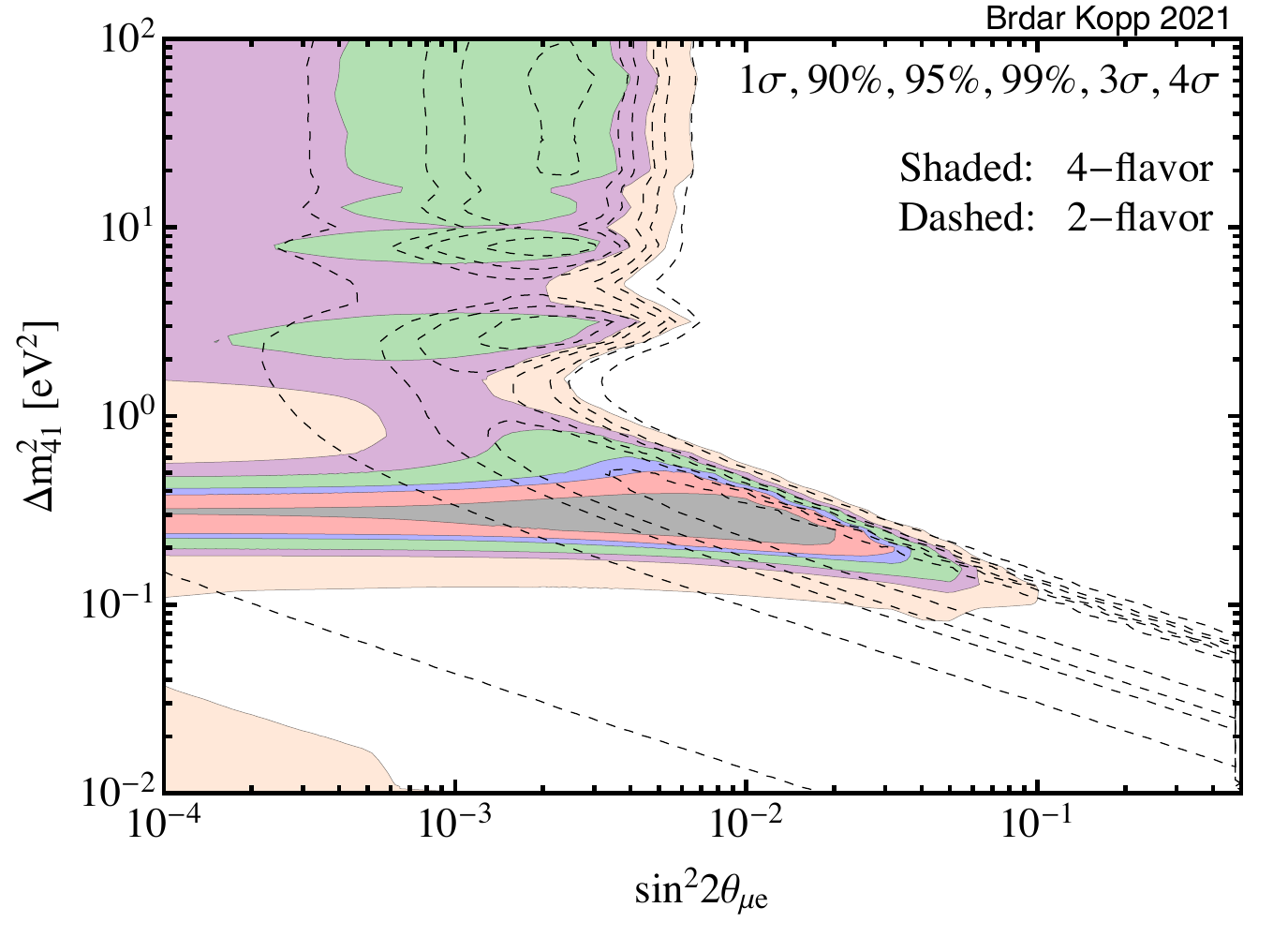} &
    \includegraphics[width=0.5\textwidth]{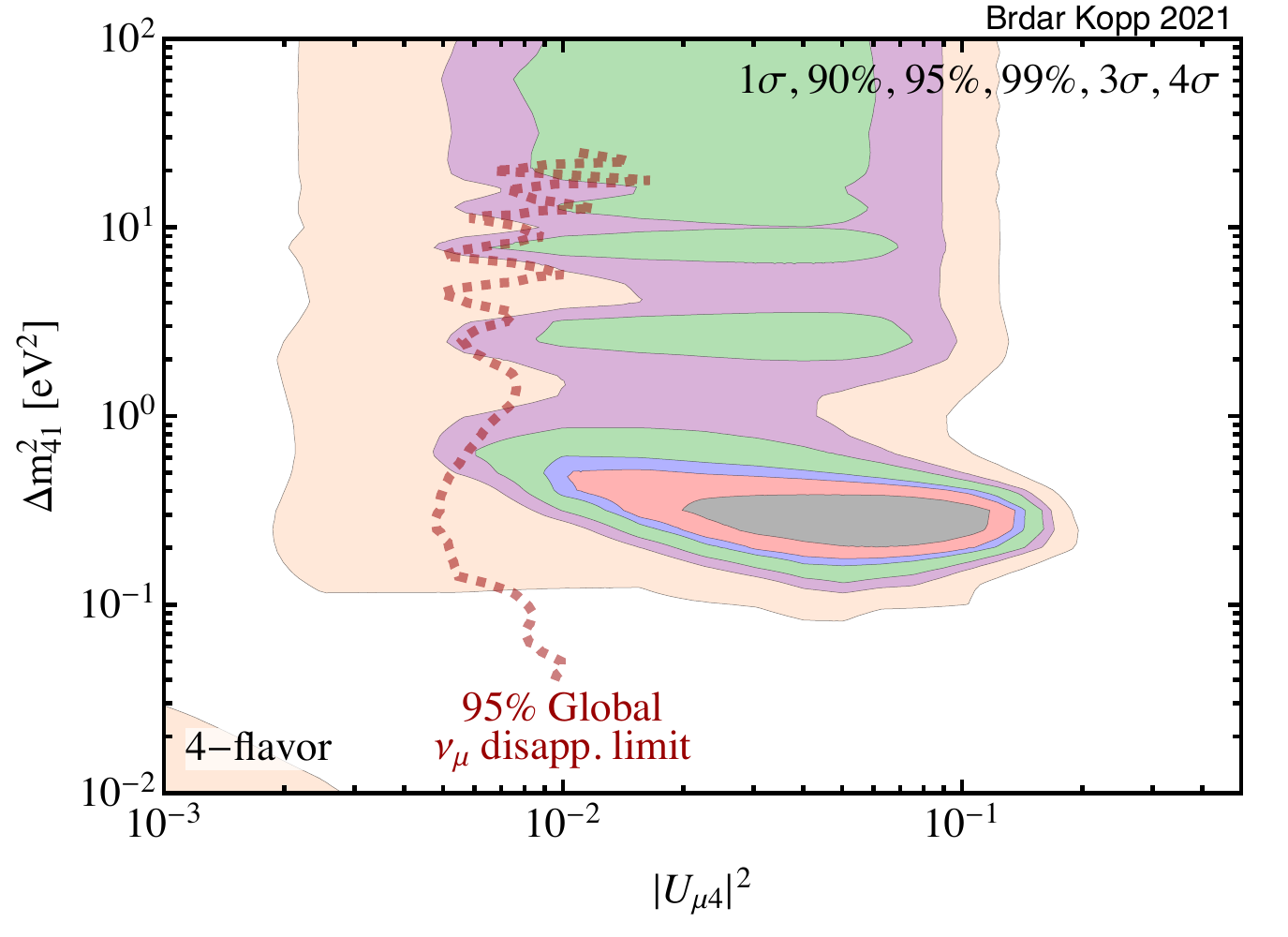} \\
    (a) & (b) \\[0.2cm]
    \includegraphics[width=0.5\textwidth]{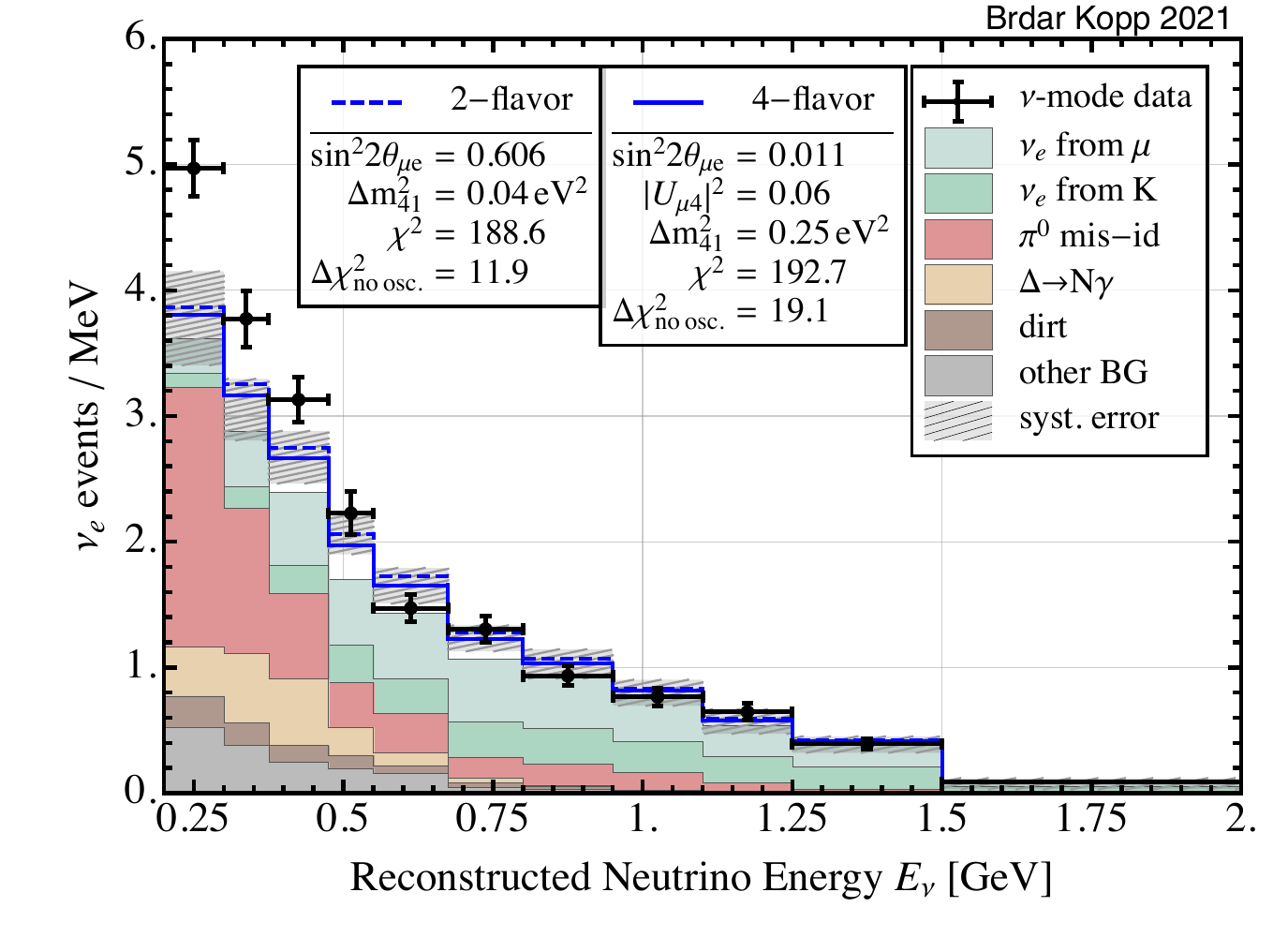} &
    \includegraphics[width=0.5\textwidth]{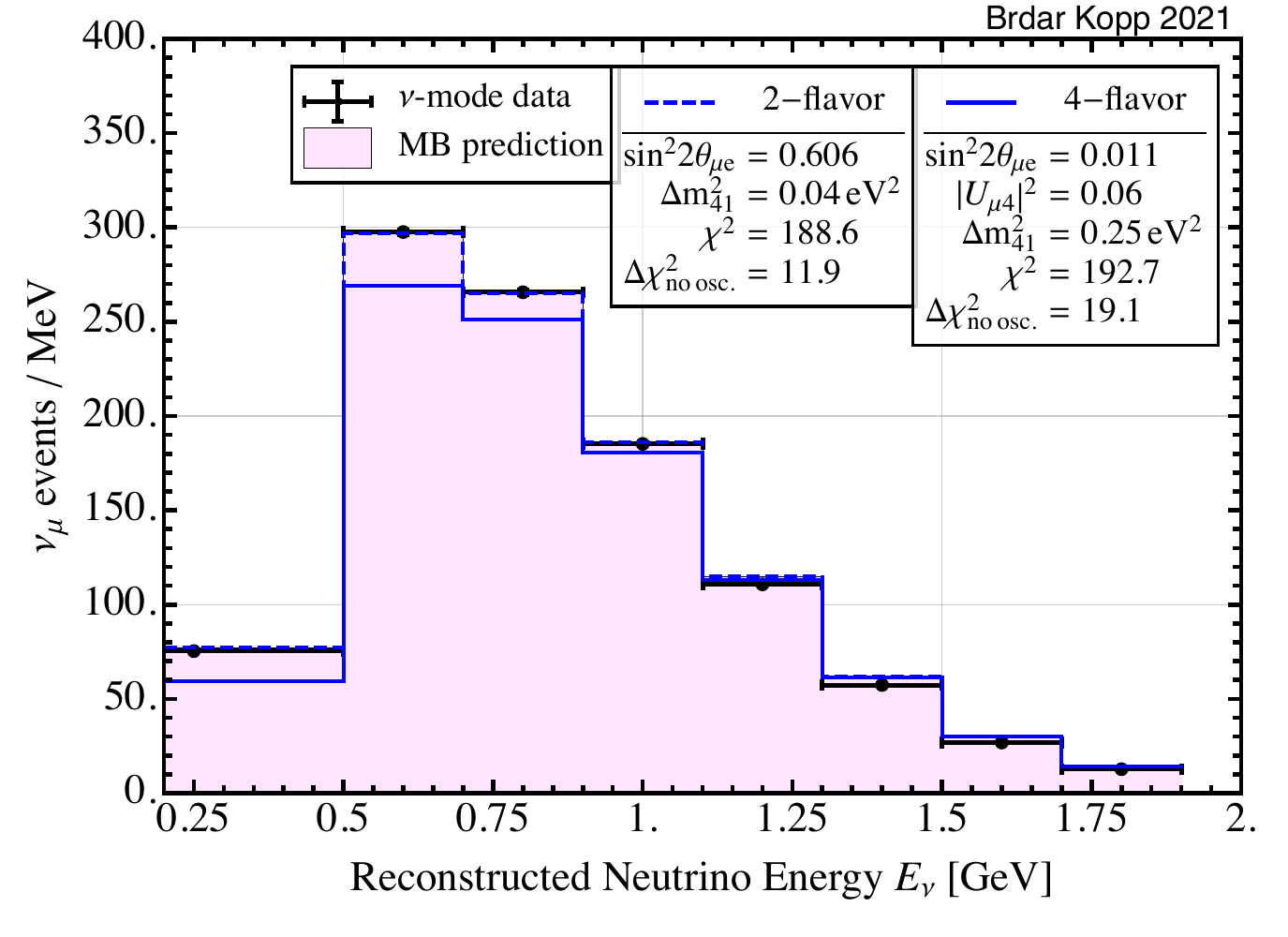} \\
    (c) & (d)
  \end{tabular}
  \caption{Comparison between two-flavor and four-flavor fits to MiniBooNE
    data.  In the two-flavor case, oscillations in the $\nu_e$
    background and in the $\nu_\mu$ control sample are negligible, while in the
    four-flavor case, $\nu_e \to \nu_s$ and $\nu_\mu \to \nu_s$
    disappearance can occur.  Because of the strong correlation between the $\nu_\mu$
    control sample and the $\nu_e$ signal sample, this leads to a shift in
    the best-fit region, with
    the four-flavor framework allowing for a good fit even in the absence of
    $\nu_\mu \to \nu_e$ oscillations, but with sizeable $|U_{\mu 4}|^2$, see
    shaded exclusion contours in panels (a) ($\sin^2 2\theta_{\mu e}$ vs.\
    $\Delta m_{41}^2$) and (b) ($|U_{\mu 4}|^2$ vs.\ $\Delta m_{41}^2$).
    The $\nu_e$ spectrum at the four-flavor best fit point (solid
    blue histogram in panel (c)) is very close to the background prediction, while
    the $\nu_\mu$ spectrum in panel (d) is suppressed by $\nu_\mu \to \nu_s$
    disappearance. In the two-flavor case (dashed histograms), on the other
    hand, the fit is driven by $\nu_\mu \to \nu_e$ appearance.
    Note that the background prediction
    shown in panel (c) is from the four-flavor fit, but the one for the two-flavor
    scenario is practically identical. Panel (b) does not contain contours for the
    two-flavor case because $|U_{\mu 4}|^2$ is not defined in that scenario.
    We show, however, the global exclusion limit on $\nu_\mu$ disappearance from
    ref.~\cite{Dentler:2018sju}.}
  \label{fig:mb-4f-fit}
\end{figure}

\noindent
To fit the $3+1$ model to MiniBooNE data, we use an adapted version of the
fitting code developed in refs.~\cite{Kopp:2011qd, Kopp:2013vaa,
Dentler:2018sju, Dentler:2019dhz}. It is based on the recommendations given by
the MiniBooNE collaboration in the supplemental material of
ref.~\cite{AguilarArevalo:2010wv} and uses the data released with
ref.~\cite{Aguilar-Arevalo:2018gpe}, see \cite{Aguilar-Arevalo:2021odc}.
However, in contrast to the fits
carried out in MiniBooNE's publications, we include the full
impact of four-flavor oscillations on the signal and background
prediction, as discussed in appendix~A of ref.~\cite{Dentler:2019dhz}.  In
particular, in a $3+1$ model, explaining MiniBooNE's $\nu_\mu \to \nu_e$
oscillation signal requires mixing between the sterile state and both $\nu_e$
and $\nu_\mu$.  $\nu_\mu \to \nu_e$ oscillations are thus necessarily
accompanied by $\nu_e \to \nu_s$ and $\nu_\mu \to \nu_s$ disappearance.  And
while the probability for the appearance signal is proportional to $|U_{e4}|^2
|U_{\mu 4}|^2$, the disappearance probabilities are only suppressed by
$|U_{e4}|^2$ and $|U_{\mu 4}|^2$.  $\nu_e$ and $\nu_\mu$ disappearance is thus
a non-negligible effect which is not captured by the two-flavor fits employed in
the official MiniBooNE analyses.  It has the following consequences:
\begin{enumerate}
  \item {\bf Oscillations in the $\nu_\mu$ control sample.}
    MiniBooNE's fit includes CC $\nu_\mu$ events as a control sample to fix
    the unoscillated neutrino flux and spectrum in a data-driven way.
    An oscillation-induced $\nu_\mu$ deficit, if not accounted for,
    will thus lead to too low a prediction for the intrinsic CC $\nu_e$
    background and for the $\nu_\mu \to \nu_e$ signal. In MiniBooNE's two-flavor fit,
    where a $\nu_\mu$ deficit is only due to $\nu_\mu \to \nu_e$ oscillation,
    this effect can be neglected. But in a realistic four-flavor model,
    a much larger $\nu_\mu$ deficit arises from $\nu_\mu \to \nu_s$ disappearance,
    leading to important corrections. We account for these corrections by scaling
    both the predicted oscillation signal and the CC $\nu_e$ background with the
    inverse of the $\nu_\mu$ disappearance probability.\footnote{In fact,
    following MiniBooNE's normalization strategy, the appropriate rescaling
    factor for $\nu_e$ from muon decay is the $\nu_\mu$ disappearance
    probability at MiniBooNE, while for $\nu_e$ from kaon decay, it is the
    corresponding probability at the SciBooNE baseline of \SI{100}{m}.} That
    way, we compensate the bias that is introduced when MiniBooNE calibrate
    these backgrounds to the observed $\nu_\mu$ rate. We do not rescale the
    other backgrounds (mostly $\pi^0$ and $\Delta \to \gamma$) because they
    are not normalized to the $\nu_\mu$ control sample, but to single-pion
    control samples.

  \item {\bf $\nu_e$ disappearance.} By the same reasoning as for $\nu_\mu$,
    also the oscillation-induced deficit of $\nu_e$ is much larger in a $3+1$
    model than in MiniBooNE's two-flavor scenario. This affects in particular
    the intrinsic CC $\nu_e$ background; we take this effect into account by
    rescaling said background with the $\nu_e$ disappearance probability.
\end{enumerate}
An interesting outcome of the four-flavor fit -- and in particular of the
inclusion of $\nu_\mu$ disappearance -- is a significant distortion in
MiniBooNE's best fit regions, as illustrated in \cref{fig:mb-4f-fit}~(a). In
particular, because of the strong correlation between the $\nu_e$ signal sample
and the $\nu_\mu$ control sample that is used to normalize the signal, a good
fit can be achieved not only by enhancing the $\nu_e$ flux, but also by
suppressing the $\nu_\mu$ flux.  Therefore the four-flavor best fit regions
(shaded contours in \cref{fig:mb-4f-fit})~(a) extend to much smaller $\sin^2
2\theta_{\mu e}=4\,|U_{e4}|^2 |U_{\mu 4}|^2 $ than in the two-flavor scenario
(unshaded dashed contours), at the expense of relatively large $|U_{\mu 4}|^2$,
see panel (b).  Correspondingly, the $\nu_e$ spectrum at the four-flavor
best-fit point (solid blue histogram in \cref{fig:mb-4f-fit}~(c)) is closer
to the background prediction, while the $\nu_\mu$ spectrum in panel (d)
is suppressed.  In the two-flavor scenario, on the other hand, the $\nu_e$ flux
needs to be enhanced as the $\nu_\mu$ flux remains unsuppressed.

Note that the four-flavor treatment \emph{increases} the significance of the
anomaly: our two-flavor fit disfavors the no-oscillation hypothesis at
$3.0\sigma$, while the four-flavor fit disfavors it at $4.0\sigma$. The lower
significance of our own fit compared to MiniBooNE's -- even when following
MiniBooNE's recommended approach (including the assumption of two-flavor
oscillations) as given in the supplemental material \cite{Aguilar-Arevalo:2021odc} to
refs.~\cite{AguilarArevalo:2010wv, Aguilar-Arevalo:2018gpe} --
has been noted before~\cite{Kopp:2011qd,
Kopp:2013vaa, Dentler:2018sju}. It implies that our results will be erring
slightly on the side of being conservative.

It should be kept in mind that $\nu_\mu$ disappearance is strongly
constrained by dedicated measurements, including measurements by MiniBooNE
itself \cite{AguilarArevalo:2009yj, Cheng:2012yy, MiniBooNE:2009ozf},
as well as IceCube \cite{TheIceCube:2016oqi,Jones:2015, Arguelles:2015},
DeepCore \cite{Aartsen:2014yll, deepcore:2016},
CDHS \cite{Dydak:1983zq}, SuperKamiokande \cite{Wendell:2010md, Wendell:2014dka},
NO$\nu$A, \cite{Adamson:2017zcg}, and MINOS/MINOS+ \cite{Adamson:2017uda}.
In \cref{fig:mb-4f-fit}~(b), we show as a dotted red line the combined
$\nu_\mu$ disappearance limit from ref.~\cite{Dentler:2018sju}.

In the following, we will always use the full four-flavor framework when
fitting MiniBooNE data.

\subsection{Dependence on Background Predictions}
\label{sec:fits-generator-dependence}

\begin{figure}
  \centering
  \begin{tabular}{cc}
    \includegraphics[width=0.5\textwidth]{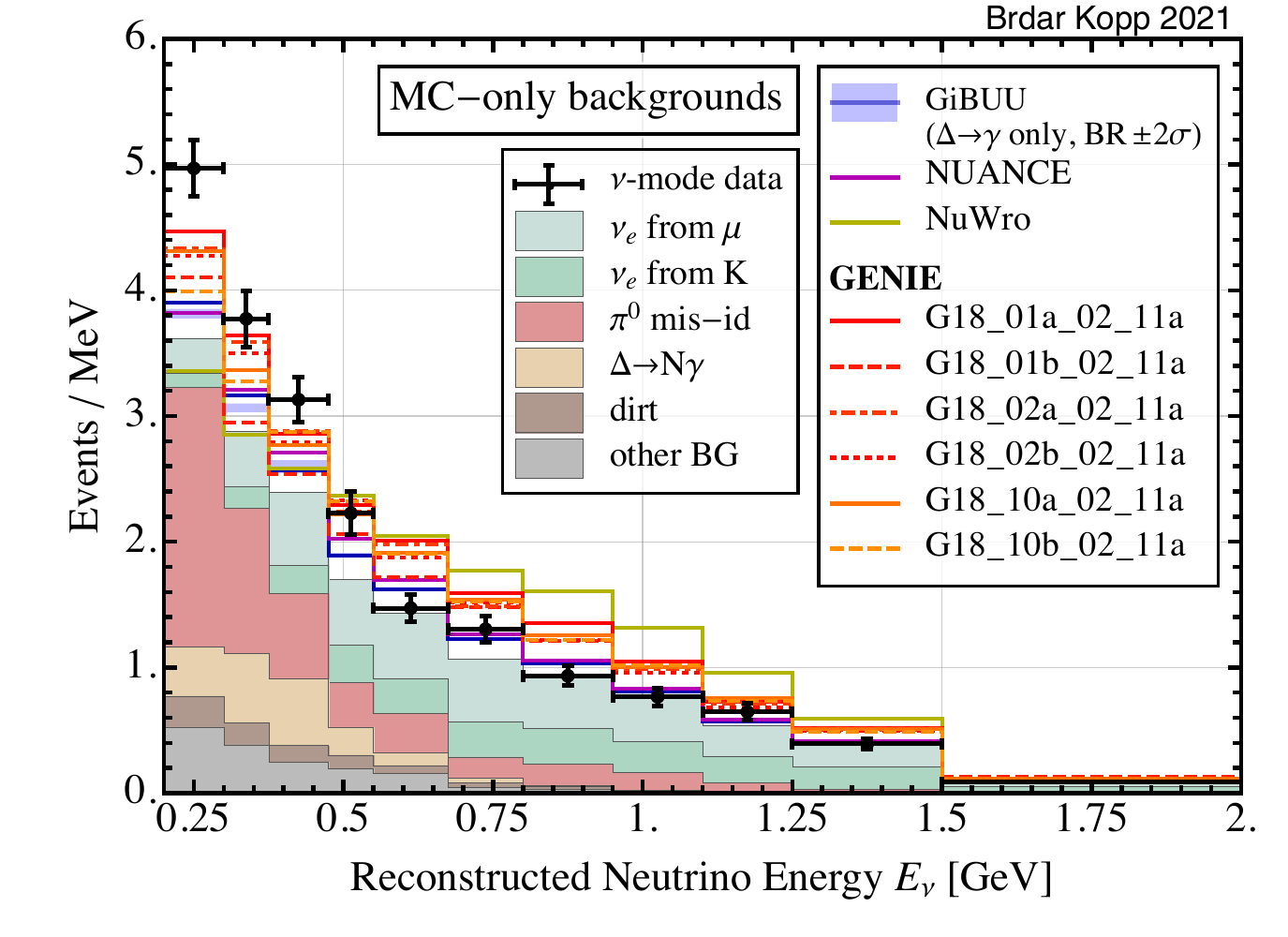} &
    \includegraphics[width=0.5\textwidth]{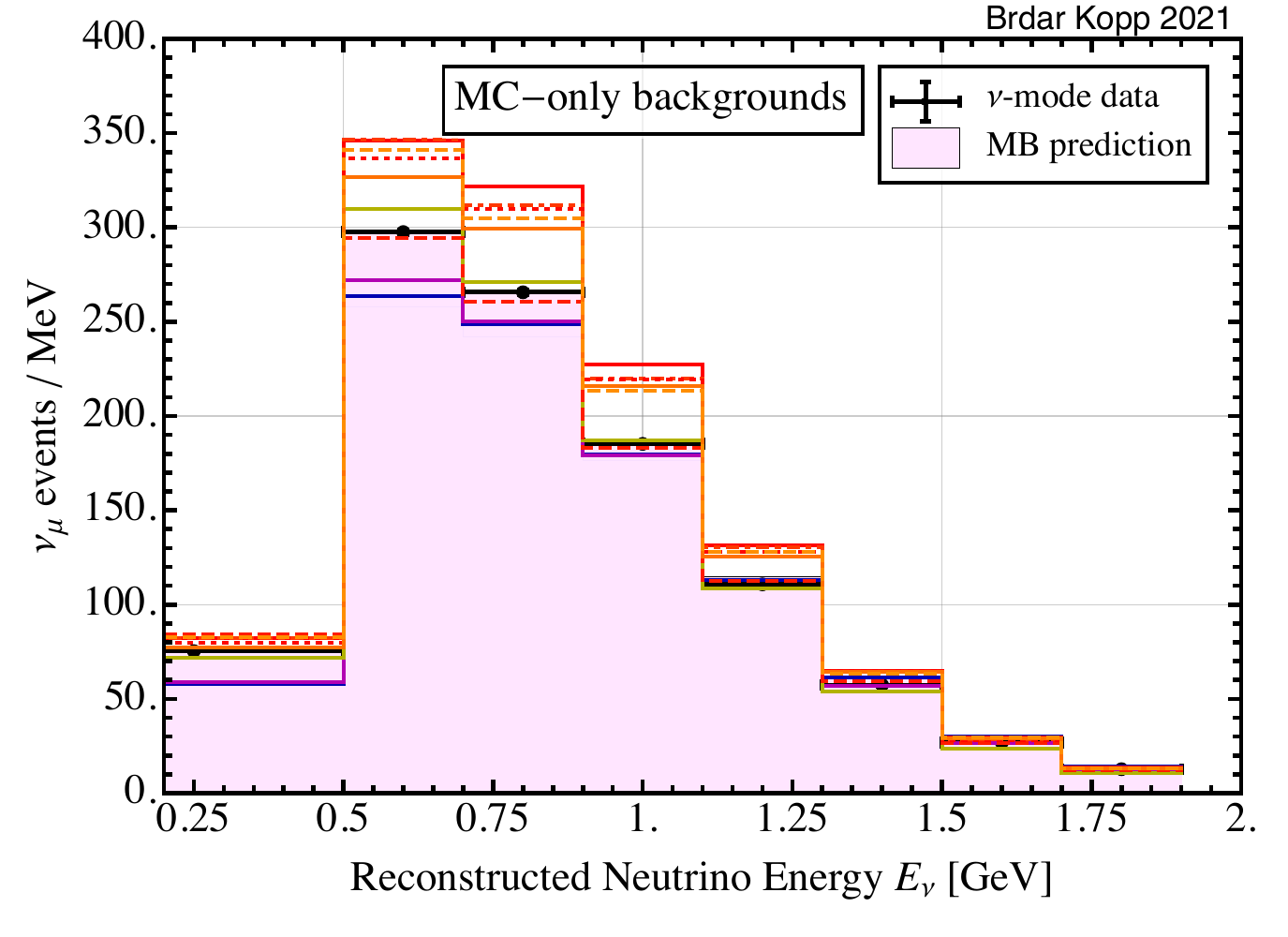}
    \\
    (a) & (b) \\[0.2cm]
    \includegraphics[width=0.5\textwidth]{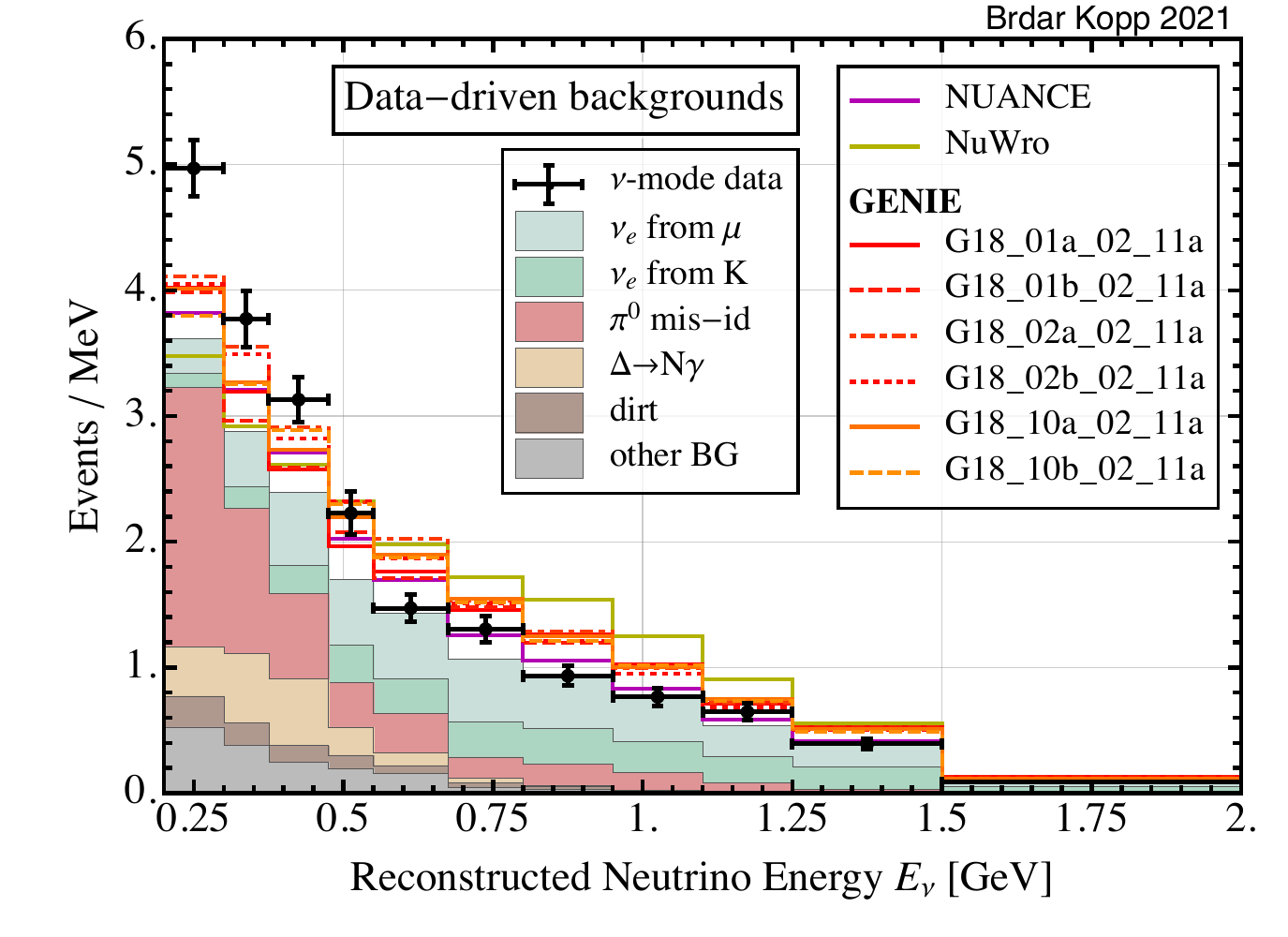} &
    \includegraphics[width=0.5\textwidth]{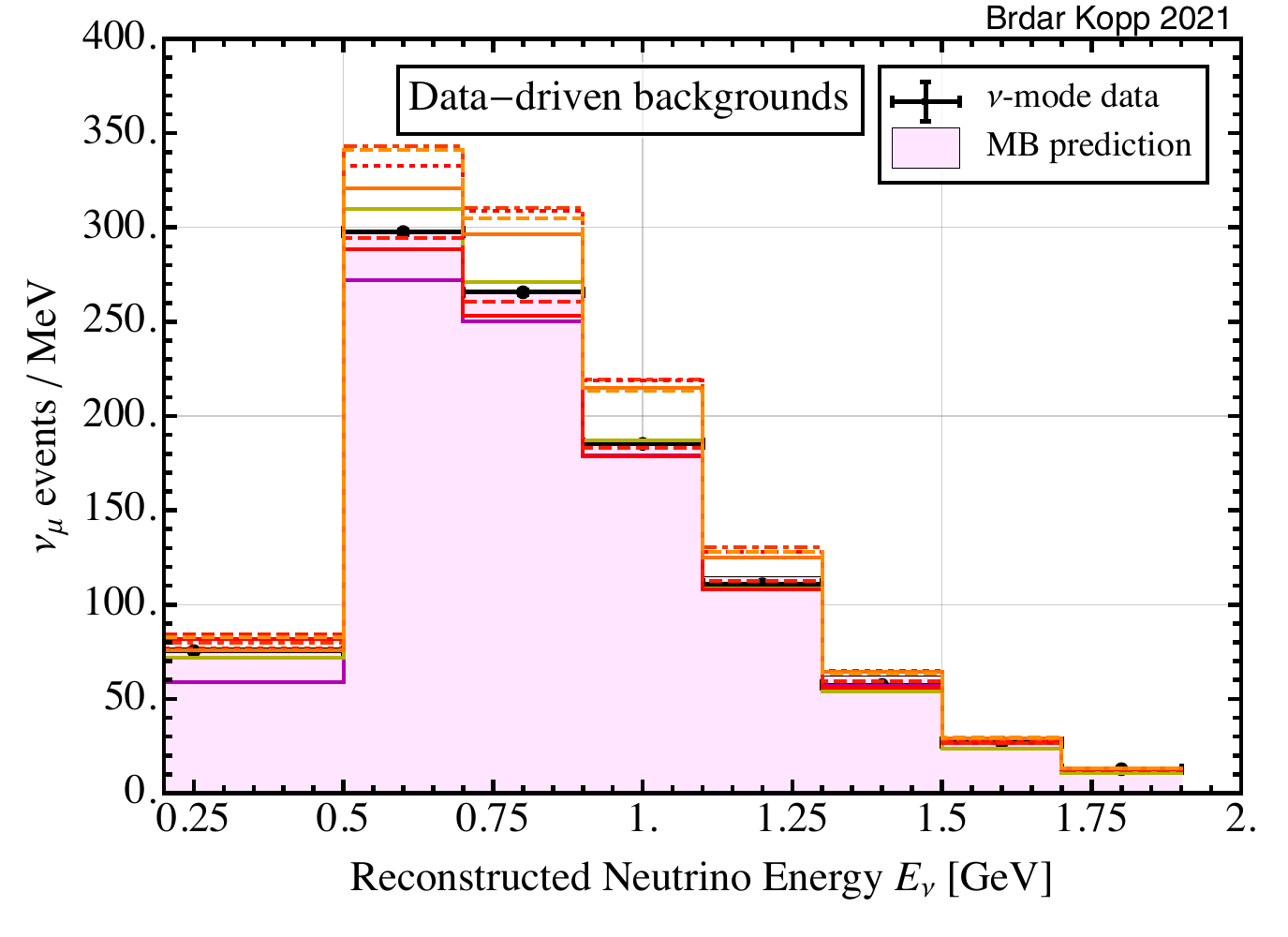}
    \\
    (c) & (d) \\[0.2cm]
  \end{tabular}
  \caption{Predicted MiniBooNE event spectra at the best-fit point of a 
    $3+1$ model, using different event generators and tunes to predict
    the $\nu_e$, $\pi^0$ and $\Delta \to N\gamma$ backgrounds. The panels on the left
    are for $\nu_e$-like events, while the panels on the right show the
    $\nu_\mu$
    spectra.  In the top panels (a) and (b), background predictions are based
    on the Monte Carlo-only simulations described in \cref{sec:channels},
    while panels (c) and (d) at the bottom are based on the data-driven predictions
    from \cref{sec:dd}.
  }
  \label{fig:mb-spectra-comparison}
\end{figure}

\noindent
We now investigate how the interpretation of MiniBooNE's results in the context
of the $3+1$ scenario depends on the event generator used for predicting the
backgrounds.  In \cref{fig:mb-spectra-comparison}, we show the
resulting best-fit spectra in the $\nu_e$ channel (left) and the $\nu_\mu$
channel (right) for different generators.  In the fits, the $\nu_e$, $\pi^0$,
and single-photon (``$\Delta \to N\gamma$'') backgrounds in the $\nu_e$
channel, as well as the prediction for the $\nu_\mu$ channel, are based on our own
prediction discussed in the previous sections.\footnote{Note that we do not take into
  account the possible impact of misreconstruction of CCQE events on the signal
  events. It has been shown in ref.~\cite{Ericson:2016yjn} that this effect is small.}
For the other backgrounds we use
MiniBooNE's official predictions, see ref.~\cite{Aguilar-Arevalo:2018gpe}.
An exception is \gibuu: as discussed in \cref{sec:channels}, \gibuu
underpredicts the rate of pion production in MiniBooNE (but, curiously, not
in other experiments). \gibuu results therefore deviate significantly
from those of other generators in the CCQE channels,
the NC $\pi^0$ backgrounds, and in all data-driven predictions anchored to
the NC $\pi^0$ rate, making a fit to data problematic. In view of this, we do not show fits with
data-driven backgrounds from \gibuu at all, and for fits with Monte Carlo-only
backgrounds, we use \gibuu only for the $\Delta \to N\gamma$ channel, and
MiniBooNE's own predictions in all other channels.  On the other hand, we
indicate as a blue band how \gibuu's best fit spectrum varies if the radiative
branching ratios of the $\Delta(1232)$ $N(1440)$, and $N(1520)$ resonances are varied
within their conservative $2\sigma$ limits, as discussed in \cref{sec:br-scans}.
Note that the backgrounds shown in  \cref{fig:mb-spectra-comparison} as colored
histograms are MiniBooNE's; see \cref{fig:ccqe-spectra,fig:NCpi,fig:NCgamma}
for comparisons between MiniBooNE's background predictions and ours for
individual generators. Finally, as we do not have a data-driven
prediction for the intrinsic $\nu_e$ backgrounds (green shaded histograms in
\cref{fig:mb-spectra-comparison}), we use the Monte Carlo-only one
even for the fits labeled ``data-driven''.

We observe significant spread
between the best-fit spectra from different generators.  Importantly, this is
the case not only for background predictions based on Monte Carlo simulations
alone (top panels), but also for our data-driven predictions (bottom panels).
For the latter, it is driven by the residual uncertainty in the single-photon
channel, see \cref{sec:dd-single-photons}.  Some generators, in particular some
\genie tunes, are able to accommodate fairly large event rates in the low-energy
bins, consistent with the observed excess. However, they tend to also overpredict
the rate at higher energies a bit, suggesting that the goodness of fit will not be
too different compared to MiniBooNE's official fit.

\begin{table}
  \centering
  {\bf Monte Carlo-only background predictions} \\[0.2cm]
  \begin{ruledtabular}
  \begin{tabular}{lllllccc}
    Generator & Tune & $\Delta m_{41}^2$ [eV$^2]$ & $\sin^2 2\theta_{\mu e}$
                     & $|U_{\mu 4}|^2$
                     & $\chi^2$/dof & $\Delta\chi^2_\text{no osc.}$ & Significance \\
    \hline
    MB official &                                 & 0.25   &   0.01   &   0.062   &   12.0   &   19.1   &   $4.0 \sigma$ \\
    \gibuu      & default                         & 0.25   &   0.01   &   0.076   &   12.0   &   24.6   &   $4.6 \sigma$ \\
                & $\BR(\Delta\to\gamma) -2\sigma$ & 0.32   &   0.0063 &   0.076   &   12.2   &   28.1   &   $4.9 \sigma$ \\
                & $\BR(\Delta\to\gamma) +2\sigma$ & 0.32   &   0.0050 &   0.076   &   12.0   &   21.1   &   $4.2 \sigma$ \\
    \nuance     &        --                       & 0.32   &   0.0079   &   0.051   &   12.3   &   19.3   &   $4.0 \sigma$ \\
    \nuwro      &        --                       & 3.2   &   0.0020   &   0.040   &   13.7   &   15.6   &   $3.5 \sigma$ \\
    \genie      & G18\_01a\_02\_11a               & 0.13   &   0.079   &   0.16   &   12.2   &   21.6   &   $4.3 \sigma$ \\
                & G18\_01b\_02\_11a               & 0.79   &   0.0001   &   0.12   &   12.2   &   16.1   &   $3.6 \sigma$ \\
                & G18\_02a\_02\_11a               & 0.13   &   0.050   &   0.16   &   12.0   &   15.1   &   $3.5 \sigma$ \\
                & G18\_02b\_02\_11a               & 0.13   &   0.050   &   0.18   &   12.1   &   15.0   &   $3.5 \sigma$ \\
                & G18\_10a\_02\_11a               & 0.25   &   0.016   &   0.051   &   12.1   &   11.2   &   $2.9 \sigma$ \\
                & G18\_10b\_02\_11a               & 0.40   &   0.013   &   0.016   &   12.1   &   17.9   &   $3.8 \sigma$ \\
  \end{tabular}
  \end{ruledtabular}
  \\[0.5cm]
  {\bf data-driven backgrounds} \\[0.2cm]
  \begin{ruledtabular}
  \begin{tabular}{llcccccc}
    Generator & Tune & $\Delta m_{41}^2$ & $\sin^2 2\theta_{\mu e}$ & $|U_{\mu 4}|^2$
                     & $\chi^2$/dof & $\Delta\chi^2_\text{no osc.}$ & Significance \\
    \hline
    MB official &                   & 0.25   &   0.01   &   0.062   &   12.0   &   19.1   &   $4.0 \sigma$ \\  
    \nuance     &        --         & 0.32   &   0.0079   &   0.051   &   12.3   &   19.3   &   $4.0 \sigma$ \\
    \nuwro      &        --         & 3.2   &   0.0016   &   0.040   &   13.3   &   12.7   &   $3.1 \sigma$ \\ 
    \genie      & G18\_01a\_02\_11a & 0.79   &   0.00020   &   0.14   &   12.2   &   23.3   &   $4.4 \sigma$ \\
                & G18\_01b\_02\_11a & 0.79   &   0.0001   &   0.12   &   12.2   &   15.5   &   $3.5 \sigma$ \\ 
                & G18\_02a\_02\_11a & 0.13   &   0.063   &   0.18   &   12.2   &   19.2   &   $4.0 \sigma$ \\  
                & G18\_02b\_02\_11a & 0.13   &   0.050   &   0.20   &   12.3   &   16.9   &   $3.7 \sigma$ \\  
                & G18\_10a\_02\_11a & 0.25   &   0.016   &   0.062   &   12.3   &   15.1   &   $3.5 \sigma$ \\ 
                & G18\_10b\_02\_11a & 0.40   &   0.013   &   0.016   &   12.1   &   19.5   &   $4.0 \sigma$ \\ 
  \end{tabular}
  \end{ruledtabular}
  \caption{Results of fitting a $3+1$ sterile neutrino model to MiniBooNE data,
    using different event generators and tunes to predict the $\nu_e$, $\pi^0$,
    and single-photon backgrounds.  Besides the parameter values at the
    respective best-fit points, we also list the $\Delta\chi^2$ at which the
    no-oscillation hypothesis is excluded, and we convert this number into a
    statistical significance for an anomaly, assuming a $\chi^2$ distribution
    with one degree of freedom. Note that we do not use \gibuu's Monte
    Carlo-only predictions for CCQE-like events and NC $\pi^0$ events
    because \gibuu predicts far fewer pions than observed. For
    the same reason, we also do not show result for \gibuu with
    data-driven backgrounds at all. Here, \gibuu's pion deficit would affect
    the $\pi^0$ control sample.}
  \label{tab:bfp}
\end{table}

\begin{figure}
  \centering
  \begin{tabular}{cc}
    \includegraphics[width=0.5\textwidth]{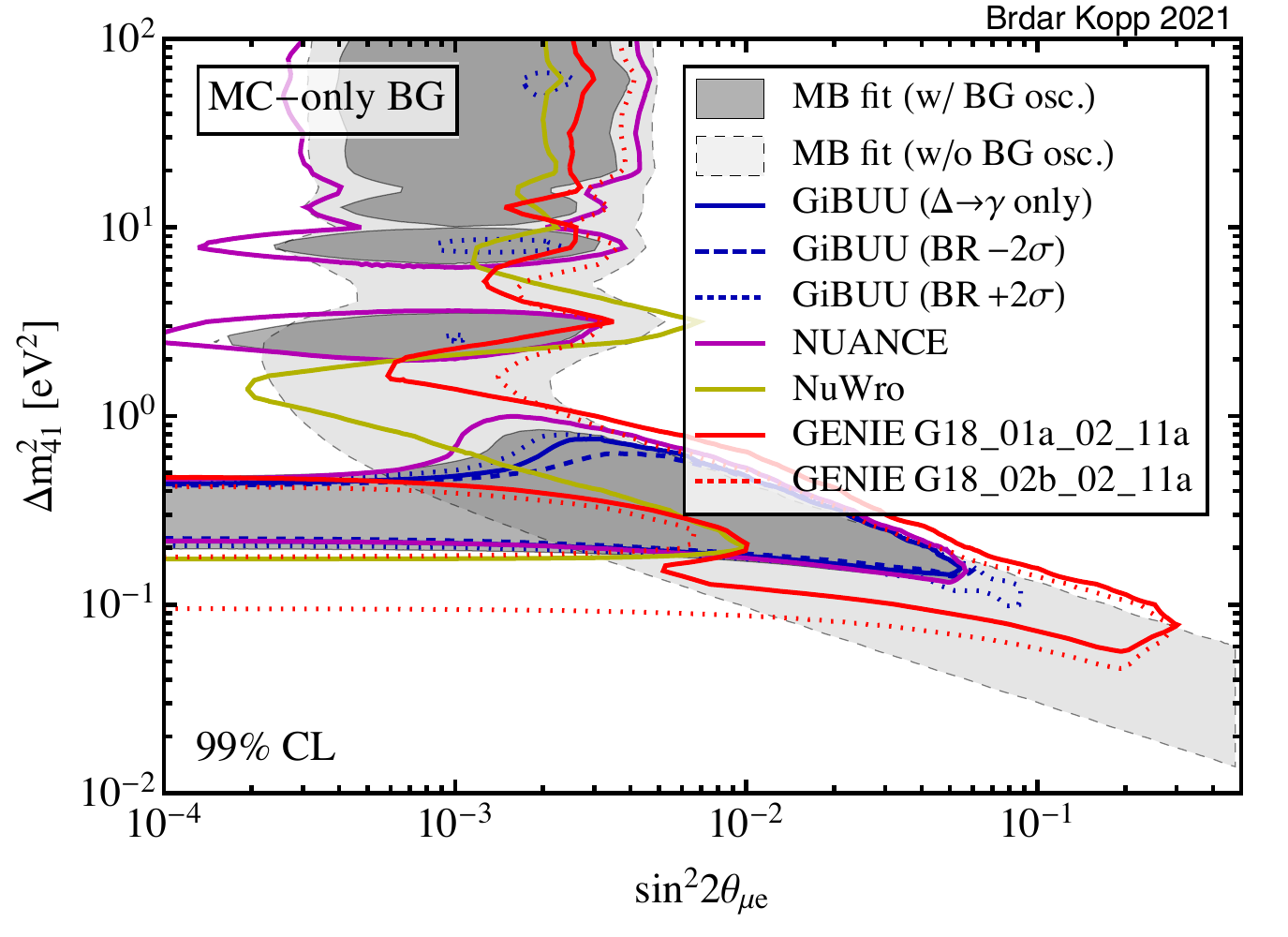} &
    \includegraphics[width=0.5\textwidth]{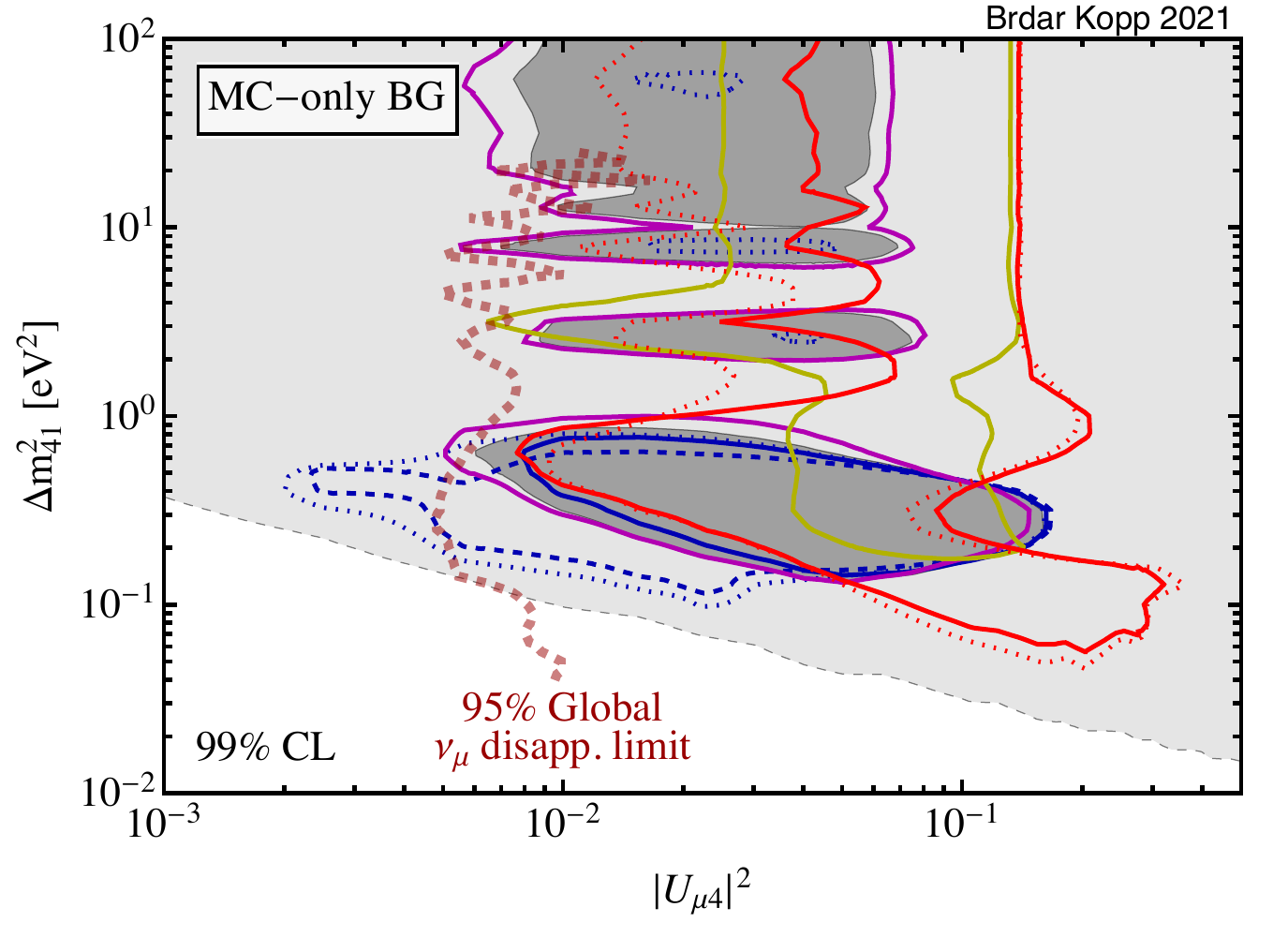}
    \\
    (a) & (b) \\[0.2cm]
    \includegraphics[width=0.5\textwidth]{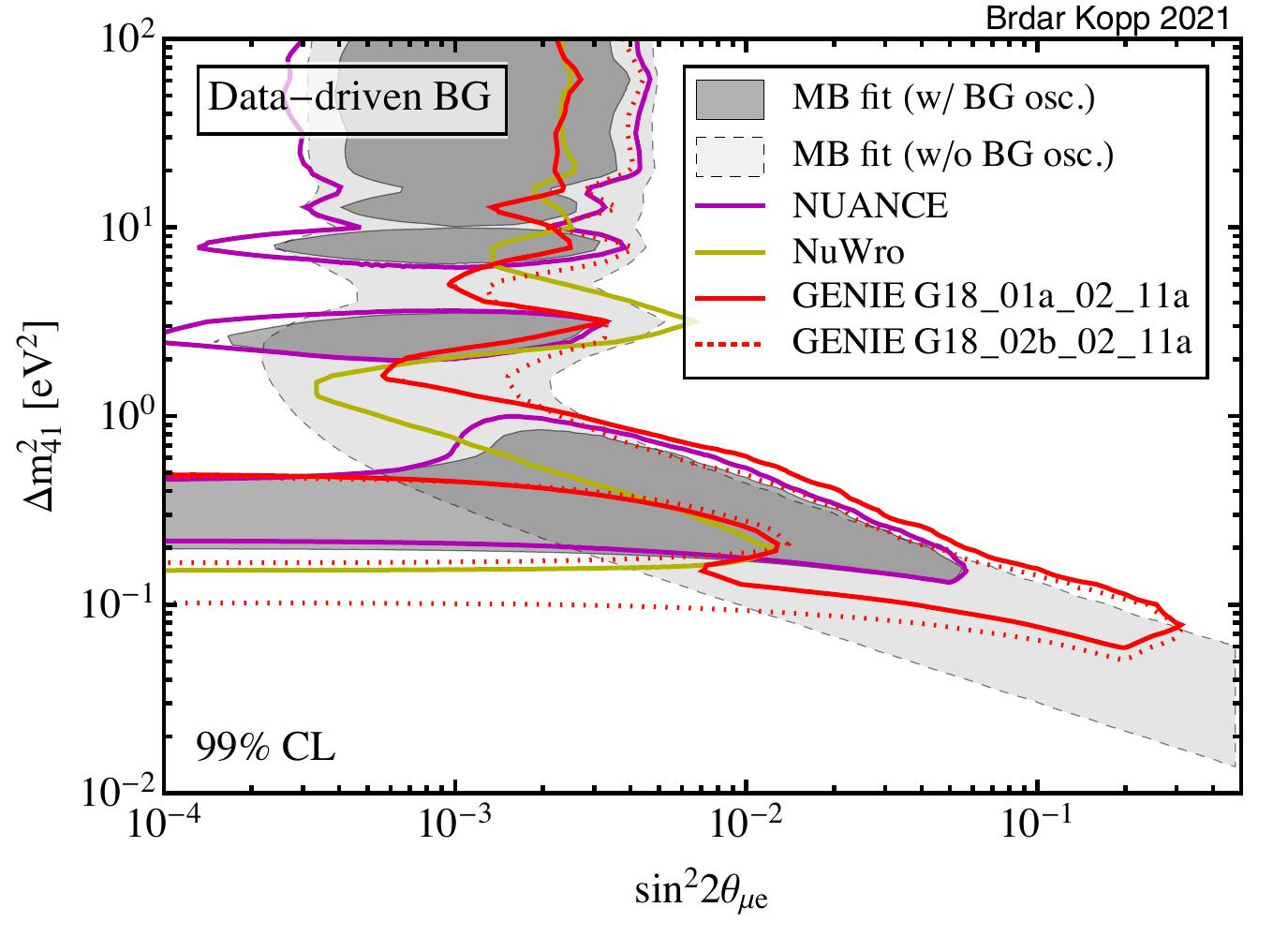} &
    \includegraphics[width=0.5\textwidth]{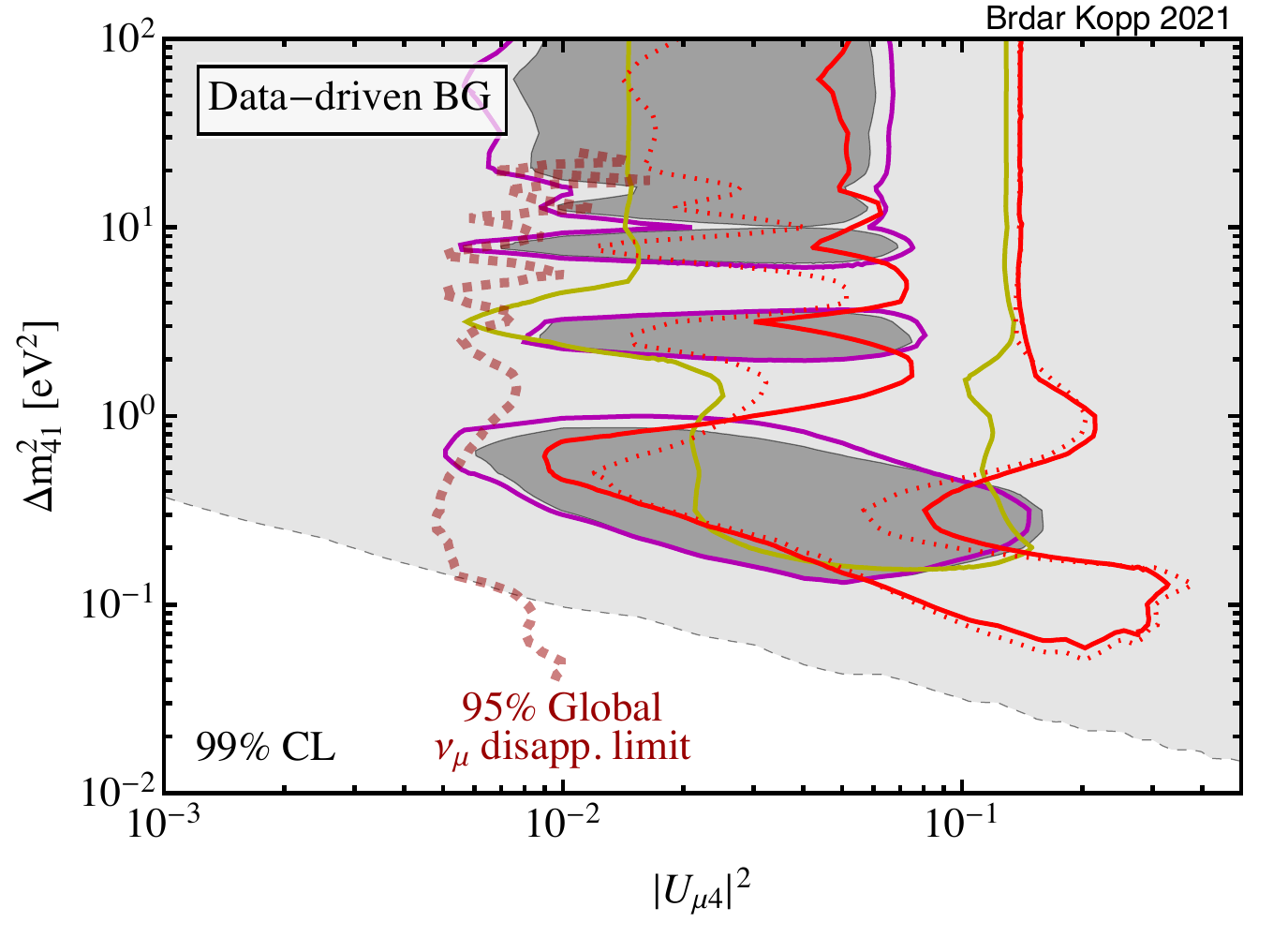}
    \\
    (c) & (d) \\[0.2cm]
  \end{tabular}
  \caption{MiniBooNE 99\% exclusion contours for a $3+1$ sterile neutrino model
    using different event generators and tunes for the prediction of the
    $\nu_e$, $\pi^0$ and $\Delta \to N\gamma$ backgrounds. (For clarity, we
    show results for only two \genie tunes here; exclusion contours for the
    others can be found in \cref{sec:mb-contours}.) The panels on the left show
    the $\sin^2 2\theta_{\mu e}$--$\Delta m_{41}^2$ plane ($\nu_e$ appearance);
    the ones on the right show the $|U_{\mu 4}|^2$--$\Delta m_{41}^2$ plane
    ($\nu_\mu$ disappearance).  In the top panels (a) and (b), background
    predictions are based on the Monte Carlo-only calculations described in
    \cref{sec:channels}, while panels (c) and (d) at the bottom are based on
    the data-driven predictions from \cref{sec:dd}.
  }
  \label{fig:mb-contours-comparison}
\end{figure}

\Cref{tab:bfp} reveals that differences between event generators translate into
some differences in the best-fit points and in the significance of the anomaly.
However, the latter remains around $3-4\sigma$, similar to the significance we
obtain when using MiniBooNE's background predictions.  We find the lowest
significance ($2.9\sigma$) for one of the ``theory-driven'' \genie tunes,
namely the G18\_10a\_02\_11a one. But differences with respect to other tunes
or generators are marginal, testifying to the robustness of the MiniBooNE excess.

This is also illustrated in \cref{fig:mb-contours-comparison}, which compares the
parameter space exclusion regions in the $3+1$ scenario between analyses based
on different generators.  This is once again done both for background
predictions based on Monte Carlo simulations alone (top panels) and for
data-driven background predictions in the $\pi^0$ and single-photon channels
(bottom panels). We show projections onto the
$\sin^2 2\theta_{\mu e}$--$\Delta m_{41}^2$ plane (panels (a) and (c)) and
onto the $|U_{\mu 4}|^2$--$\Delta m_{41}^2$ plane (panels (b) and (d)).

We observe that the contours based on the \nuance generator (purple) are, as
expected, in excellent agreement with those based on MiniBooNE's official
background predictions (which also rely on \nuance).  Significant deviations
are seen for fits using \nuwro, \genie, and \gibuu predictions, though.
(As before, we include \gibuu only in the comparison Monte Carlo-only
predictions, and even there we use its predictions only for the single-photon
channel.)  Notably, \nuwro and \genie allow $\sin^2 2\theta_{\mu e} = 0$
at the 99\%~CL over
a wide range of $\Delta m_{41}^2$ values, while \gibuu, as well as
the fit using MiniBooNE's official background predictions, do so only
in a narrow window around $\Delta m_{41}^2 \sim \SI{0.3}{eV^2}$.
Remember that allowing $\sin^2 2\theta_{\mu e} = 0$ does not mean
that the anomaly is resolved -- the fits still require non-zero $|U_{\mu 4}|^2$,
see the right-hand panels of \cref{fig:mb-contours-comparison}.
As explained in \cref{sec:fits-2f-vs-4f}, this can be understood from the strong
correlations between the $\nu_e$ and $\nu_\mu$ data. An excess of $\nu_\mu$
events compared to the theory prediction may thereby be sufficient to explain
an excess also in the $\nu_e$ channel even without explicit $\nu_e$ appearance.
Note, however, that the values of $|U_{\mu 4}|^2$ required to accommodate the
MiniBooNE anomaly are still in tension with the global exclusion
limit on $\nu_\mu$ disappearance (dotted red line in the right-hand panels
of \cref{fig:mb-contours-comparison}).

\section{Summary and Conclusions}
\label{sec:summary}

\noindent
The MiniBooNE anomaly is still one of the biggest mysteries in neutrino
physics.  In this paper we have revisited the most relevant backgrounds for the
MiniBooNE $\nu_e$ appearance analysis in which the anomalous event excess is
observed.  We have in particular studied CC interactions of beam $\nu_e$, NC
$\pi^0$ production, and single-photon production.  We have predicted the event
rates in these channels using different event generators -- namely \nuance,
\gibuu, \genie, and \nuwro\ -- and have compared the results to estimate the
theoretical uncertainties associated with our predictions.  For the $\pi^0$
background, we have found that generators agree at the 10\% level (with the
exception of \gibuu, which underpredicts the $\pi^0$ production rate by almost
a factor of two). For CC $\nu_e$ and single-photon events, discrepancies are
somewhat larger, with predictions differing by $\mathcal{O}(30\%)$.

The situation improves only slightly when we attempt to predict the $\pi^0$ and
single-$\gamma$ backgrounds in a more data-driven way by normalizing them to
MiniBooNE's measured $\pi^0$ production rate.  (The deficit of $\pi^0$-induced
events in \gibuu is, however, almost entirely removed that way.)

In addition, we have discussed the impact of uncertainties in the radiative
branching ratios of heavy hadronic resonances, most importantly the $\Delta(1232)$.
If these errors are as large as the conservative estimate from \cite{Zyla:2020zbs},
(tens of per cent), they affect in particular the single-photon background, but
are still too small to explain the anomaly.

In the final part of the paper, we have discussed fits to the MiniBooNE data in
the context of sterile neutrino models. We have highlighted the important
differences between fits in a two-flavor framework, which are often shown in
the literature, and a more careful fit that takes into account full
four-flavor oscillations. In the four-flavor case, oscillations of the
$\nu_e$ background and in the $\nu_\mu$ control sample play a crucial role.
Most notably, the anomaly could be entirely explained by $\nu_\mu$
disappearance alone, thanks to the strong correlation between the $\nu_\mu$ and
$\nu_e$ samples in the fit. The tension with the non-observation of $\nu_\mu$
disappearance in other experiments would still persist, though.

We have then studied how the choice of event generator affects the fit in a
$3+1$ sterile neutrino model, and in particular the significance of the
anomaly. We have found that all generators roughly agree on the significance of
the anomaly between $3\sigma$ and $4\sigma$, and that in none of the scenarios
we have considered the significance drops below $2.9\sigma$.

We conclude that theoretical uncertainties in MiniBooNE's background
predictions certainly deserve further study. However, it seems that with our
current understanding of neutrino interaction physics -- as implemented in
state-of-the-art event generators, the anomaly is robust.  Not even an
``Altarelli cocktail'' of several small deviations in different channels that
add up to a potentially much bigger overall discrepancy seems to be able to
fully explain the event excess.

\section*{Note Added}
\noindent
The MicroBooNE collaboration has recently released first results from several
$\nu_e$ appearance searches \cite{MicroBooNE:2021rmx, MicroBooNE:2021nxr,
MicroBooNE:2021jwr, MicroBooNE:2021sne} and from a search for single photons
from $\Delta(1232)$ decay \cite{MicroBooNE:2021zai}.  All results appear to agree
with SM predictions and therefore do not explained the MiniBooNE anomaly.
The results from ref.~\cite{MicroBooNE:2021zai} confirm our findings that radiative
decays of the $\Delta(1232)$ are unlikely to account for MiniBooNE's event excess.

\section*{Acknowledgments}

\noindent
We would like to thank Luis Alvarez Ruso, David Caratelli, Dave Casper, Teppei
Katori, Bill Louis, Xiao Luo, Ulrich Mosel, and Jan Sobczyk for very useful
discussions, as well as Pedro Machado for collaboration in the early stages of
this work.  Fermilab is operated by Fermi  Research  Alliance, LLC  under
contract  No.  DE-AC02-07CH11359  with  the  United States Department of
Energy. JK has been partly funded by the German Research Foundation (DFG) in
the framework of the PRISMA+ Cluster of Excellence and by the European Research
Council (ERC) under the European Union's Horizon 2020 research and innovation
programme (grant agreement No.\ \texttt{637506}, ``$\nu$Directions'').

\appendix
\section{Parameter Scans in the $3+1$ Sterile Neutrino Model}
\label{sec:mb-contours}

\begin{figure}
  \centering
  \vspace{-0.5cm}
  \textbf{\large Monte Carlo Backgrounds} \\[0.2cm]
  \hspace*{-1.5cm}
  {\renewcommand{\arraystretch}{0.1}
  \begin{tabular}{c@{\hspace{-0.4cm}}c|c@{\hspace{-0.4cm}}c}
    \includegraphics[width=0.30\textwidth]{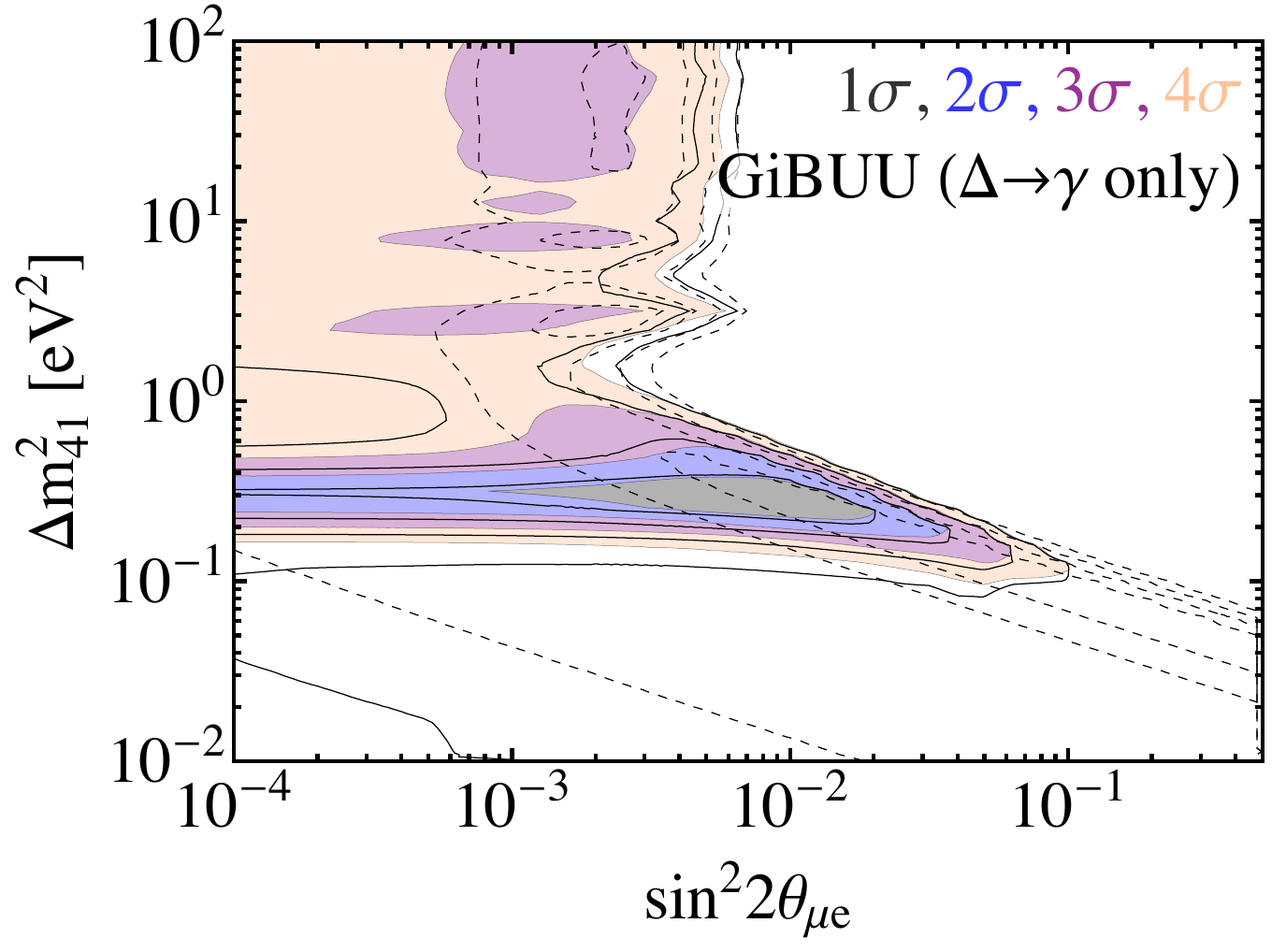} &
    \includegraphics[width=0.30\textwidth]{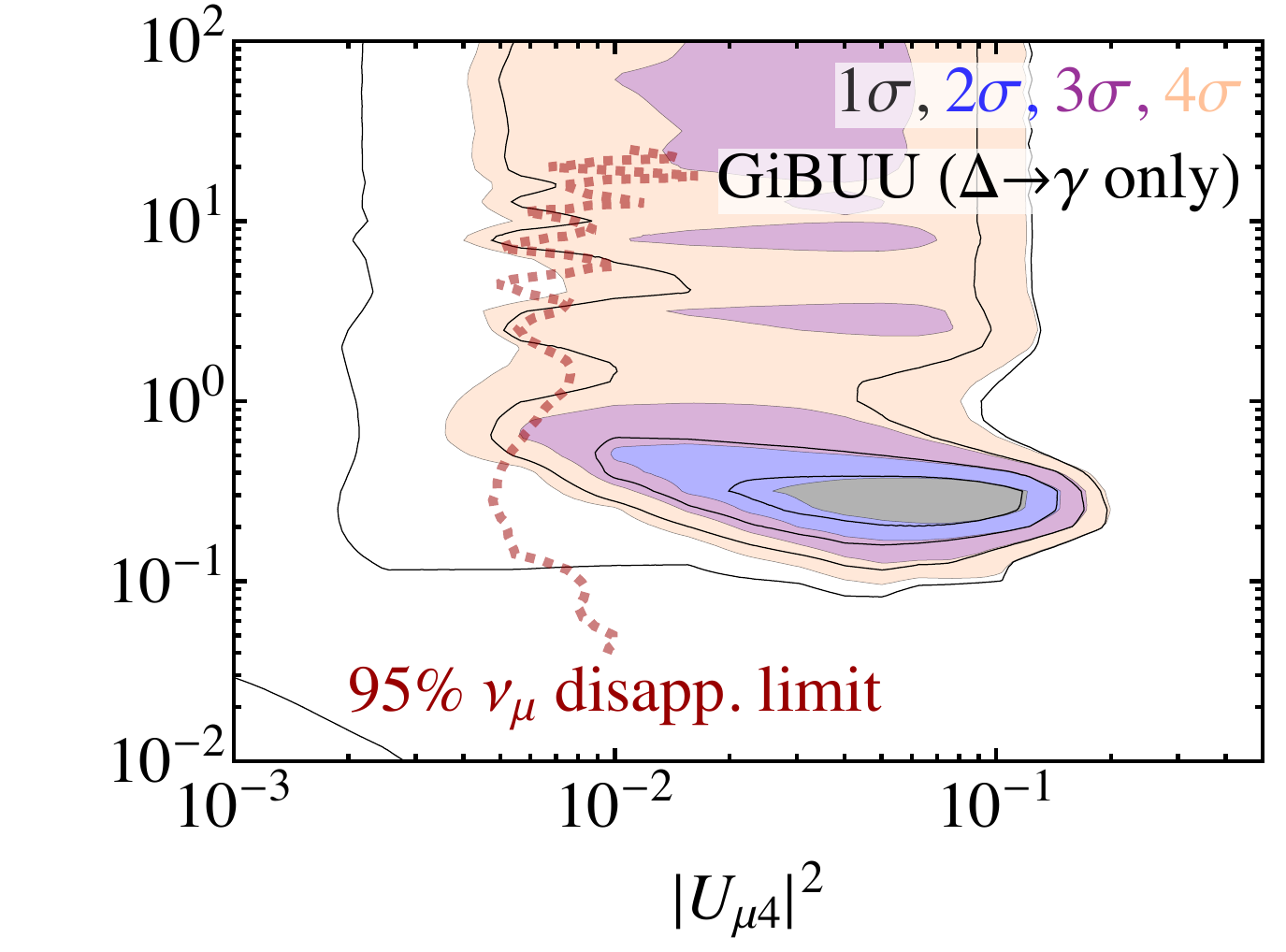} &
    \includegraphics[width=0.30\textwidth]{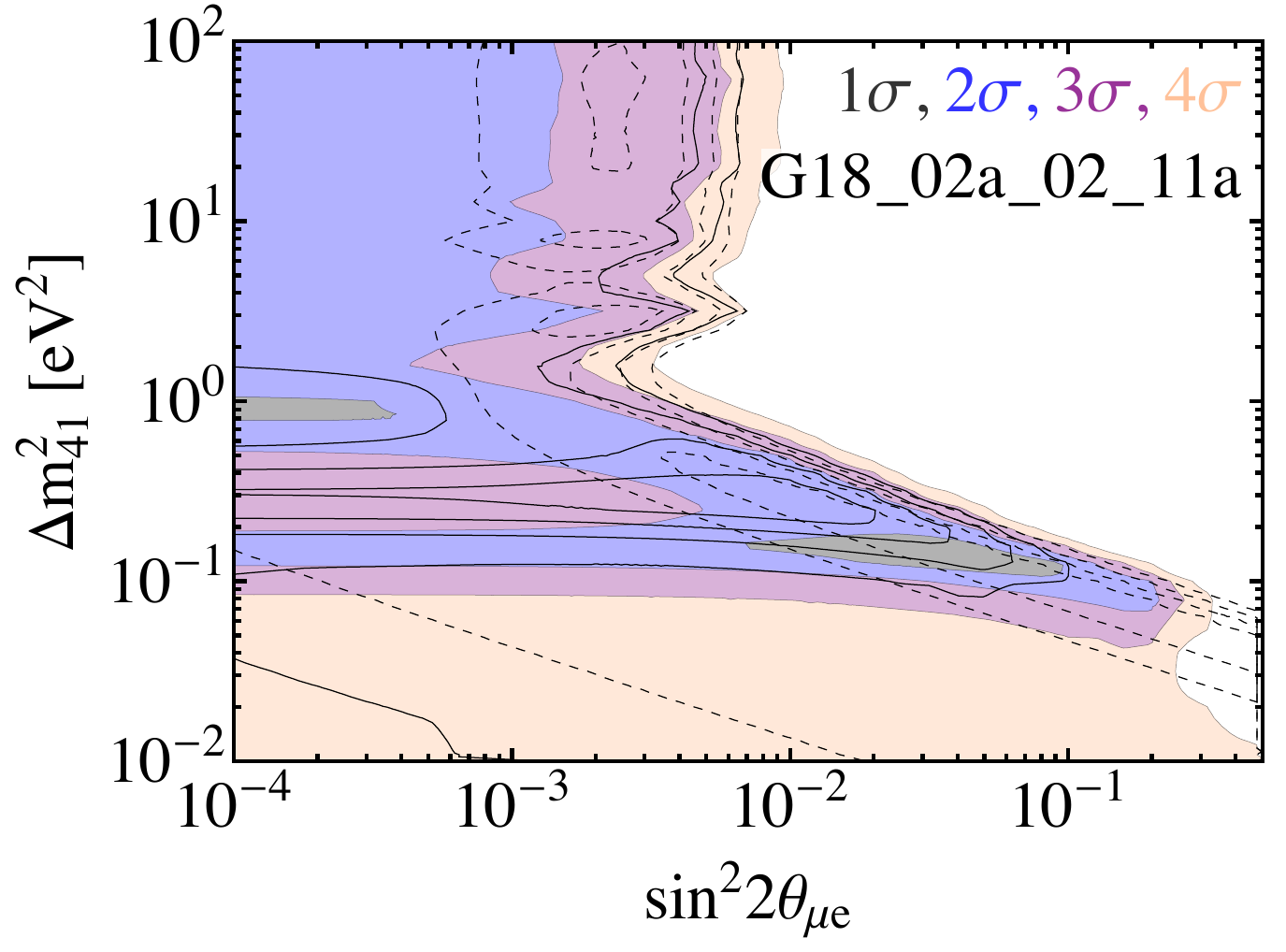} &
    \includegraphics[width=0.30\textwidth]{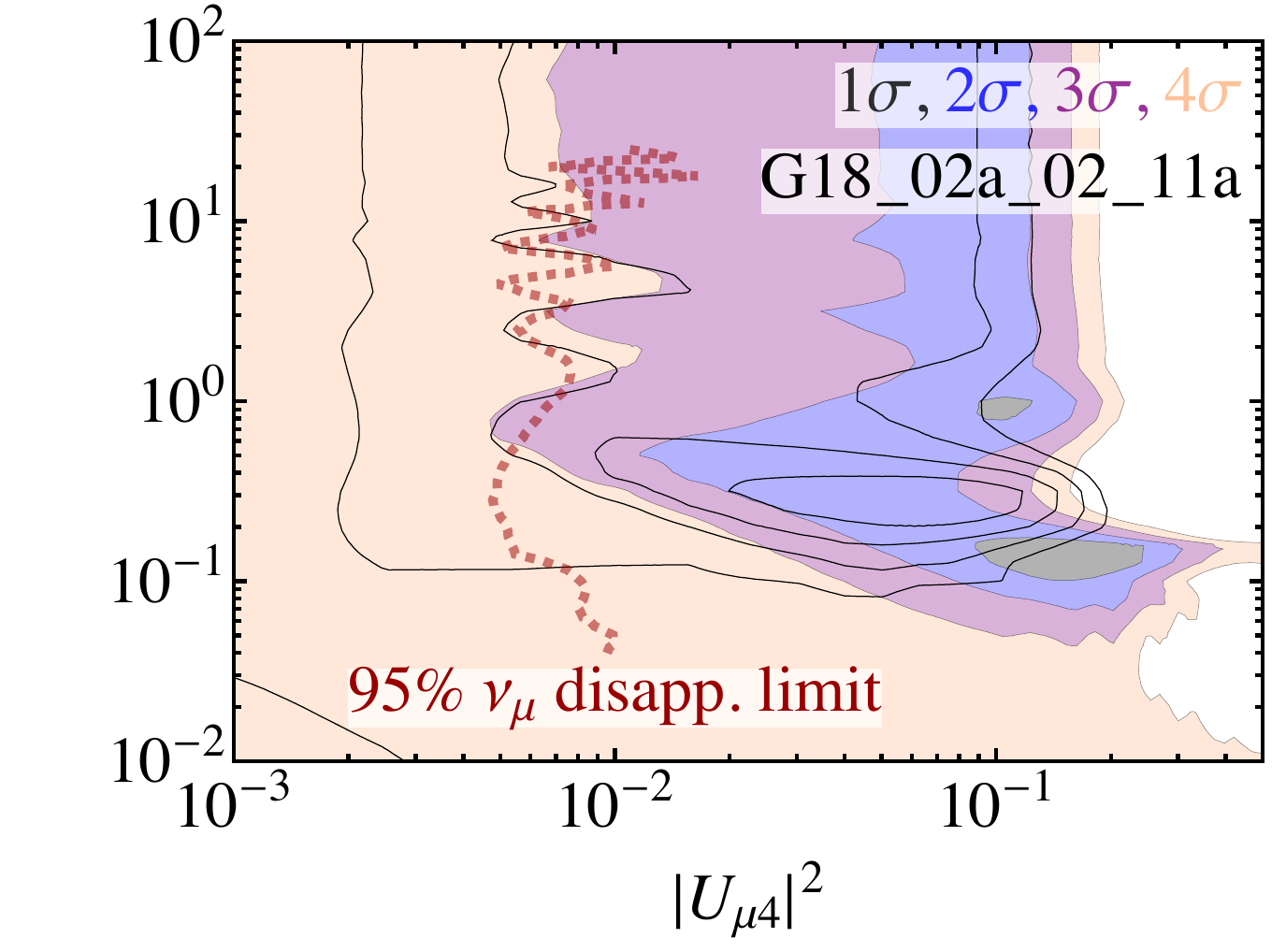} \\
    \includegraphics[width=0.30\textwidth]{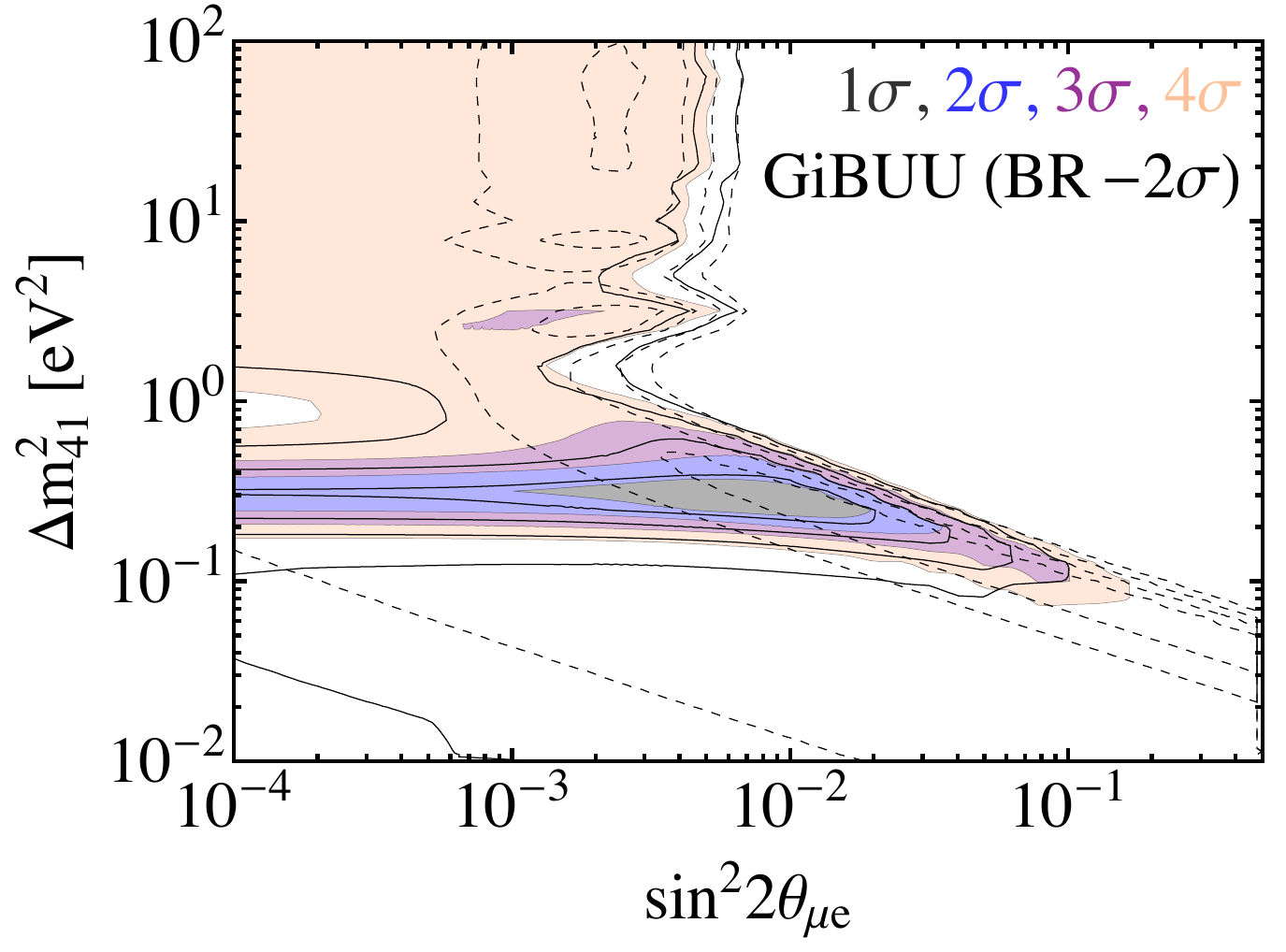} &
    \includegraphics[width=0.30\textwidth]{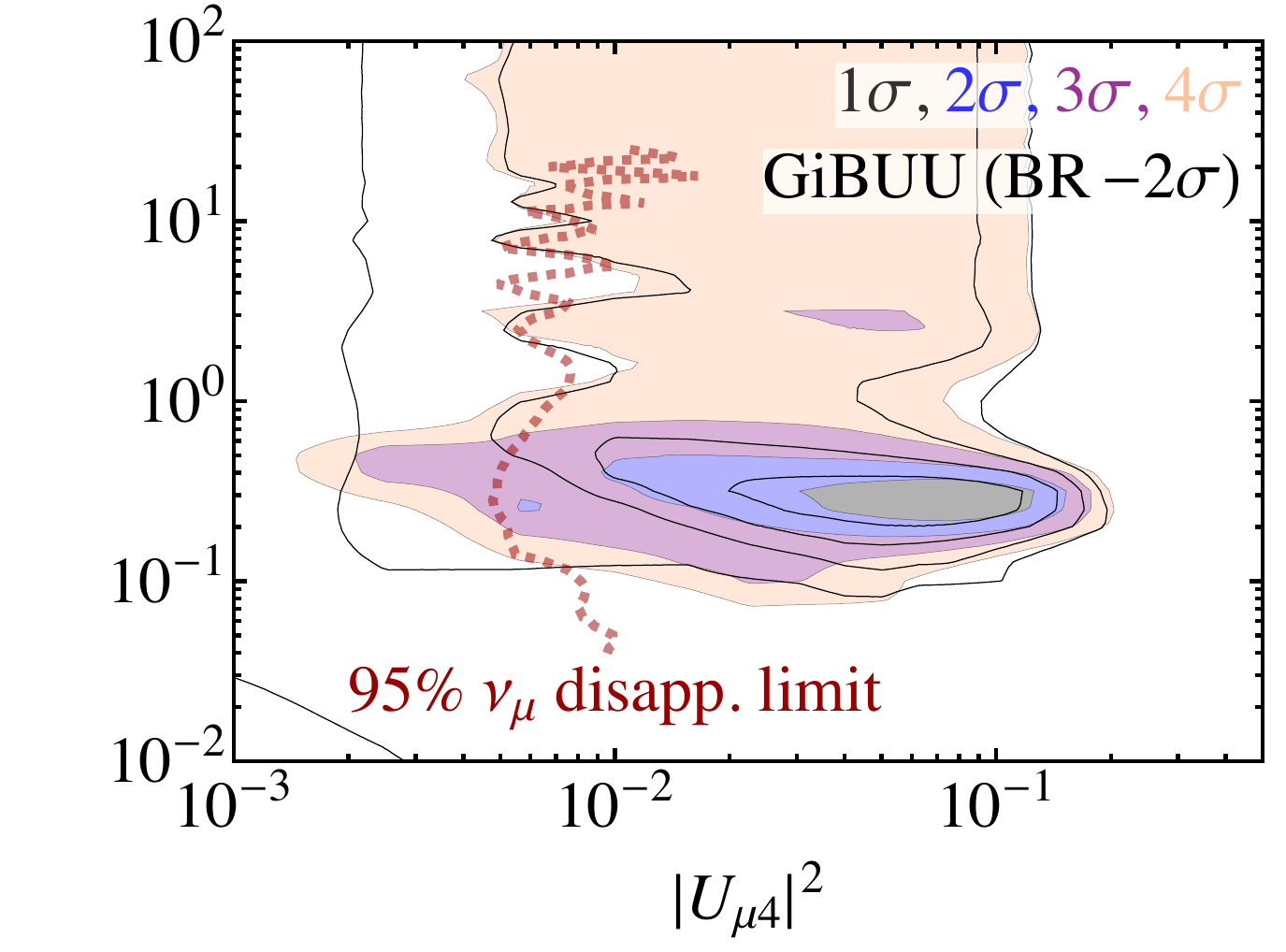} &
    \includegraphics[width=0.30\textwidth]{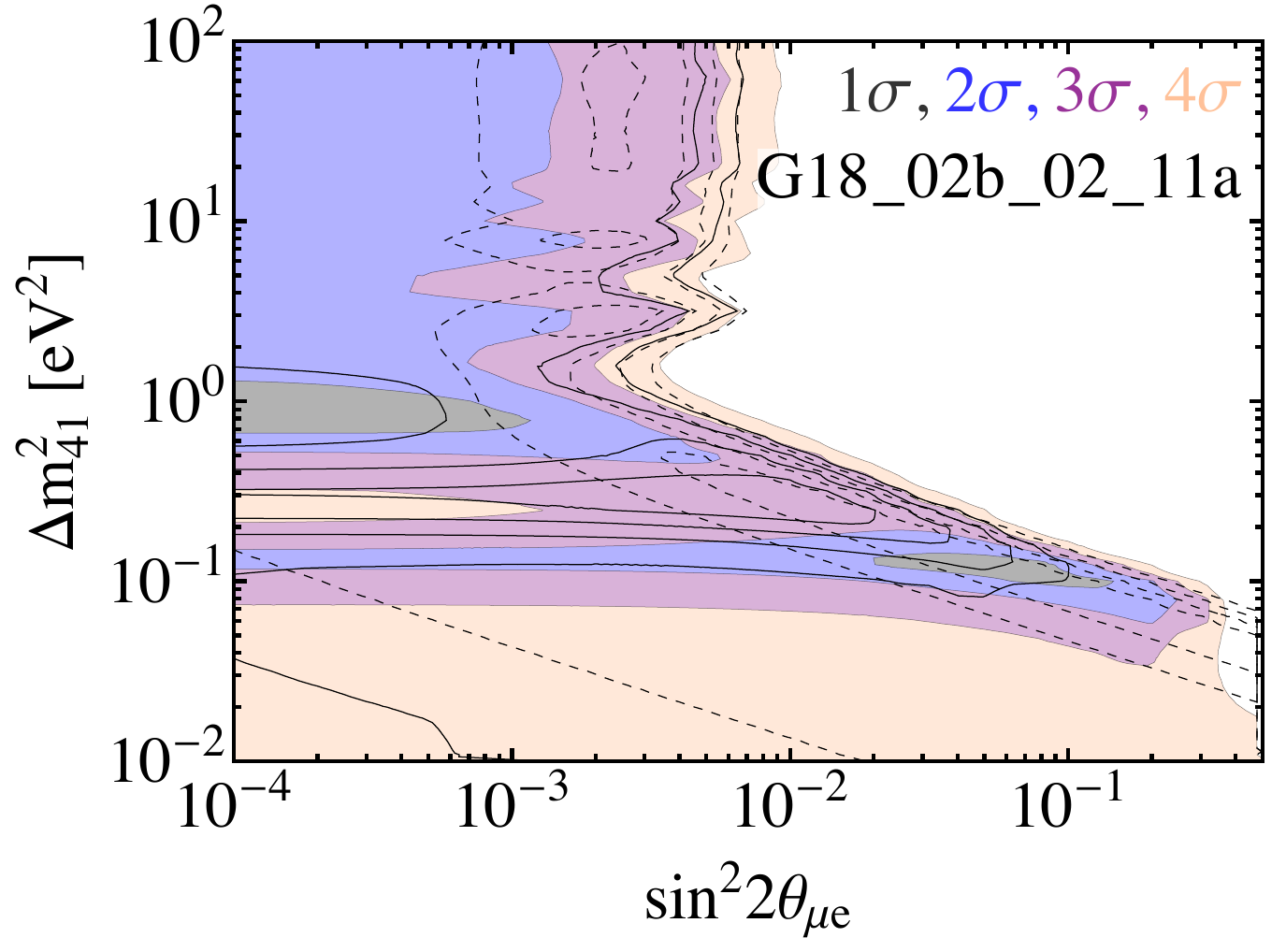} &
    \includegraphics[width=0.30\textwidth]{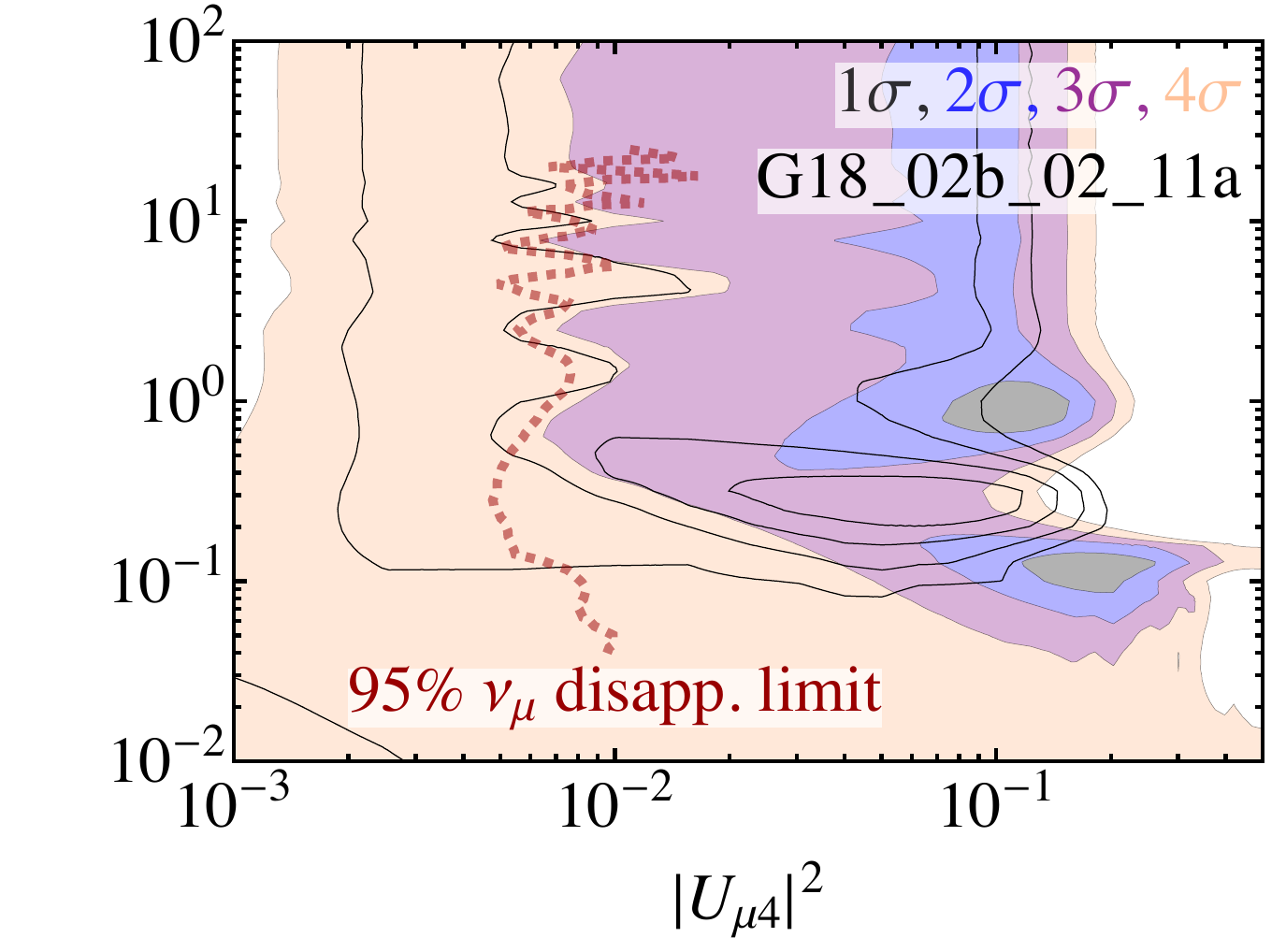} \\
    \includegraphics[width=0.30\textwidth]{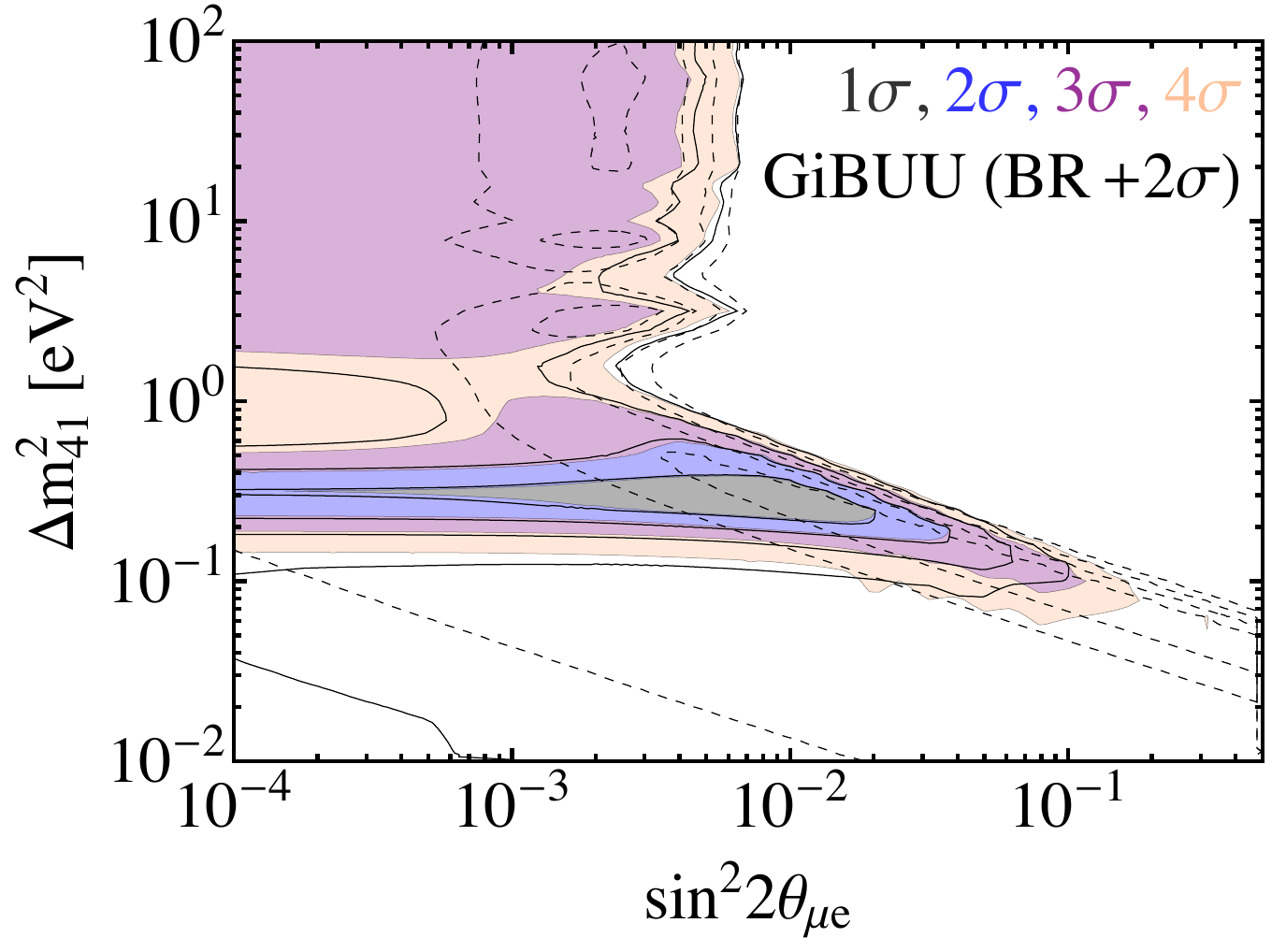} &
    \includegraphics[width=0.30\textwidth]{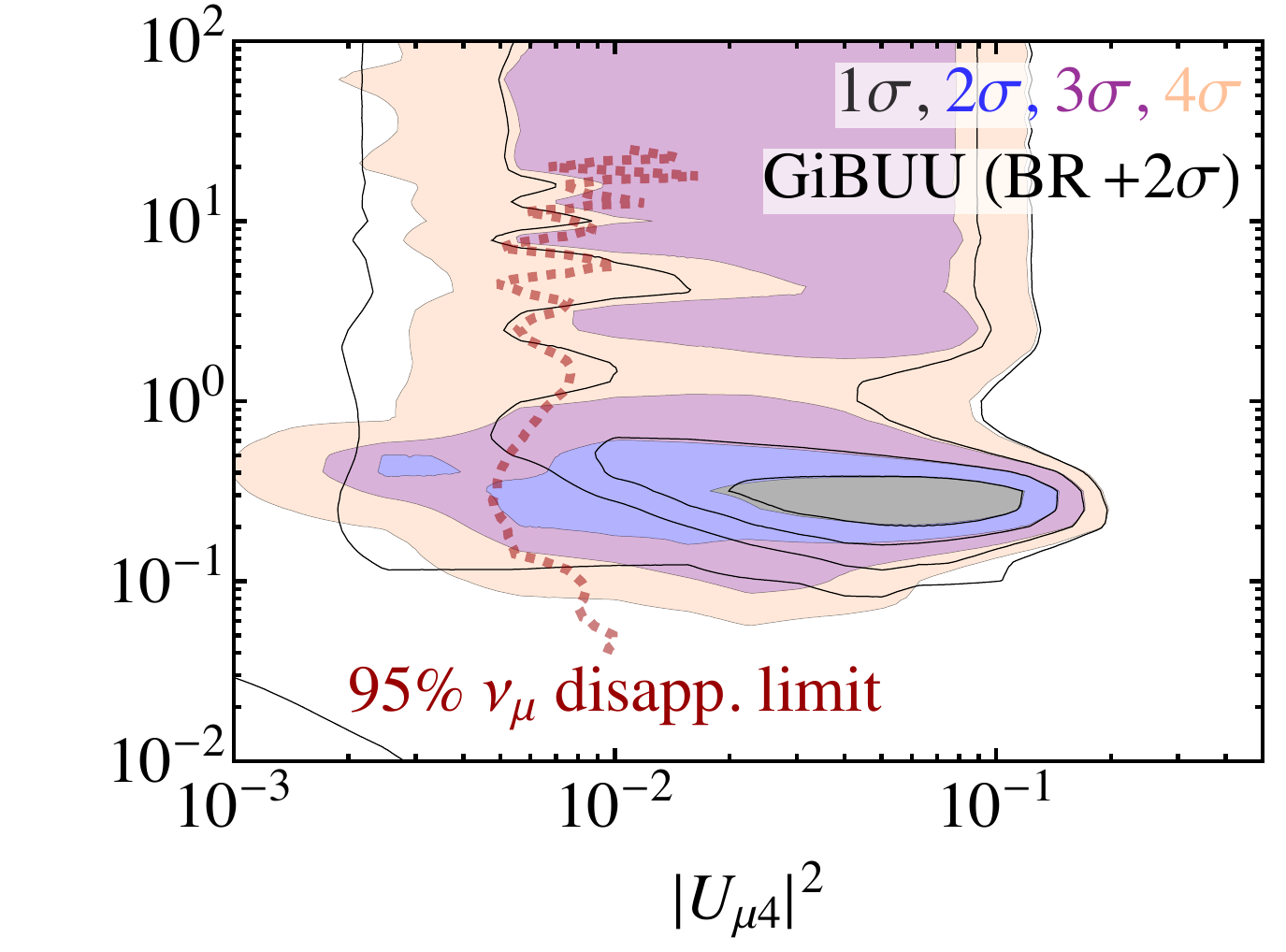} &
    \includegraphics[width=0.30\textwidth]{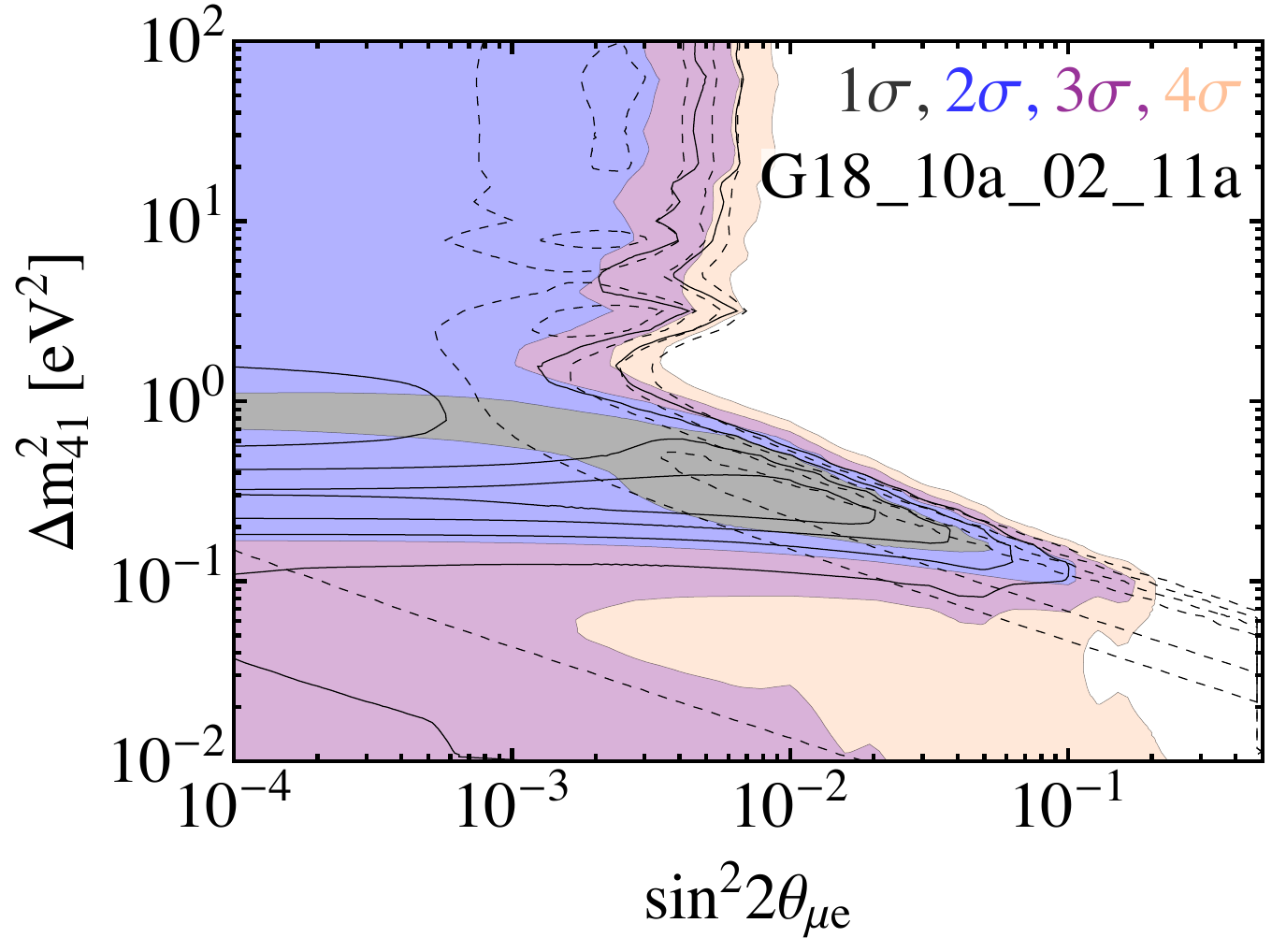} &
    \includegraphics[width=0.30\textwidth]{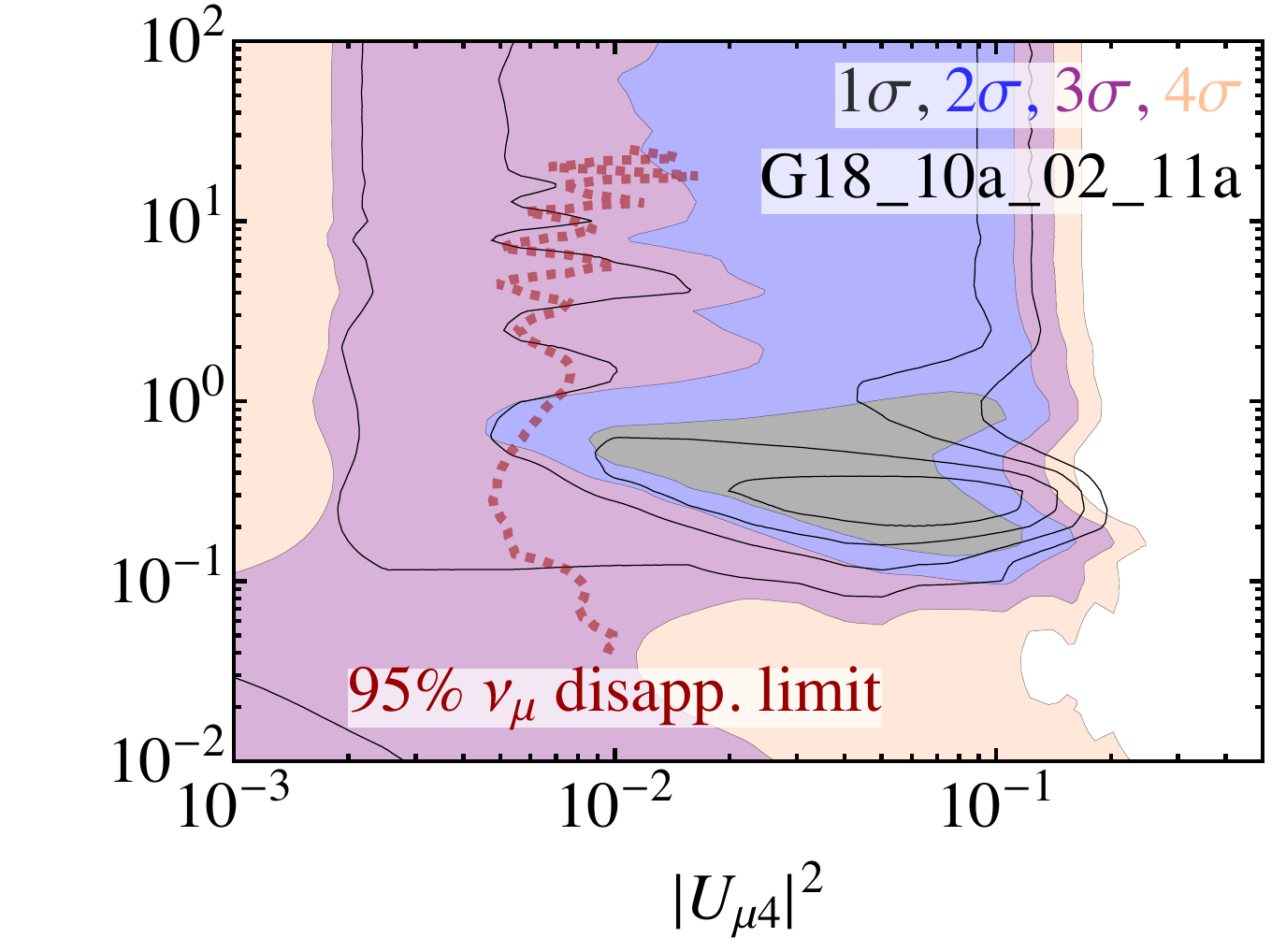} \\
    \includegraphics[width=0.30\textwidth]{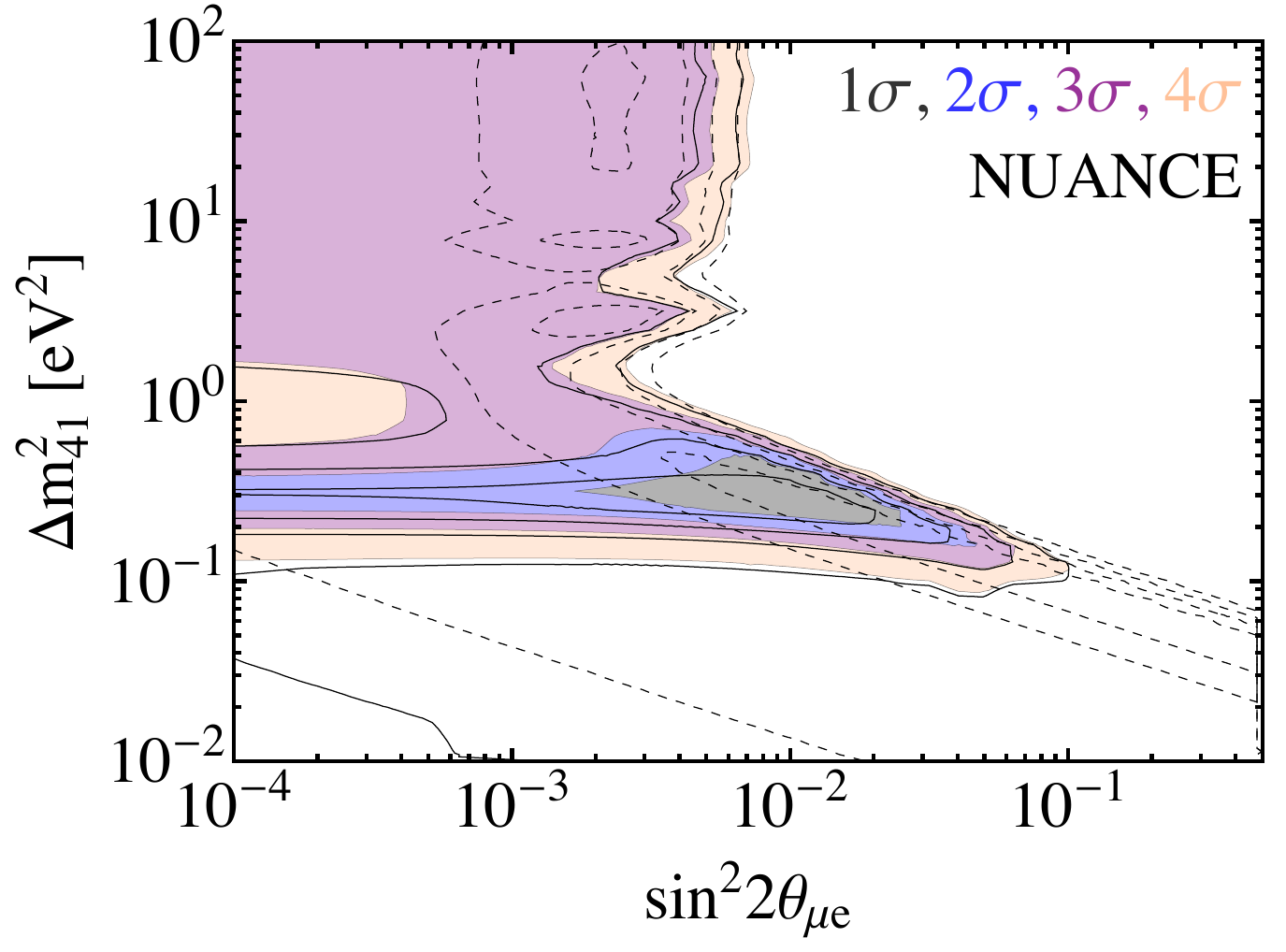} &
    \includegraphics[width=0.30\textwidth]{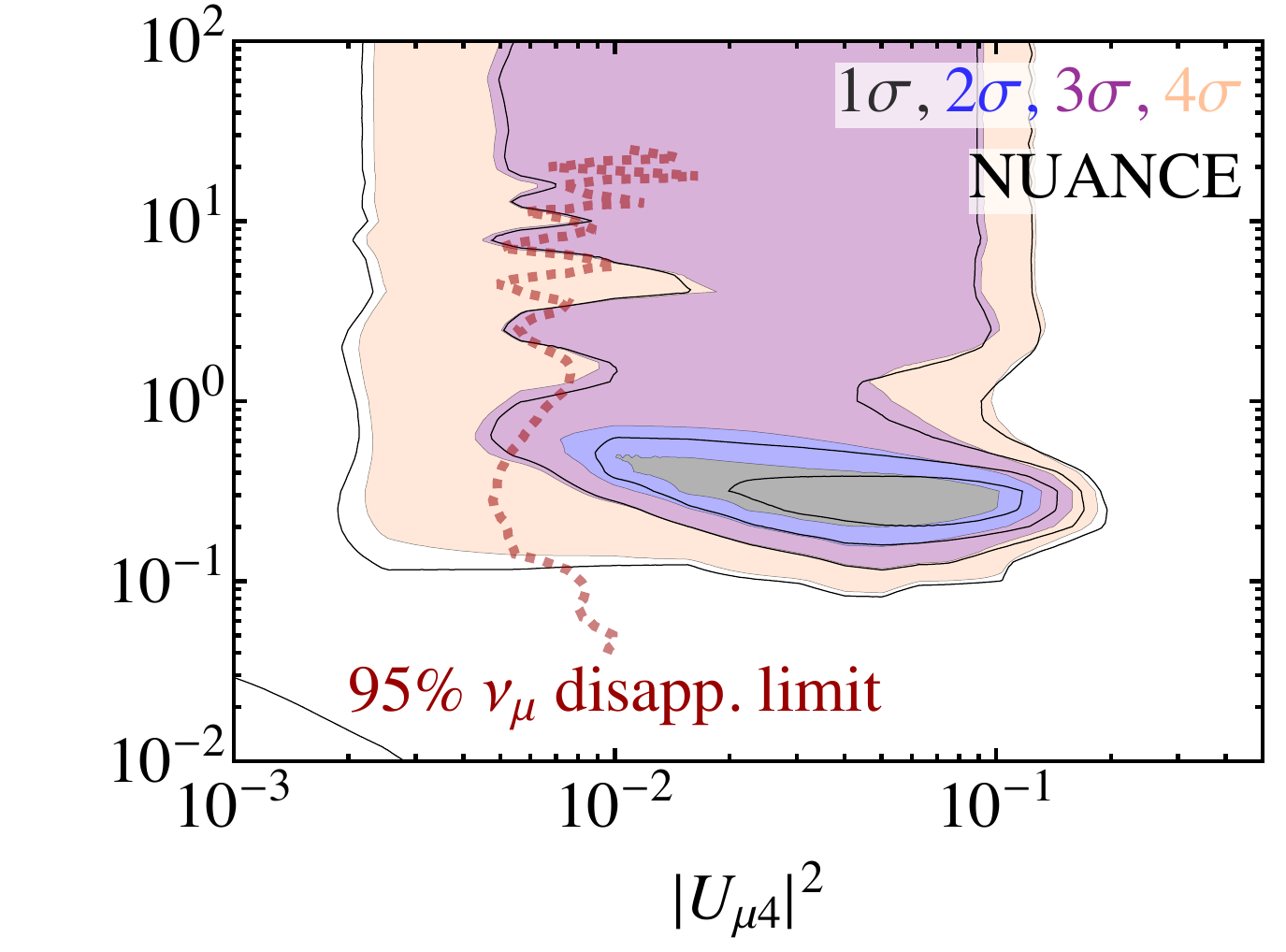} &
    \includegraphics[width=0.30\textwidth]{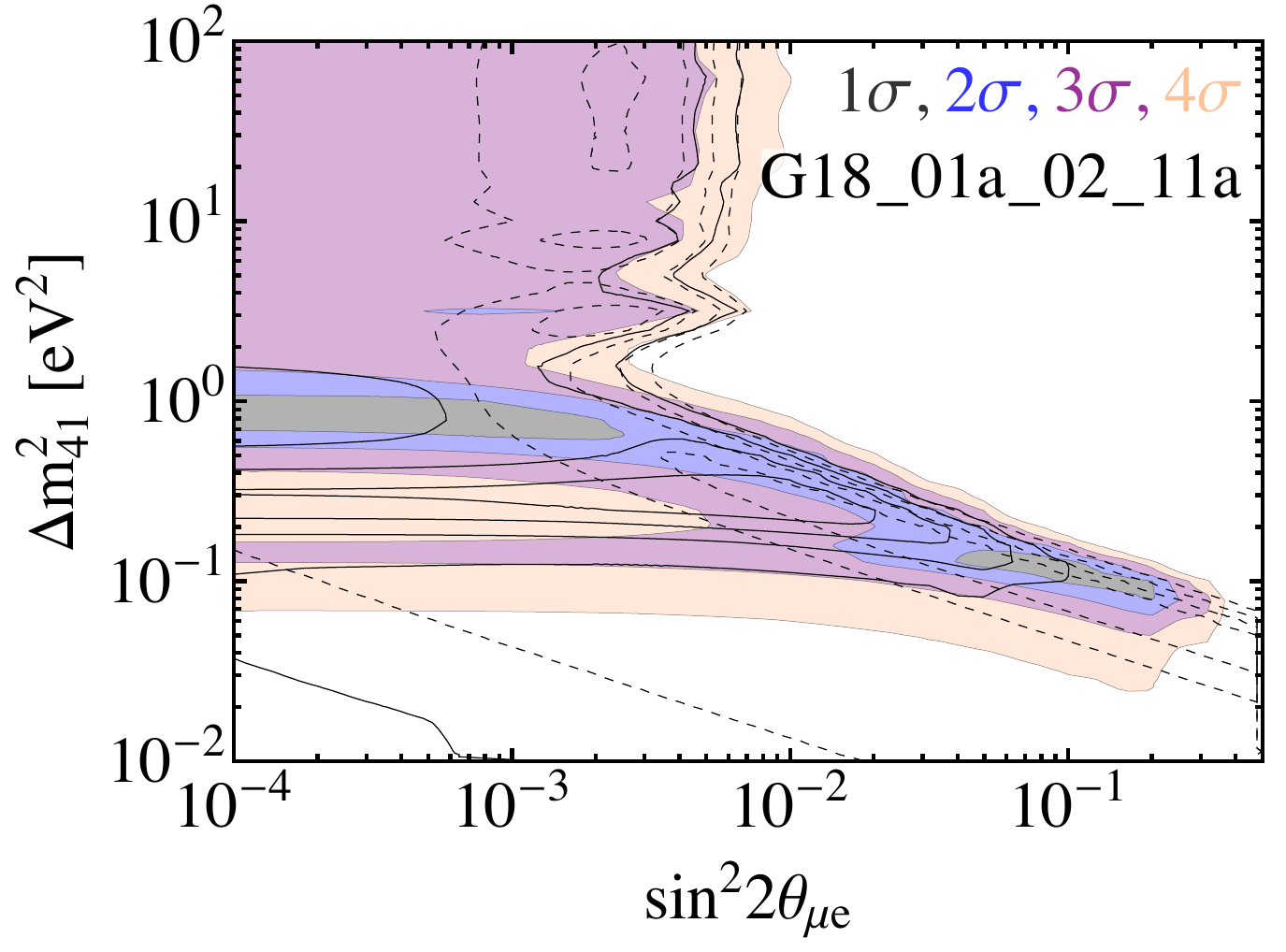} &
    \includegraphics[width=0.30\textwidth]{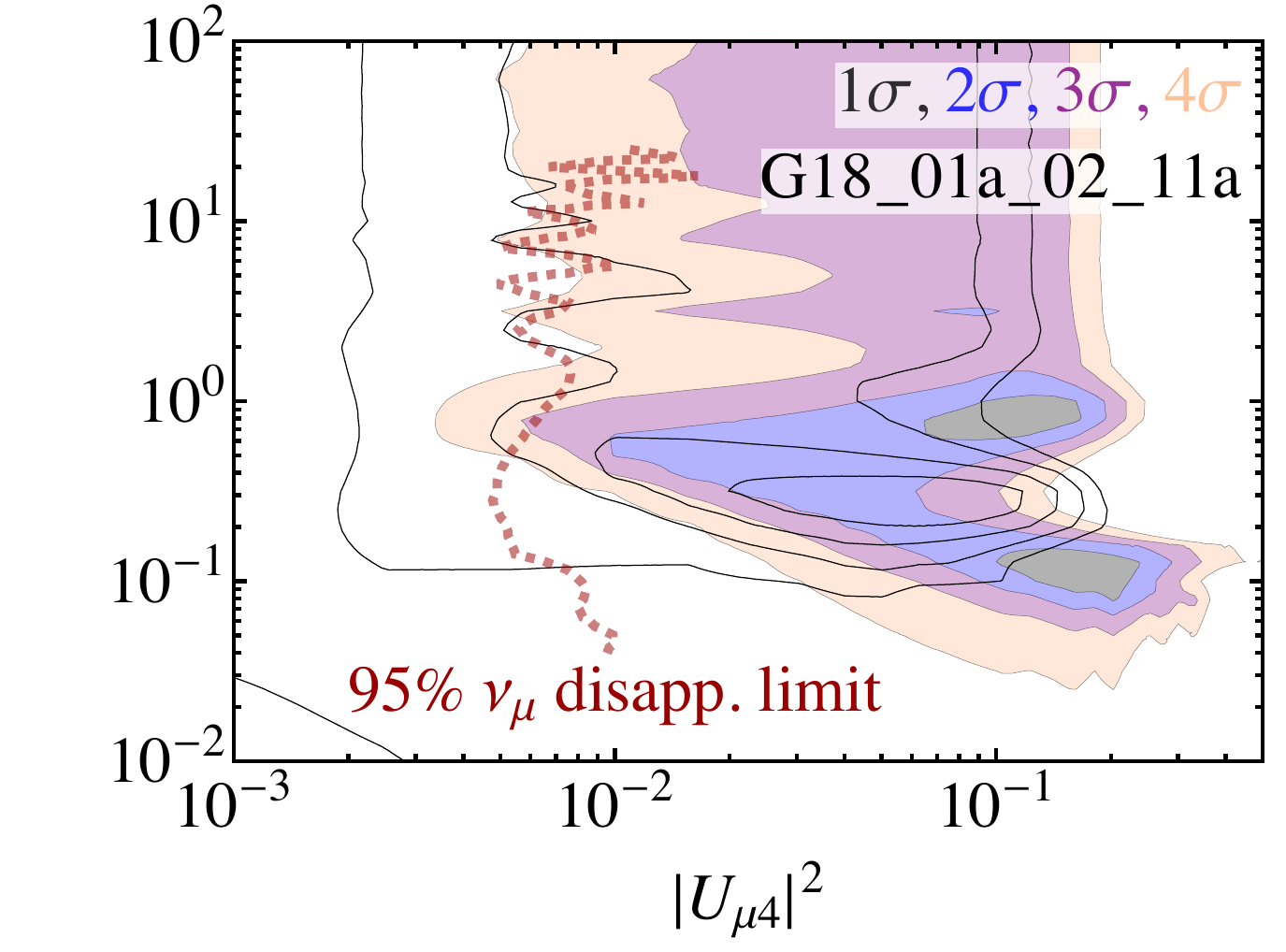} \\
    \includegraphics[width=0.30\textwidth]{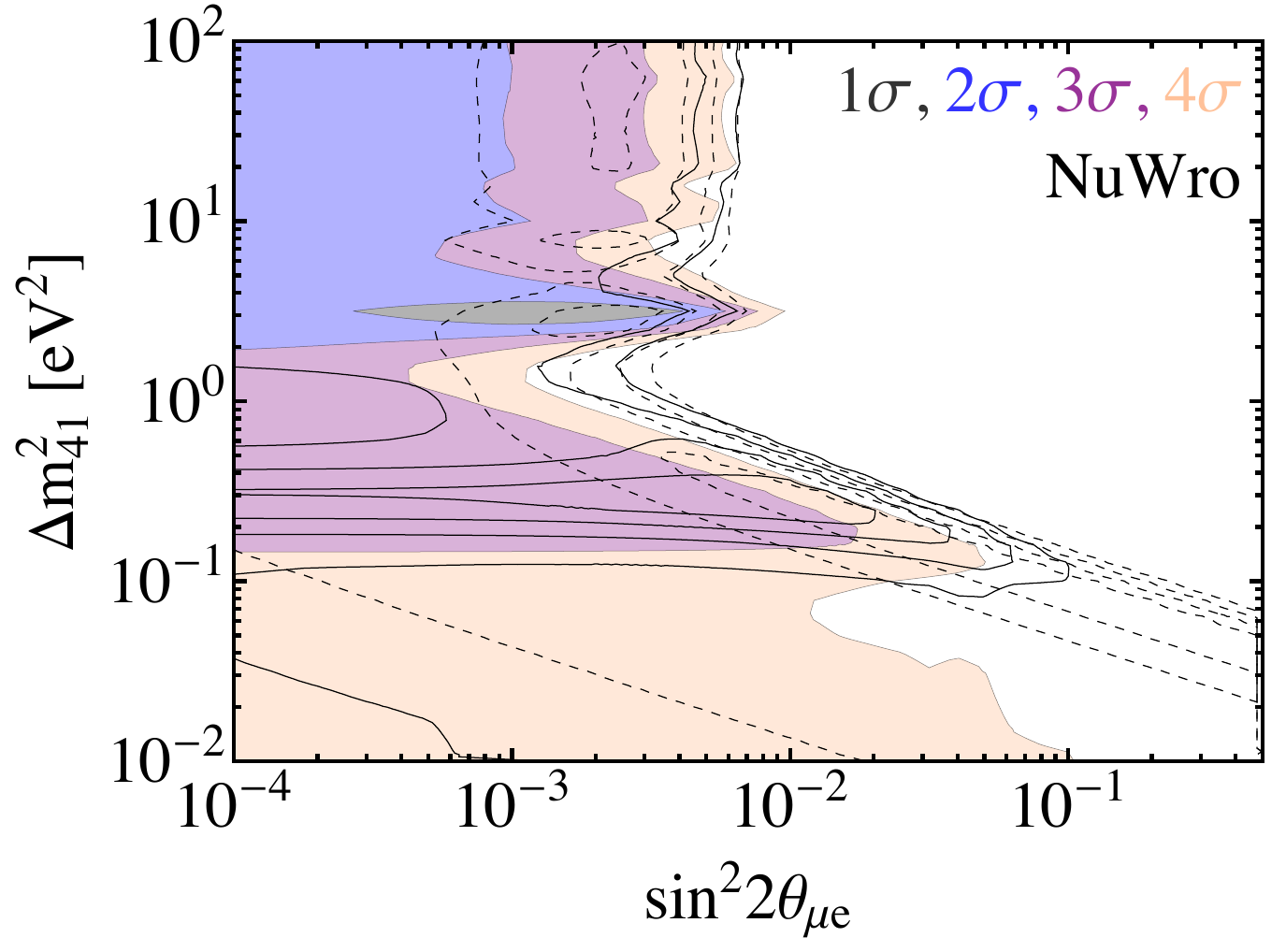} &
    \includegraphics[width=0.30\textwidth]{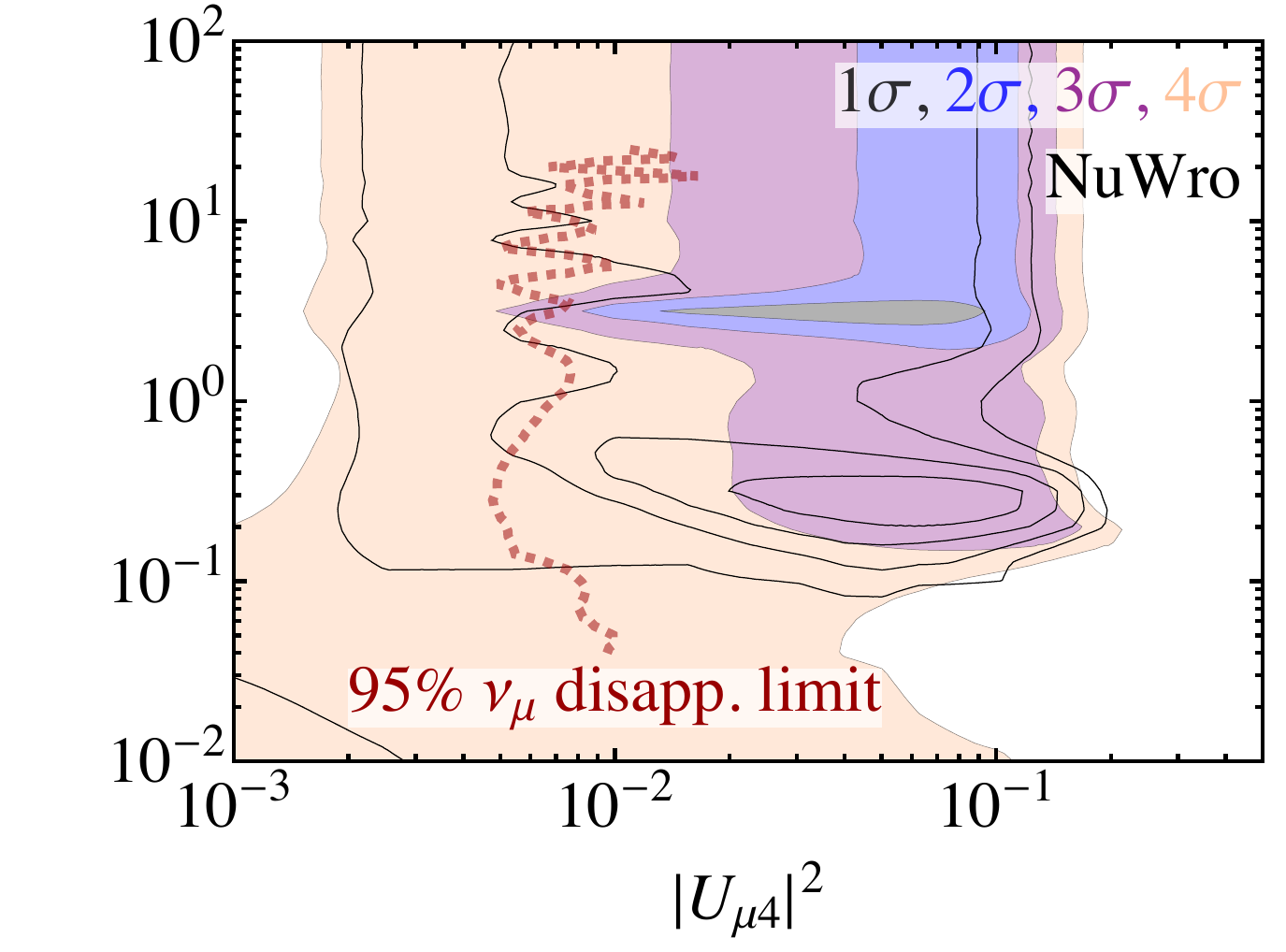} &
    \includegraphics[width=0.30\textwidth]{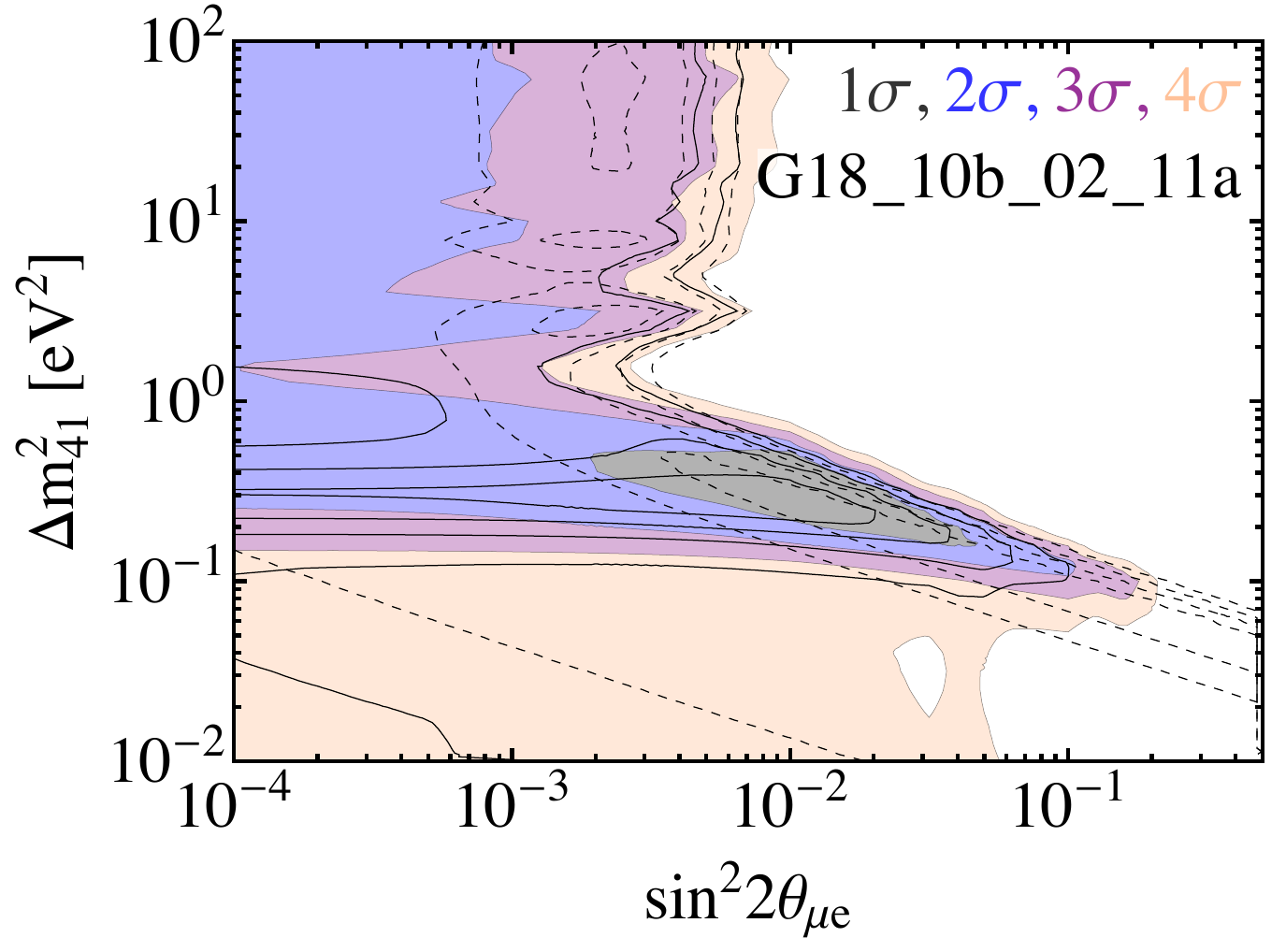} &
    \includegraphics[width=0.30\textwidth]{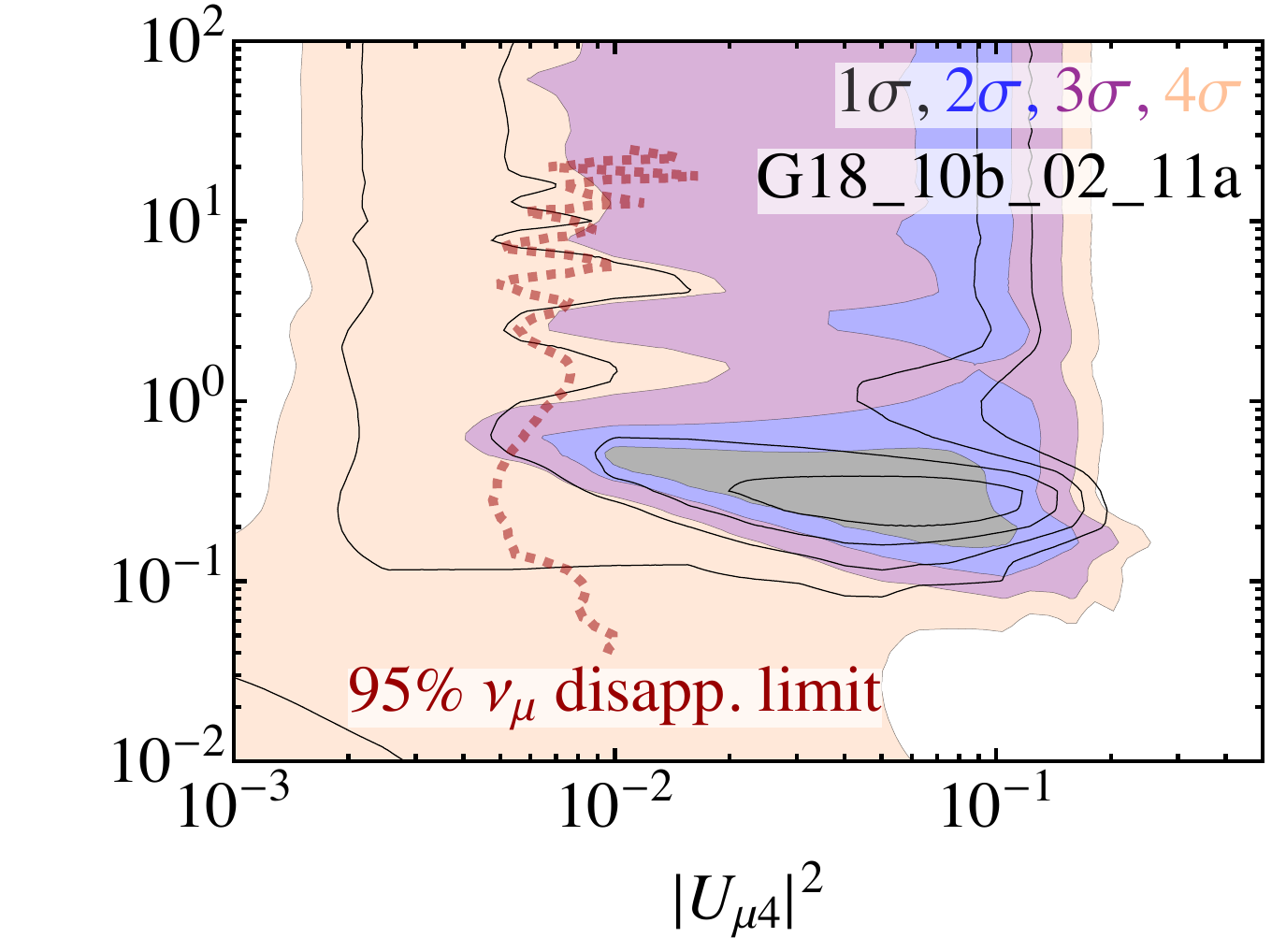} \\
    & &
    \includegraphics[width=0.30\textwidth]{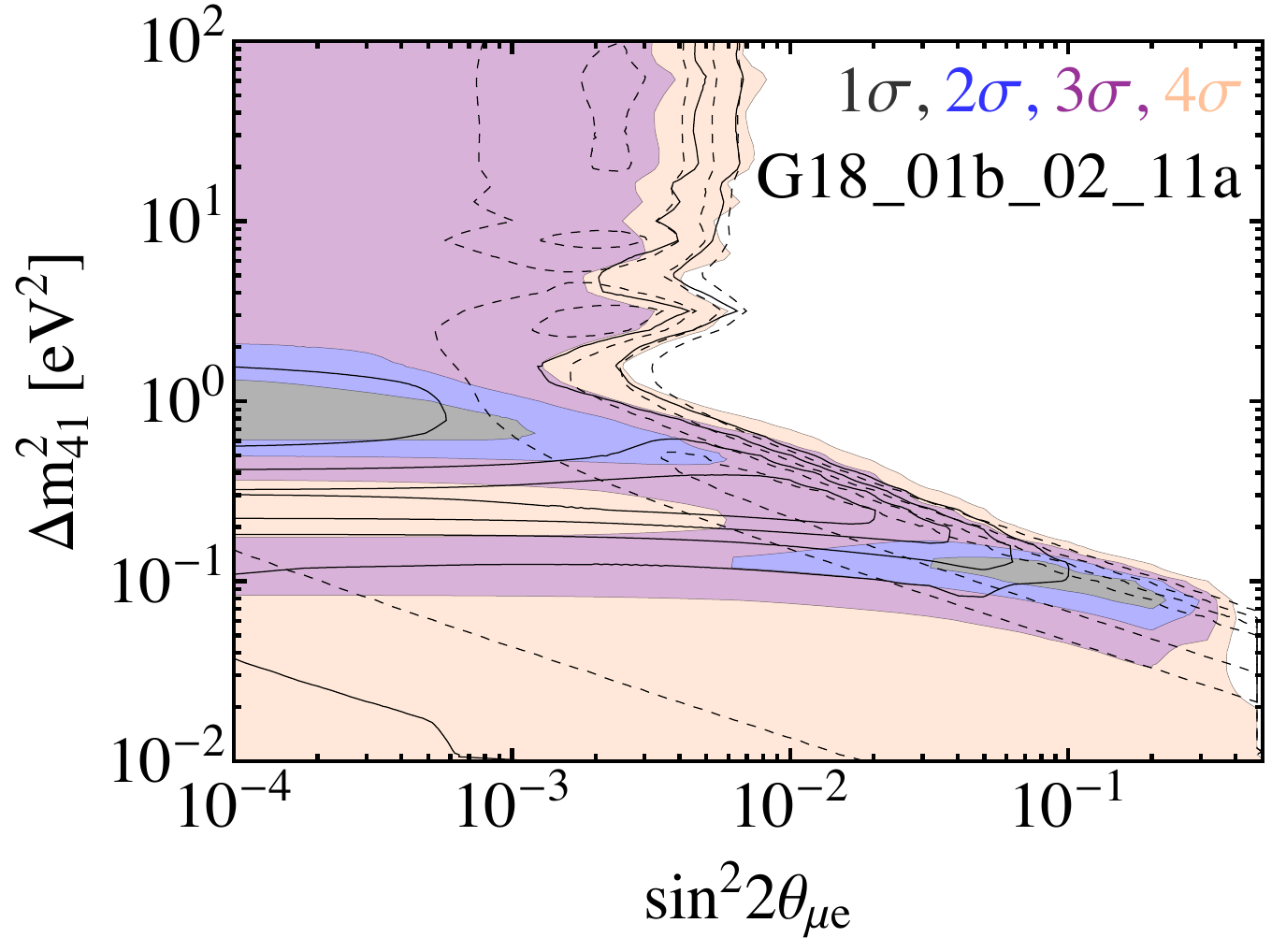} &
    \includegraphics[width=0.30\textwidth]{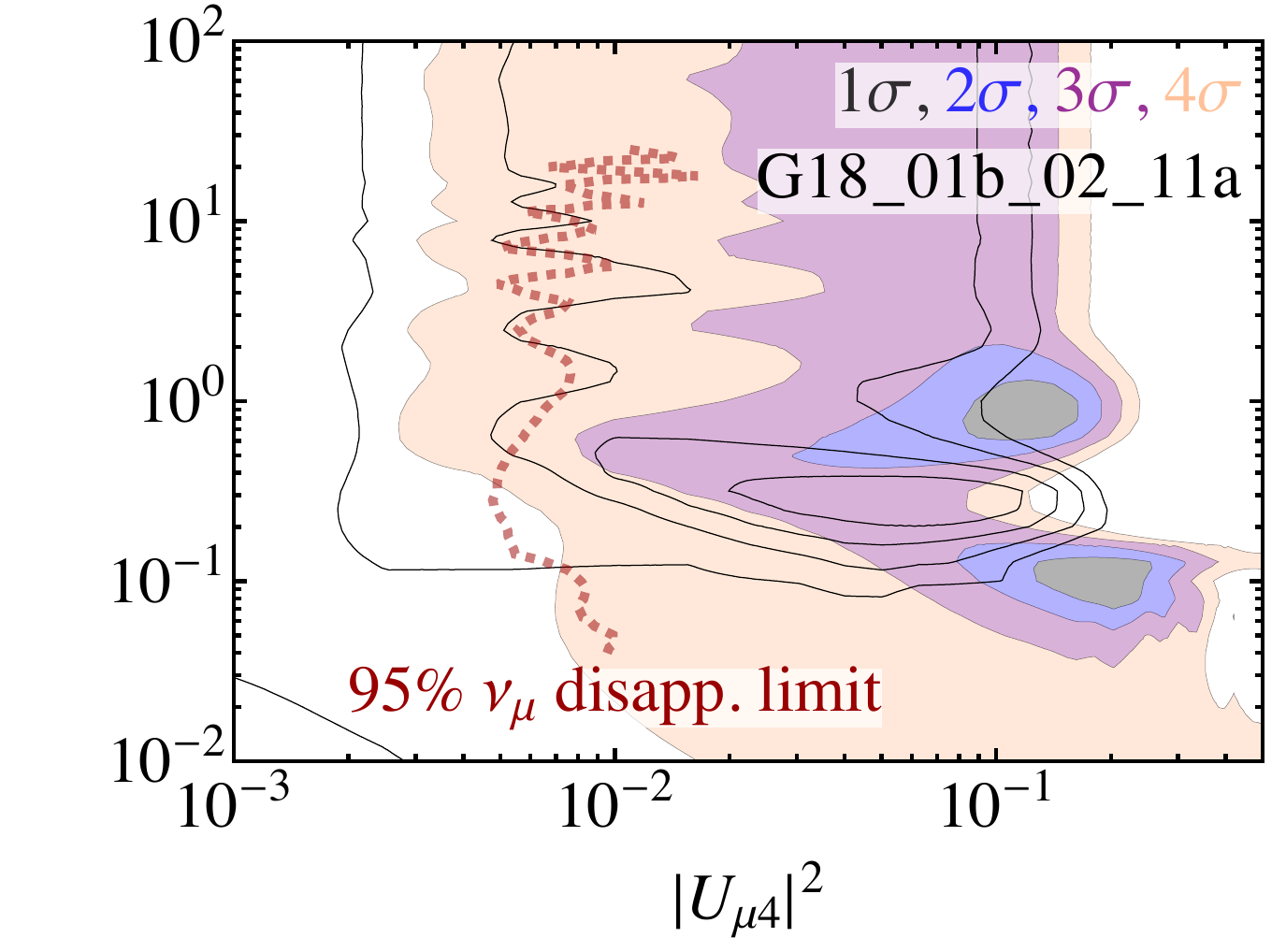}
  \end{tabular}}
  \caption{Shaded contours: MiniBooNE exclusion contours in the
    $\sin^2 2\theta_{\mu e}$--$\Delta m_{41}^2$ plane and in the $|U_{\mu
    4}|^2$--$\Delta m_{41}^2$ plane for different event generators using Monte
    Carlo predictions \emph{without} tuning to MiniBooNE $\pi^0$ data. Unfilled
    black contours: fit results using MiniBooNE's official background
    predictions.  Red dotted line in the $|U_{\mu 4}|^2$-vs.-$\Delta m_{41}^2$
    panels: global 95\% $\nu_\mu$ disappearance limit from
    ref.~\cite{Dentler:2018sju}.
  }
  \label{fig:mb-contours-fp}
\end{figure}

\begin{figure}
  \centering
  \textbf{\large Data-Driven Backgrounds} \\[0.2cm]
  \hspace*{-1.5cm}
  \begin{tabular}{c@{\hspace{-0.4cm}}c|c@{\hspace{-0.4cm}}c}
    \includegraphics[width=0.30\textwidth]{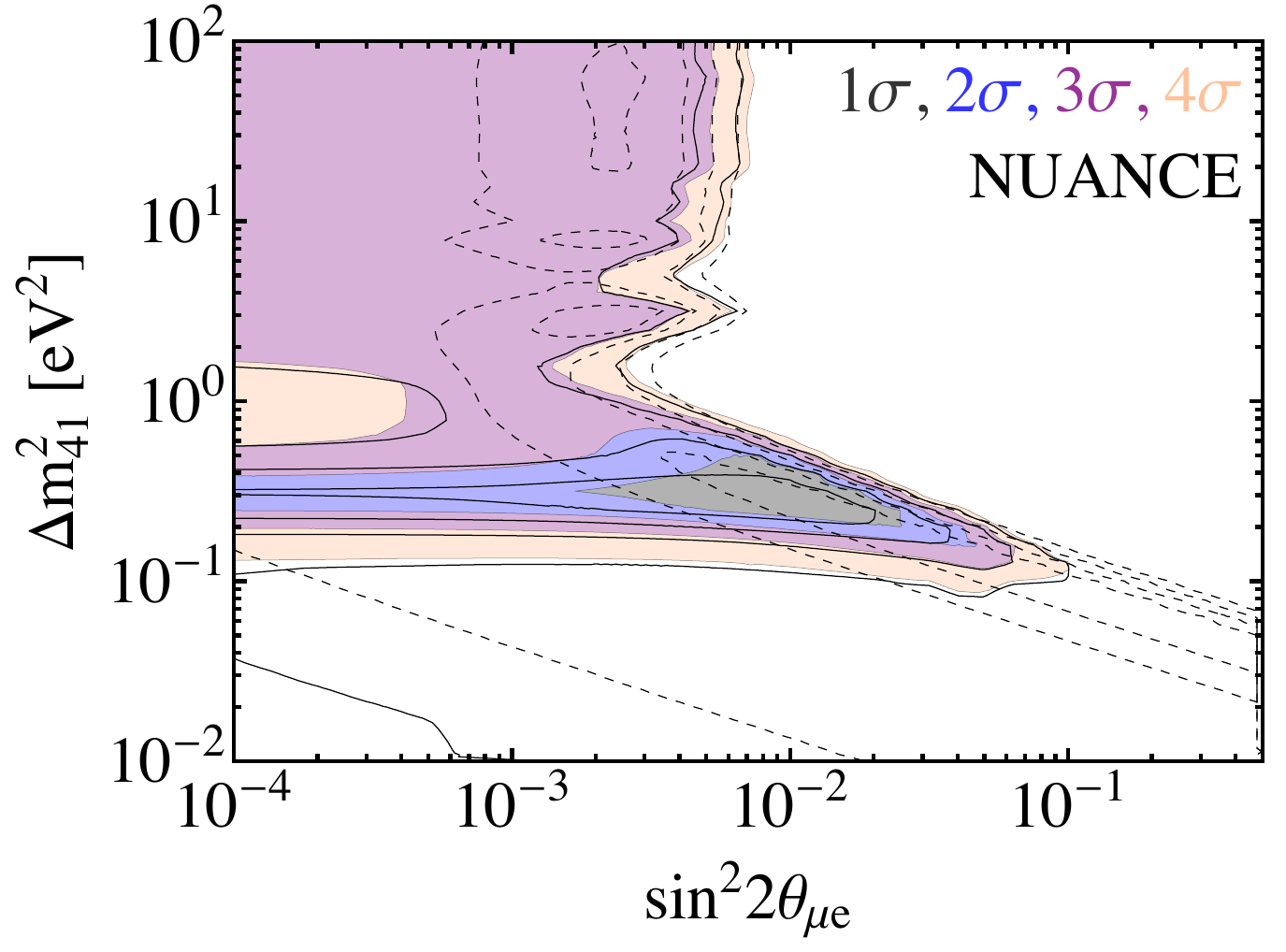} &
    \includegraphics[width=0.30\textwidth]{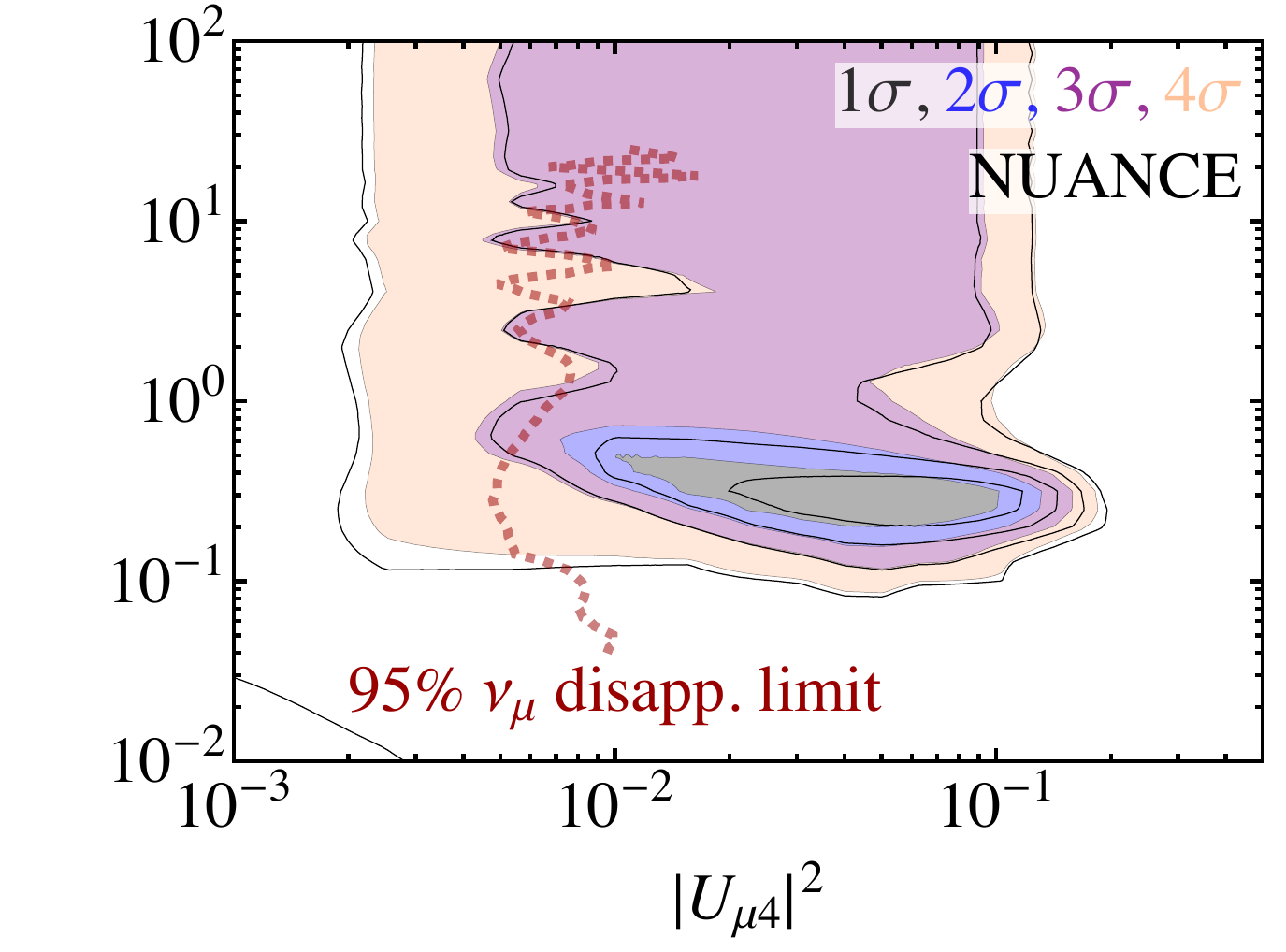} &
    \includegraphics[width=0.30\textwidth]{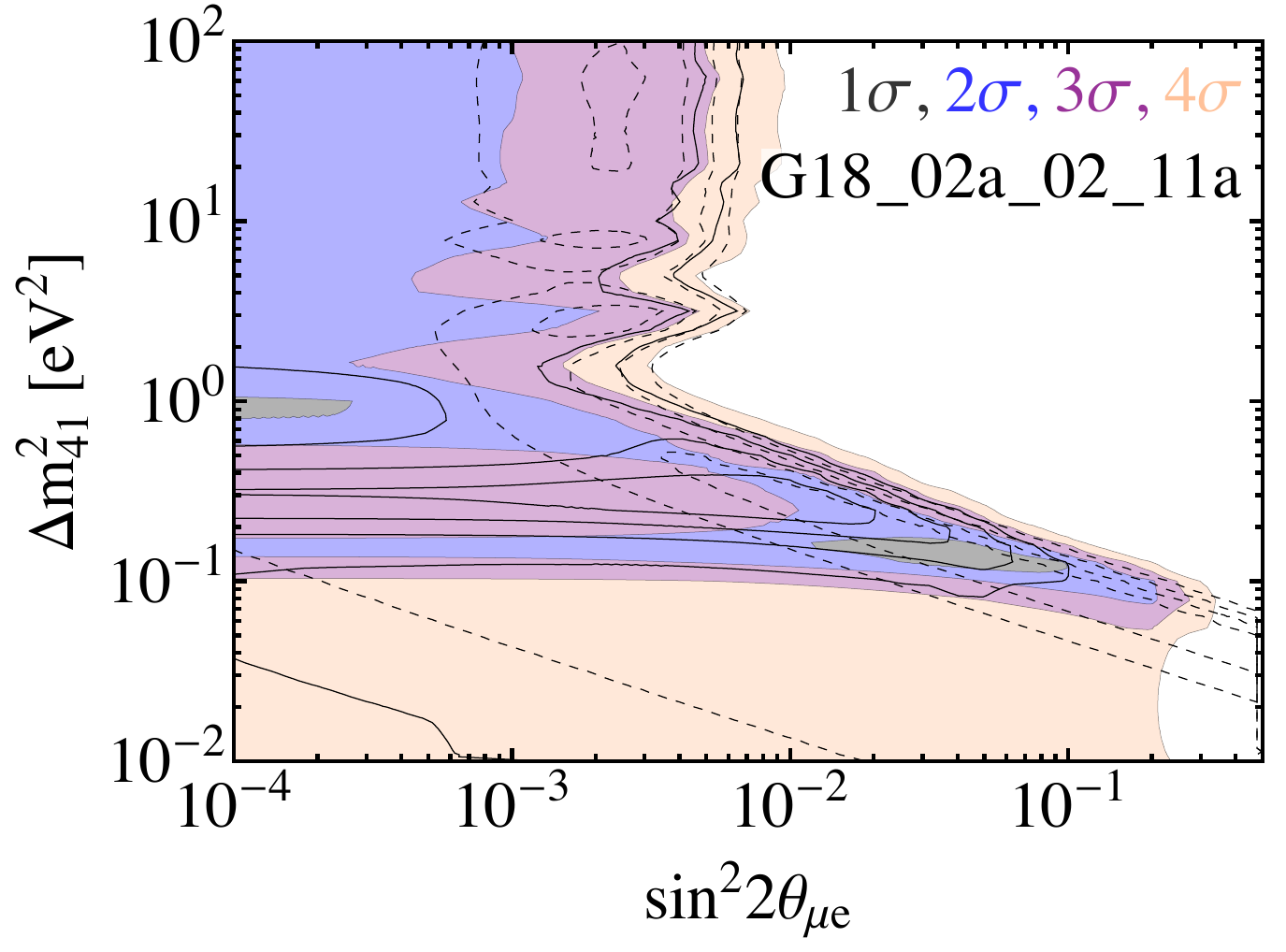} &
    \includegraphics[width=0.30\textwidth]{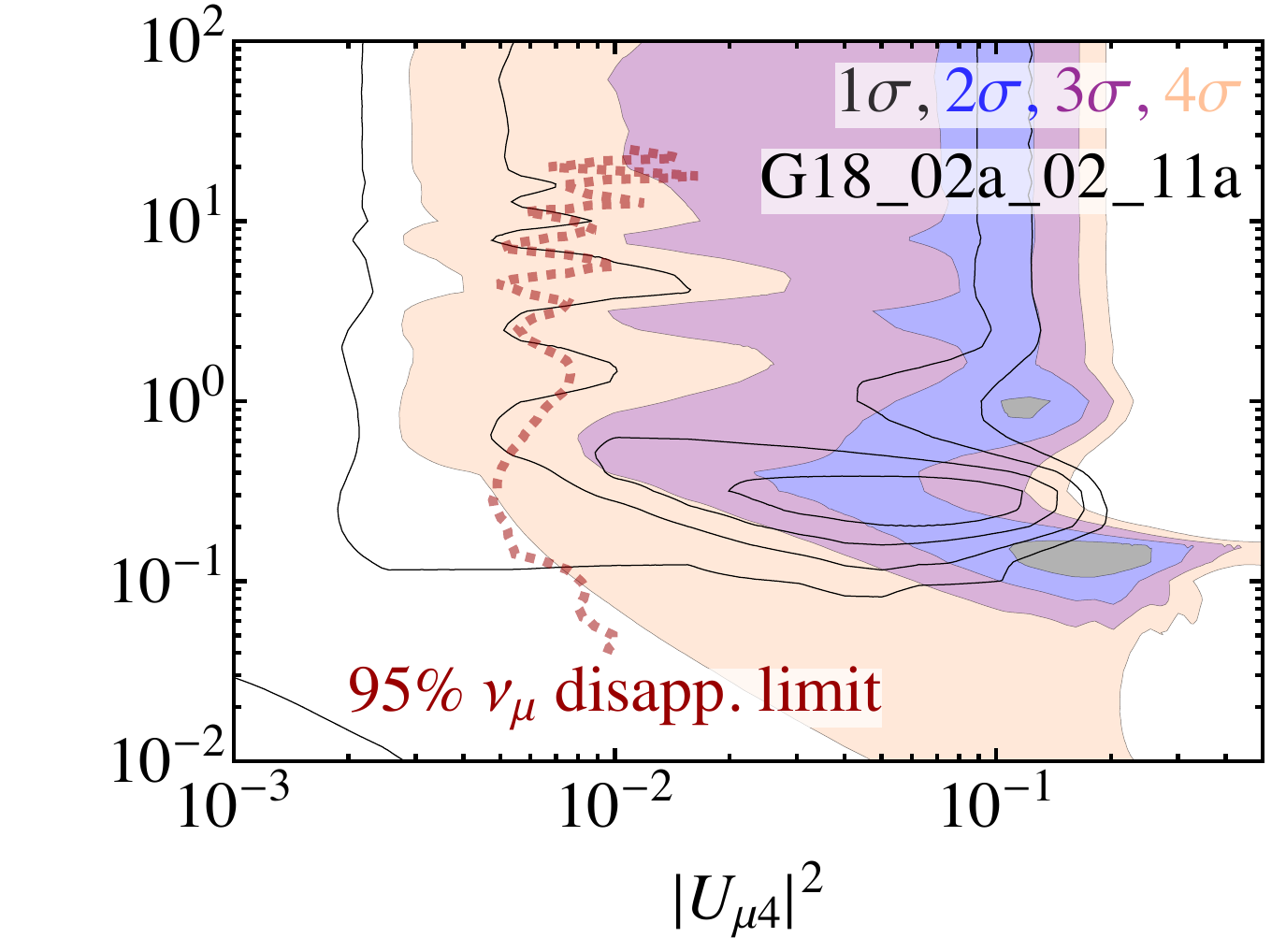} \\
    \includegraphics[width=0.30\textwidth]{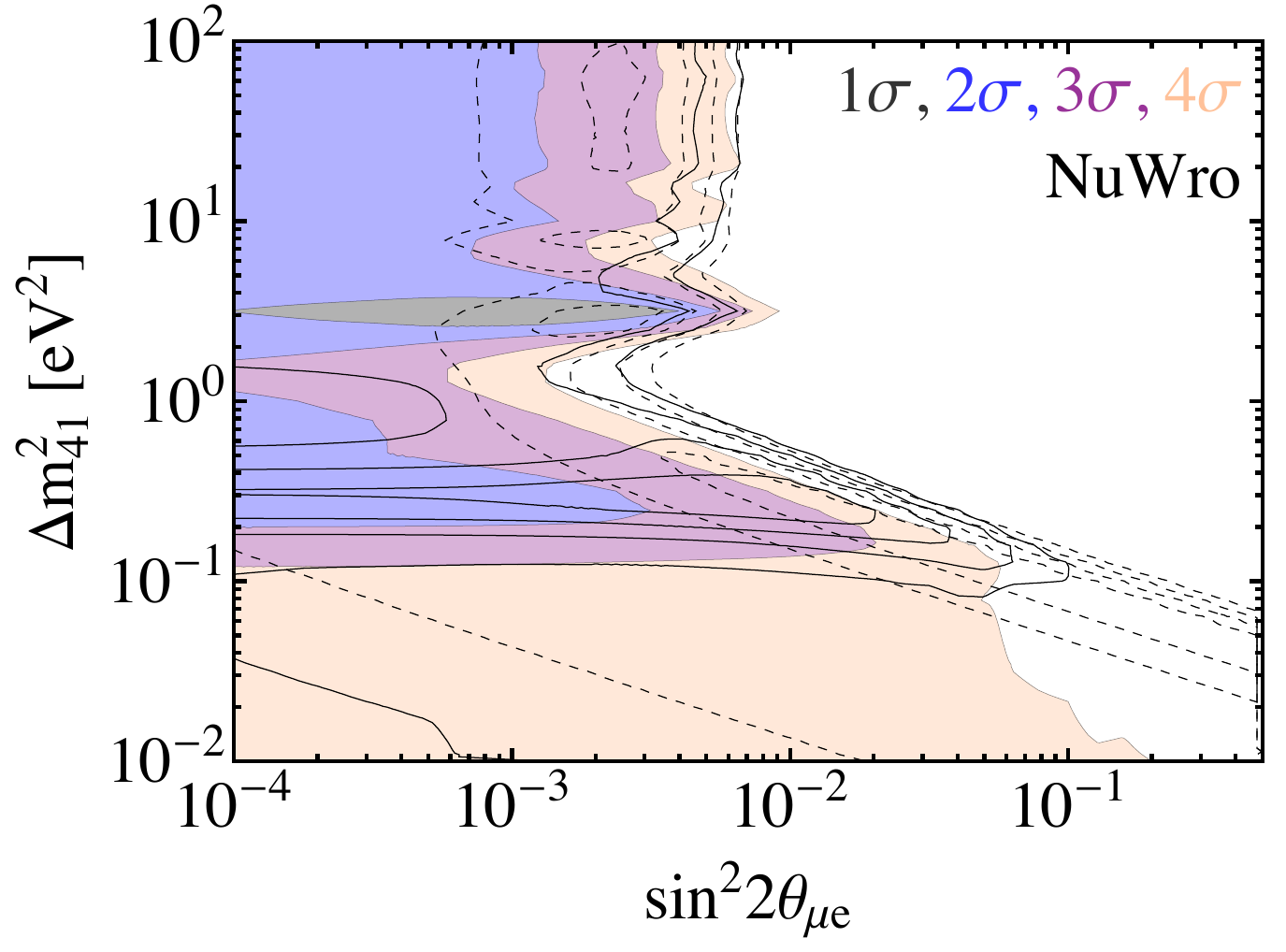} &
    \includegraphics[width=0.30\textwidth]{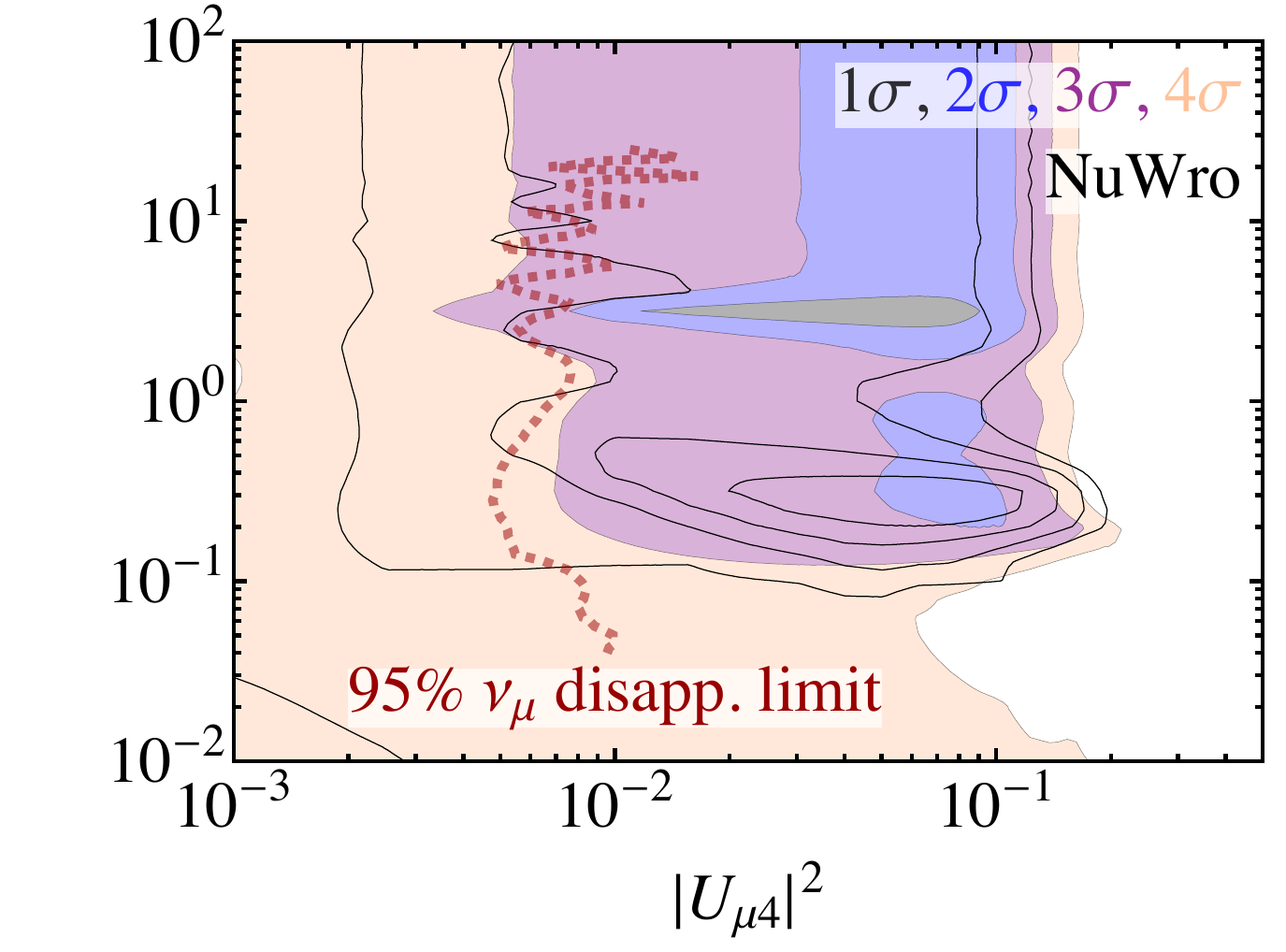} &
    \includegraphics[width=0.30\textwidth]{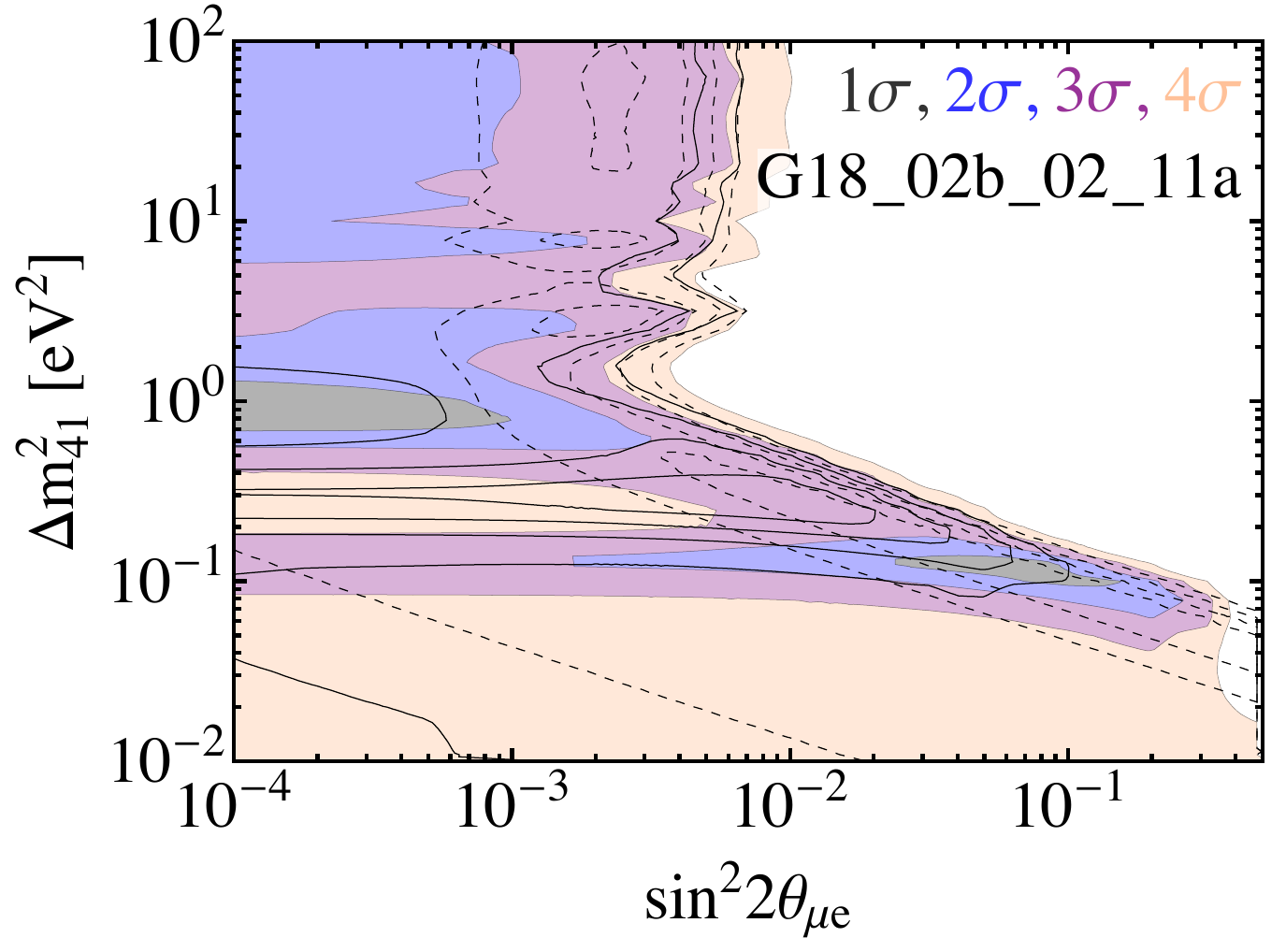} &
    \includegraphics[width=0.30\textwidth]{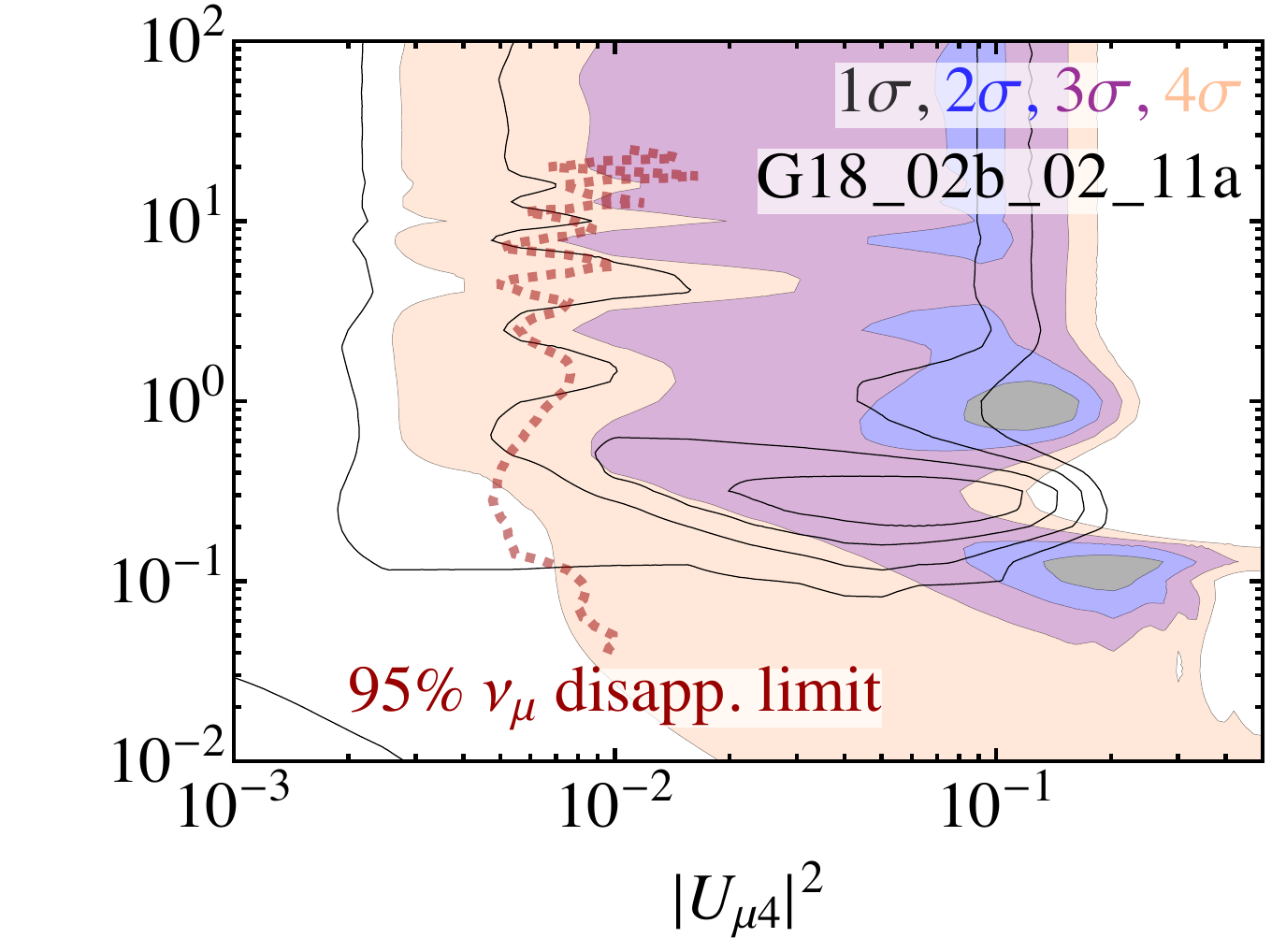} \\
    \includegraphics[width=0.30\textwidth]{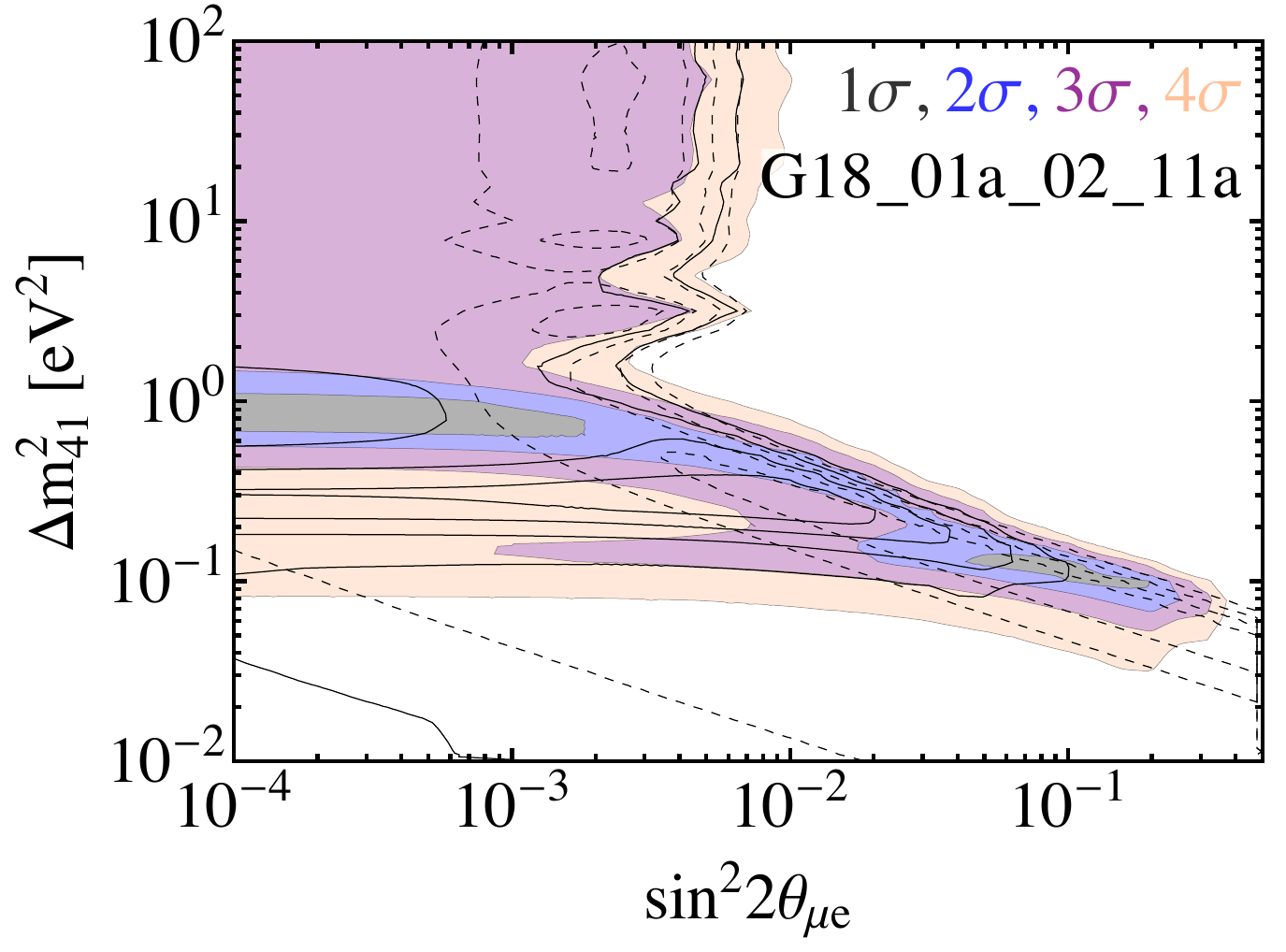} &
    \includegraphics[width=0.30\textwidth]{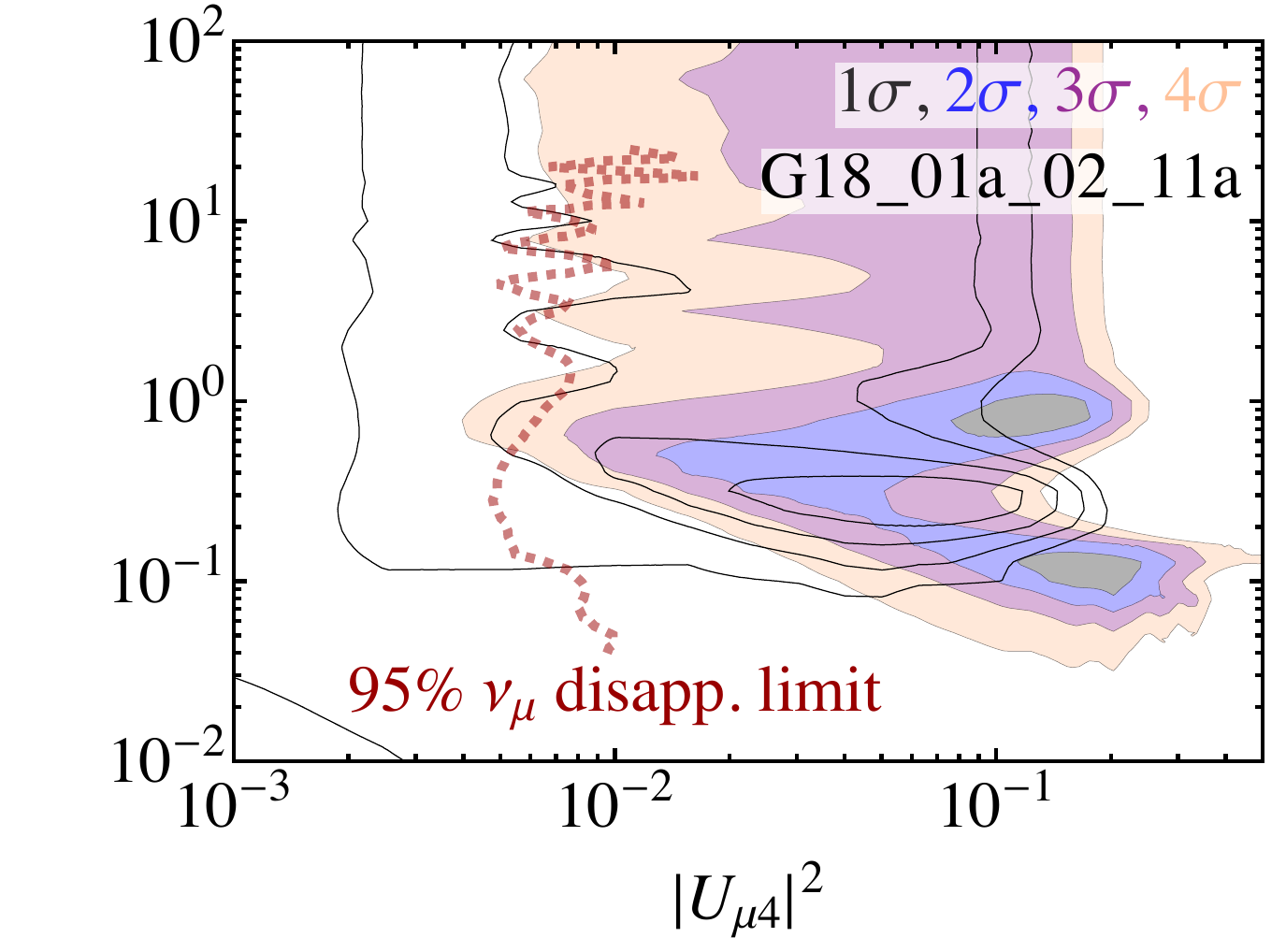} &
    \includegraphics[width=0.30\textwidth]{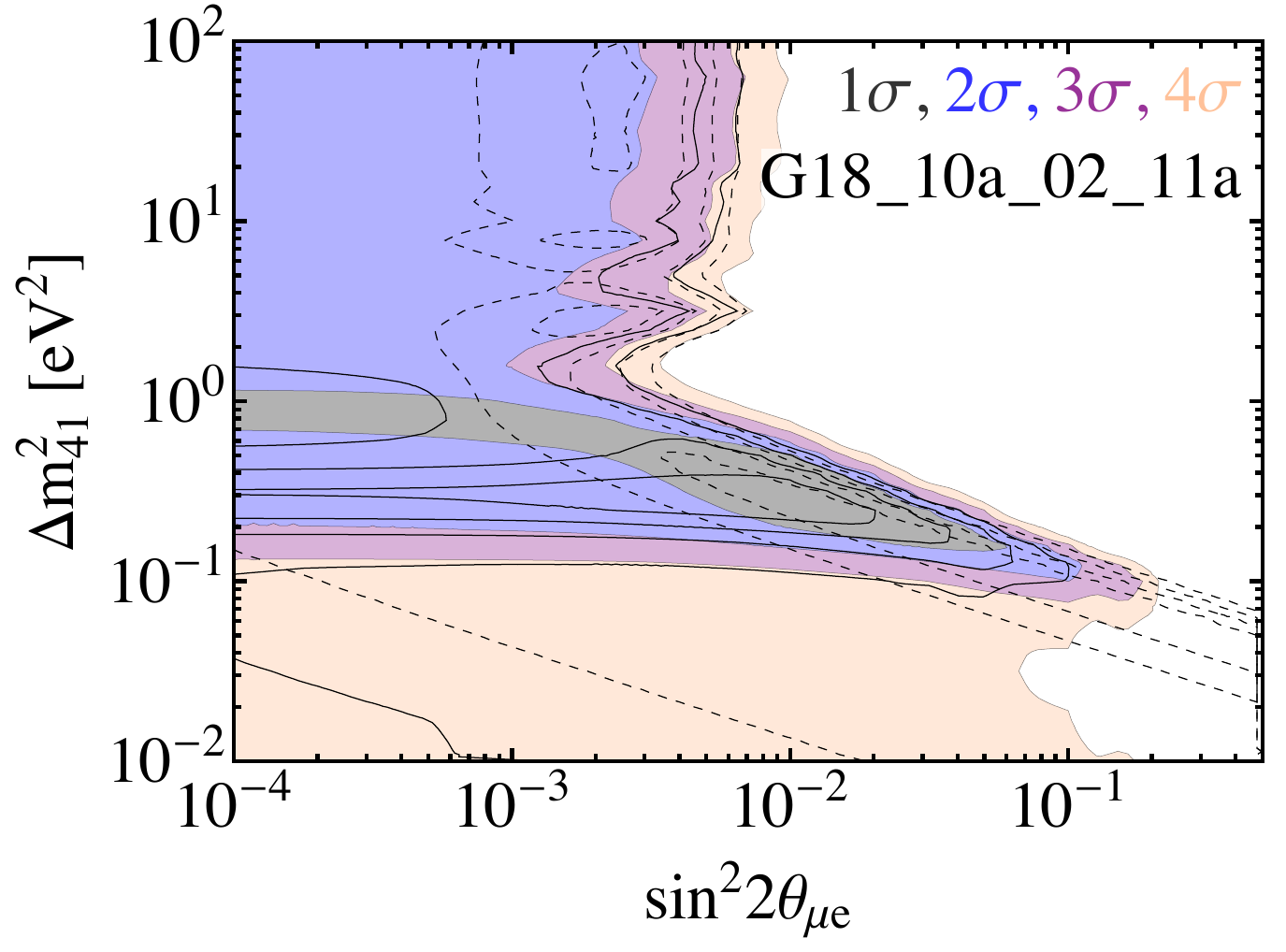} &
    \includegraphics[width=0.30\textwidth]{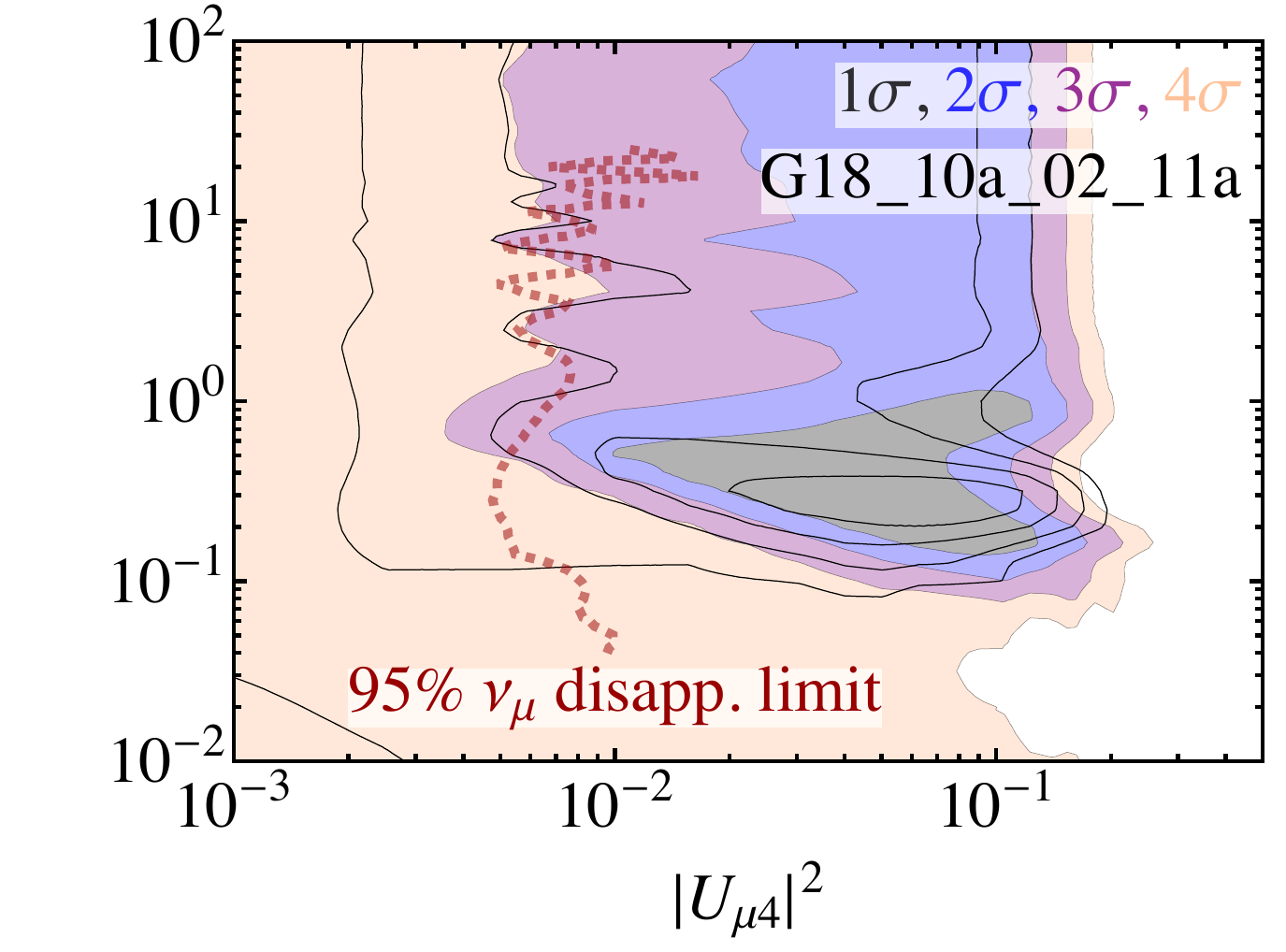} \\
    \includegraphics[width=0.30\textwidth]{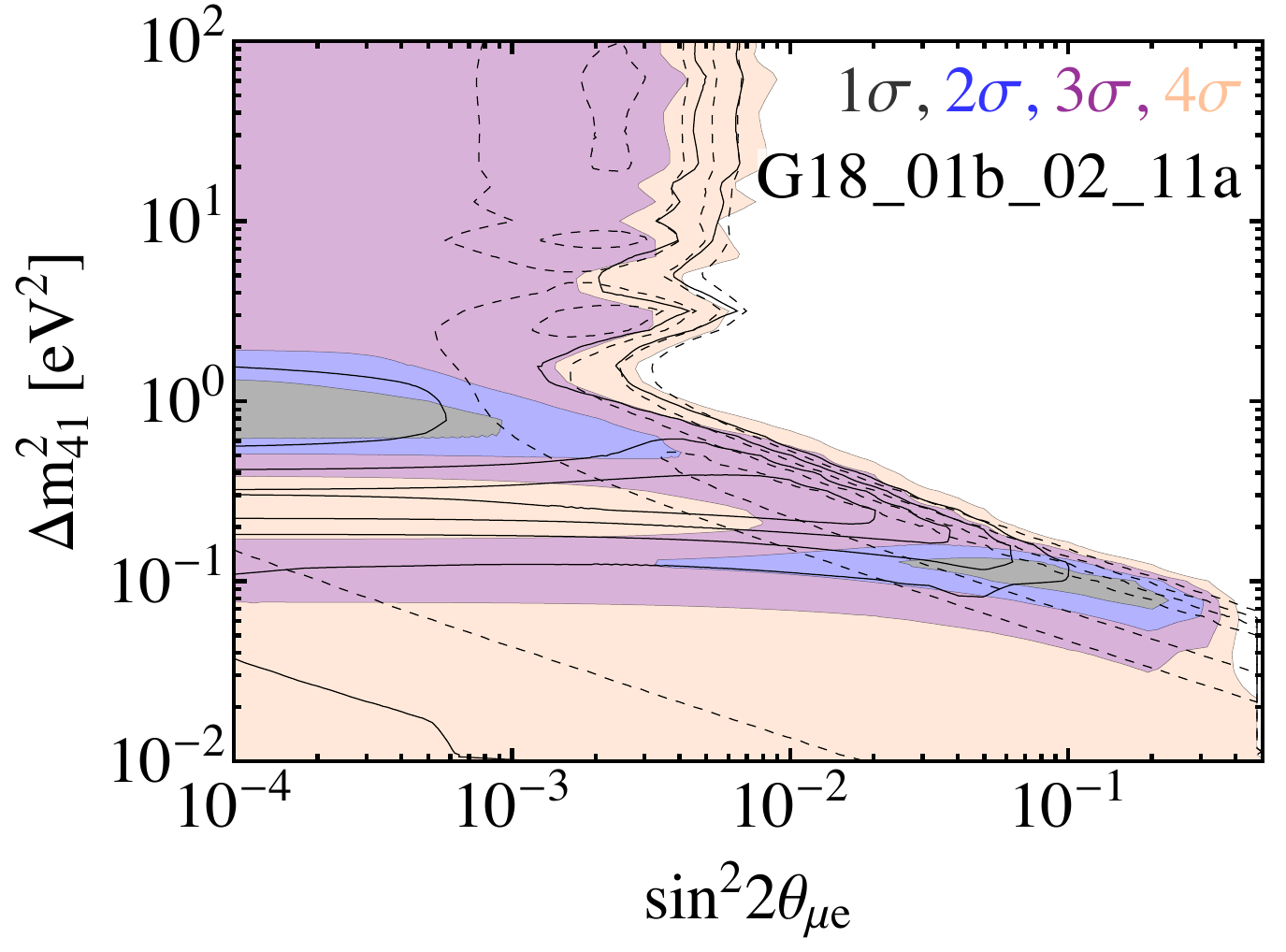} &
    \includegraphics[width=0.30\textwidth]{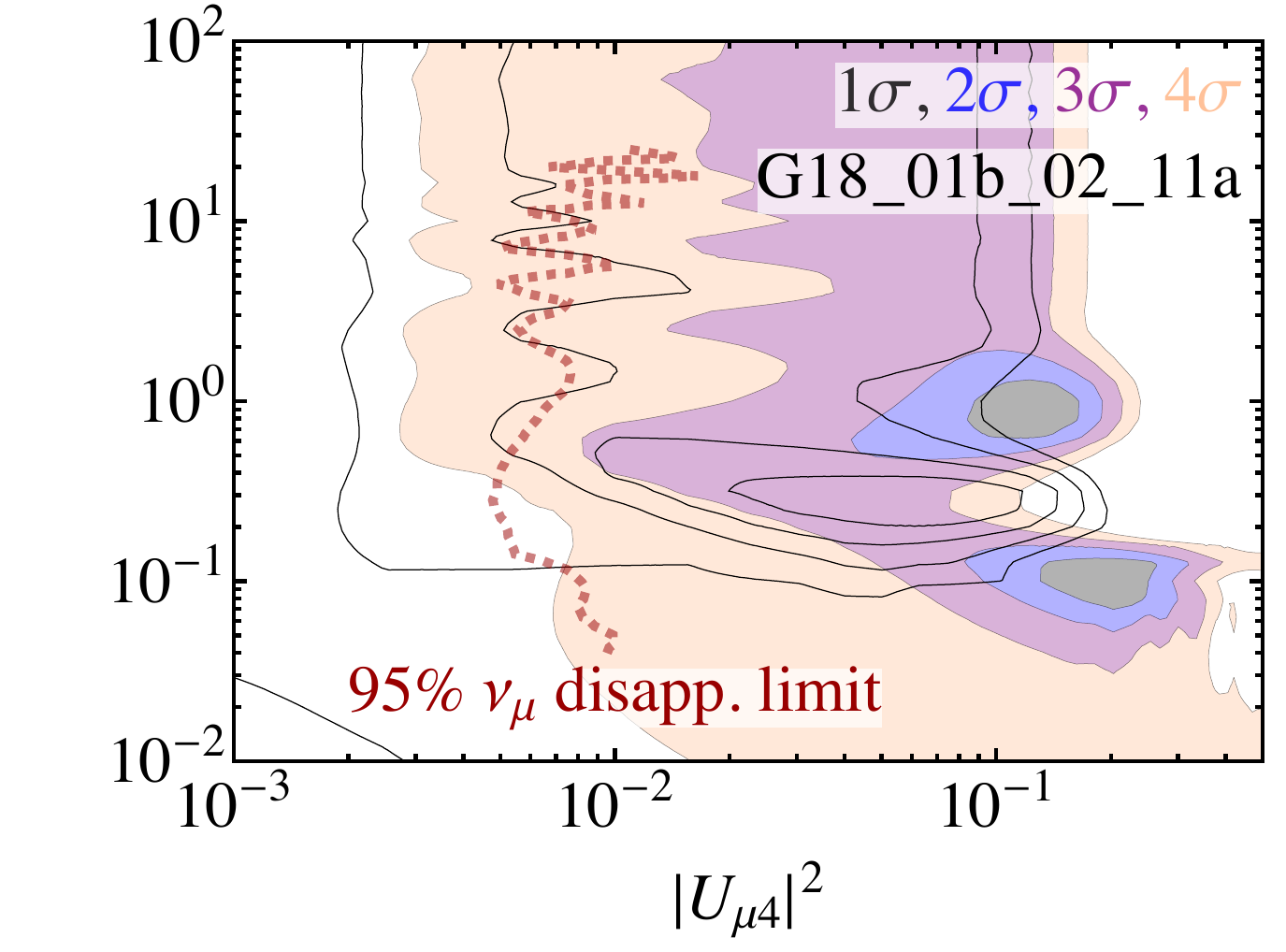} &
    \includegraphics[width=0.30\textwidth]{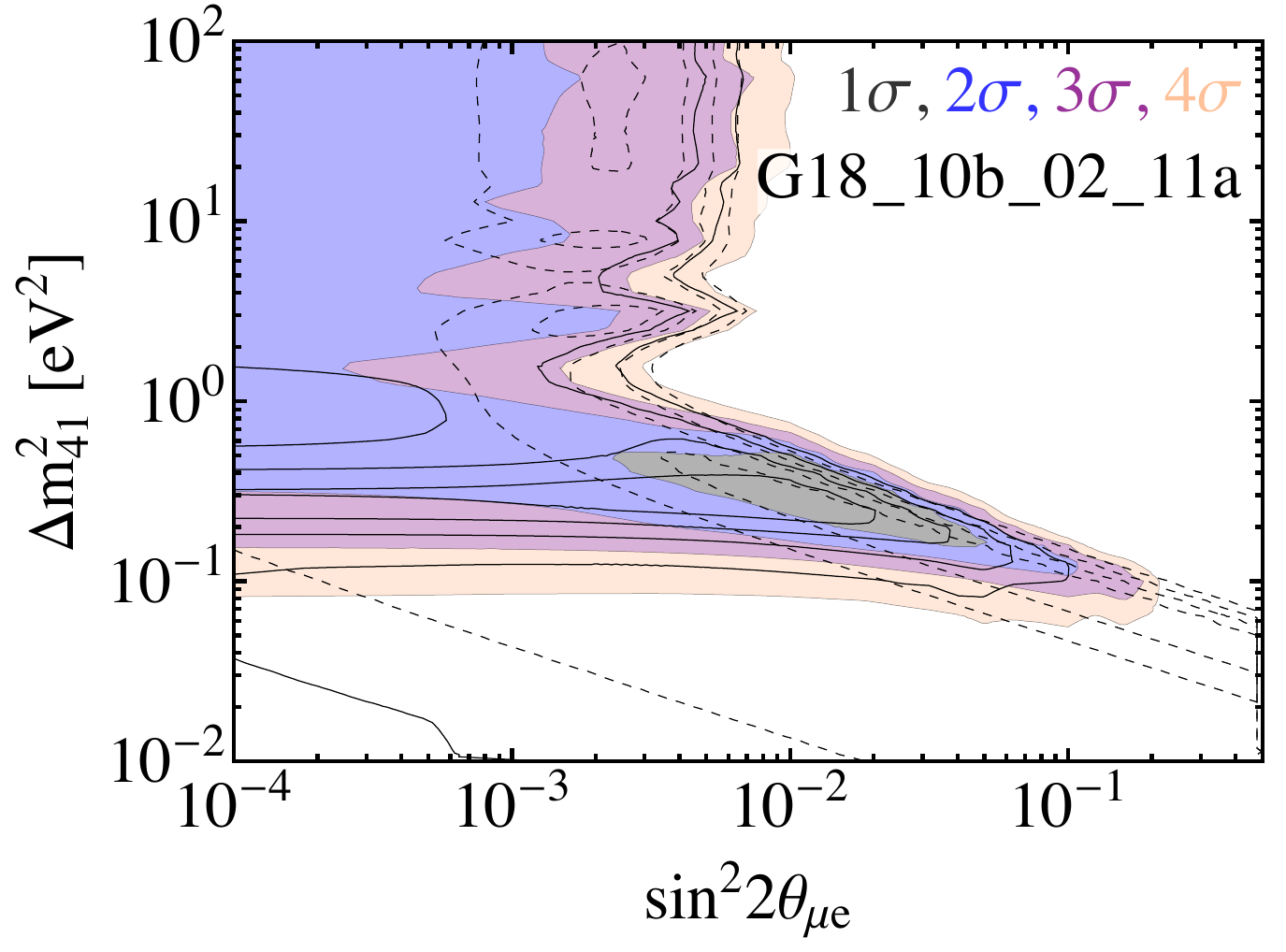} &
    \includegraphics[width=0.30\textwidth]{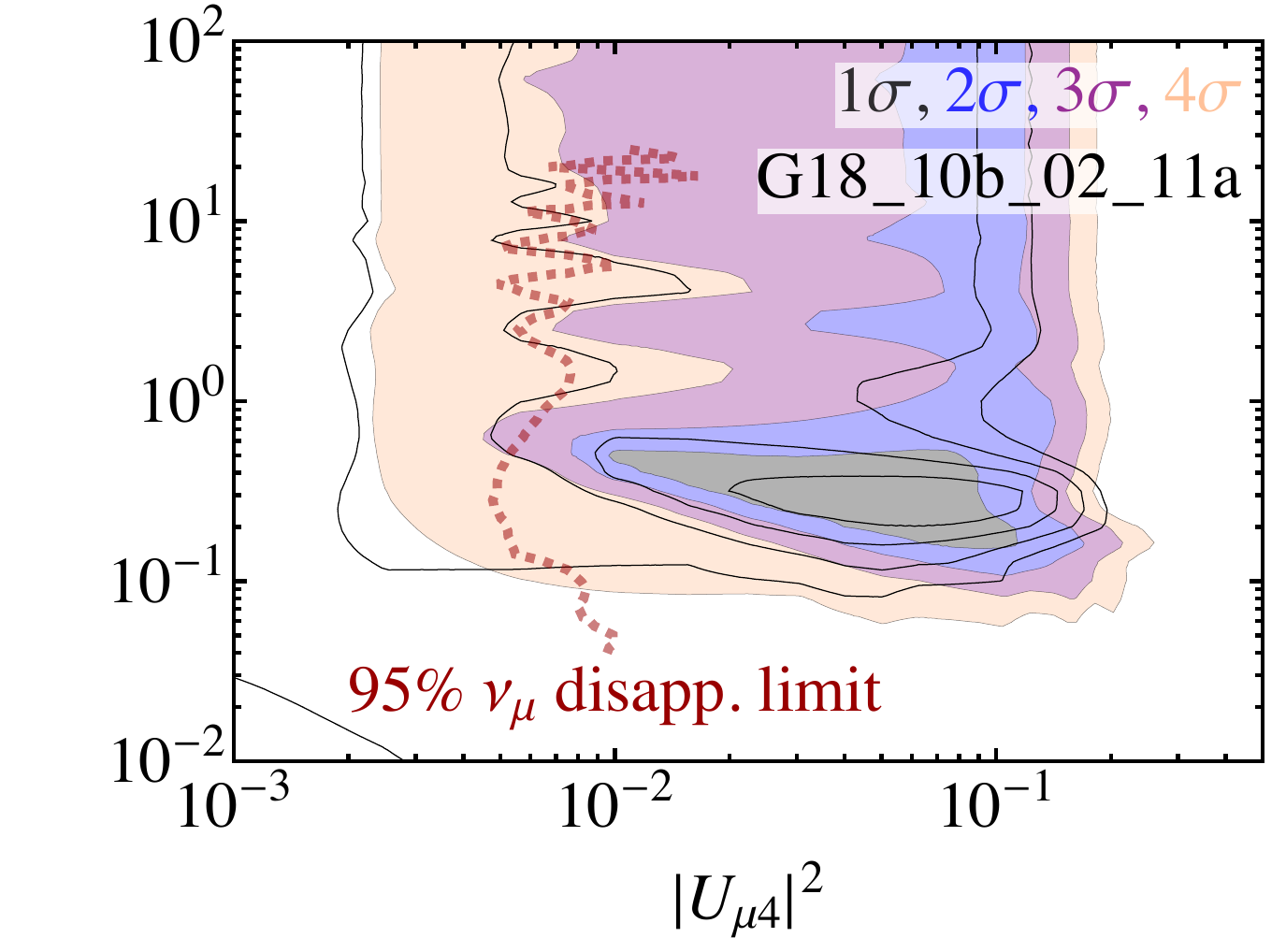}
  \end{tabular}
  \caption{Same as \cref{fig:mb-contours-fp}, but for data-driven background
    predictions in the $\pi^0$ and single-photon channels.}
  \label{fig:mb-contours-dd}
\end{figure}

\noindent
In this appendix, we supplement the discussion in \cref{sec:fits-generator-dependence}
by providing parameter space exclusion plots for all Monte Carlo
generators and tunes studied in this paper in the $3+1$ sterile neutrino
scenario. In particular, we show in \cref{fig:mb-contours-fp} our
fit results using Monte Carlo-only background predictions, that is predictions
which are not tuned to
MiniBooNE's own $\pi^0$ production data. In \cref{fig:mb-contours-dd}
we do the same using data-driven predictions backgrounds for the $\pi^0$
and single-photon channels, which have been tuned to $\pi^0$ data.
For each background model, we show the projections of the $3+1$ model's
parameter space onto the $\sin^2 2\theta_{\mu e}$--$\Delta m_{41}^2$
plane and onto the $|U_{\mu 4}|^2$--$\Delta m_{41}^2$ plane.
(The SM oscillation parameters are irrelevant here because MiniBooNE's
baseline is too short for SM oscillations to develop, and the mixing
matrix element $U_{\tau 4}$ is irrelevant due to the absence of $\tau$
neutrinos.)

\bibliographystyle{JHEP}
\bibliography{refs}

\providecommand{\href}[2]{#2}\begingroup\raggedright\begin{thebibliography}{100}

\bibitem{Church:1997ry}
{\bf BooNE} {\bf Collaboration}, E.~Church {\em et~al.}, {\it {A proposal for
  an experiment to measure $\nu_\mu \to \nu_e$ oscillations and $\nu_\mu$
  disappearance at the Fermilab Booster: BooNE}}, .

\bibitem{Aguilar-Arevalo:2020nvw}
{\bf MiniBooNE} {\bf Collaboration}, A.~A. Aguilar-Arevalo {\em et~al.}, {\it
  {Updated MiniBooNE Neutrino Oscillation Results with Increased Data and New
  Background Studies}},  \href{http://www.arxiv.org/abs/2006.16883}{{\tt
  2006.16883}}.

\bibitem{Formaggio:2012cpf}
J.~A. Formaggio and G.~P. Zeller, {\it {From eV to EeV: Neutrino Cross Sections
  Across Energy Scales}},  {\em Rev. Mod. Phys.} {\bf 84} (2012) 1307--1341,
  [\href{http://www.arxiv.org/abs/1305.7513}{{\tt 1305.7513}}].

\bibitem{Alvarez-Ruso2021}
L.~Alvarez-Ruso and E.~Saul-Sala, {\it {Neutrino interactions with matter and
  the MiniBooNE anomaly}},  {\em The European Physical Journal Special Topics}
  (2021).

\bibitem{Fischer:2019fbw}
O.~Fischer, A.~Hern\'{a}ndez-Cabezudo, and T.~Schwetz, {\it {Explaining the
  MiniBooNE excess by a decaying sterile neutrino with mass in the 250 MeV
  range}},  \href{http://www.arxiv.org/abs/1909.09561}{{\tt 1909.09561}}.

\bibitem{Gninenko:2009ks}
S.~N. Gninenko, {\it {The MiniBooNE anomaly and heavy neutrino decay}},  {\em
  Phys. Rev. Lett.} {\bf 103} (2009) 241802,
  [\href{http://www.arxiv.org/abs/0902.3802}{{\tt 0902.3802}}].

\bibitem{Bertuzzo:2018itn}
E.~Bertuzzo, S.~Jana, P.~A.~N. Machado, and R.~Zukanovich~Funchal, {\it {Dark
  Neutrino Portal to Explain MiniBooNE excess}},  {\em Phys. Rev. Lett.} {\bf
  121} (2018), no.~24 241801, [\href{http://www.arxiv.org/abs/1807.09877}{{\tt
  1807.09877}}].

\bibitem{Dentler:2019dhz}
M.~Dentler, I.~Esteban, J.~Kopp, and P.~Machado, {\it {Decaying Sterile
  Neutrinos and the Short Baseline Oscillation Anomalies}},
  \href{http://www.arxiv.org/abs/1911.01427}{{\tt 1911.01427}}.

\bibitem{Ballett:2018ynz}
P.~Ballett, S.~Pascoli, and M.~Ross-Lonergan, {\it {U(1)' mediated decays of
  heavy sterile neutrinos in MiniBooNE}},  {\em Phys. Rev.} {\bf D99} (2019)
  071701, [\href{http://www.arxiv.org/abs/1808.02915}{{\tt 1808.02915}}].

\bibitem{deGouvea:2019qre}
A.~de~Gouvêa, O.~L.~G. Peres, S.~Prakash, and G.~V. Stenico, {\it {On The
  Decaying-Sterile Neutrino Solution to the Electron (Anti)Neutrino Appearance
  Anomalies}},  \href{http://www.arxiv.org/abs/1911.01447}{{\tt 1911.01447}}.

\bibitem{Abdallah:2020biq}
W.~Abdallah, R.~Gandhi, and S.~Roy, {\it {Understanding the MiniBooNE and the
  muon $g-2$ anomalies with a light $Z'$ and a second Higgs doublet}},
  \href{http://www.arxiv.org/abs/2006.01948}{{\tt 2006.01948}}.

\bibitem{Dutta:2020scq}
B.~Dutta, S.~Ghosh, and T.~Li, {\it {Explaining $(g-2)_{\mu,e}$, KOTO anomaly
  and MinibooNE excess in an extended Higgs model with sterile neutrinos}},
  \href{http://www.arxiv.org/abs/2006.01319}{{\tt 2006.01319}}.

\bibitem{Datta:2020auq}
A.~Datta, S.~Kamali, and D.~Marfatia, {\it {Dark sector origin of the KOTO and
  MiniBooNE anomalies}},  \href{http://www.arxiv.org/abs/2005.08920}{{\tt
  2005.08920}}.

\bibitem{Abdallah:2020vgg}
W.~Abdallah, R.~Gandhi, and S.~Roy, {\it {A two-Higgs doublet solution to the
  LSND, MiniBooNE and muon $g-2$ anomalies}},
  \href{http://www.arxiv.org/abs/2010.06159}{{\tt 2010.06159}}.

\bibitem{Abdullahi:2020nyr}
A.~Abdullahi, M.~Hostert, and S.~Pascoli, {\it {A Dark Seesaw Solution to Low
  Energy Anomalies: MiniBooNE, the muon $(g-2)$, and BaBar}},
  \href{http://www.arxiv.org/abs/2007.11813}{{\tt 2007.11813}}.

\bibitem{Brdar:2020tle}
V.~Brdar, O.~Fischer, and A.~Y. Smirnov, {\it {Model Independent Bounds on the
  Non-Oscillatory Explanations of the MiniBooNE Excess}},
  \href{http://www.arxiv.org/abs/2007.14411}{{\tt 2007.14411}}.

\bibitem{Acciarri:2016smi}
{\bf MicroBooNE} {\bf Collaboration}, R.~Acciarri {\em et~al.}, {\it {Design
  and Construction of the MicroBooNE Detector}},  {\em JINST} {\bf 12} (2017),
  no.~02 P02017, [\href{http://www.arxiv.org/abs/1612.05824}{{\tt
  1612.05824}}].

\bibitem{McConkey:2017dsv}
{\bf SBND} {\bf Collaboration}, N.~McConkey, {\it {SBND: Status of the Fermilab
  Short-Baseline Near Detector}},  {\em J. Phys. Conf. Ser.} {\bf 888} (2017),
  no.~1 012148.

\bibitem{Rubbia:2011ft}
C.~Rubbia {\em et~al.}, {\it {Underground operation of the ICARUS T600 LAr-TPC:
  first results}},  {\em JINST} {\bf 6} (2011) P07011,
  [\href{http://www.arxiv.org/abs/1106.0975}{{\tt 1106.0975}}].

\bibitem{Machado:2019oxb}
P.~A. Machado, O.~Palamara, and D.~W. Schmitz, {\it {The Short-Baseline
  Neutrino Program at Fermilab}},  {\em Ann. Rev. Nucl. Part. Sci.} {\bf 69}
  (2019) 363--387, [\href{http://www.arxiv.org/abs/1903.04608}{{\tt
  1903.04608}}].

\bibitem{Ellis:2017itw}
R.~K. Ellis, {\it {Guido Altarelli and the evolution of QCD}},  {\em Nuovo Cim.
  C} {\bf 39} (2017), no.~4 355,
  [\href{http://www.arxiv.org/abs/1608.05574}{{\tt 1608.05574}}].

\bibitem{Martini:2009uj}
M.~Martini, M.~Ericson, G.~Chanfray, and J.~Marteau, {\it {A Unified approach
  for nucleon knock-out, coherent and incoherent pion production in neutrino
  interactions with nuclei}},  {\em Phys. Rev.} {\bf C80} (2009) 065501,
  [\href{http://www.arxiv.org/abs/0910.2622}{{\tt 0910.2622}}].

\bibitem{Nieves:2011pp}
J.~Nieves, I.~Ruiz~Simo, and M.~J. Vicente~Vacas, {\it {Inclusive
  Charged--Current Neutrino--Nucleus Reactions}},  {\em Phys. Rev. C} {\bf 83}
  (2011) 045501, [\href{http://www.arxiv.org/abs/1102.2777}{{\tt 1102.2777}}].

\bibitem{Sobczyk:2012ms}
J.~T. Sobczyk, {\it {Multinucleon ejection model for Meson Exchange Current
  neutrino interactions}},  {\em Phys. Rev.} {\bf C86} (2012) 015504,
  [\href{http://www.arxiv.org/abs/1201.3673}{{\tt 1201.3673}}].

\bibitem{Meucci:2012yq}
A.~Meucci and C.~Giusti, {\it {Relativistic descriptions of final-state
  interactions in charged-current quasielastic antineutrino-nucleus scattering
  at MiniBooNE kinematics}},  {\em Phys. Rev.} {\bf D85} (2012) 093002,
  [\href{http://www.arxiv.org/abs/1202.4312}{{\tt 1202.4312}}].

\bibitem{Lalakulich:2012ac}
O.~Lalakulich, K.~Gallmeister, and U.~Mosel, {\it {Many-Body Interactions of
  Neutrinos with Nuclei - Observables}},  {\em Phys. Rev.} {\bf C86} (2012),
  no.~1 014614, [\href{http://www.arxiv.org/abs/1203.2935}{{\tt 1203.2935}}].
  [Erratum: Phys. Rev.C90,no.2,029902(2014)].

\bibitem{Meloni:2012fq}
D.~Meloni and M.~Martini, {\it {Revisiting the T2K data using different models
  for the neutrino–nucleus cross sections}},  {\em Phys. Lett.} {\bf B716}
  (2012) 186--192, [\href{http://www.arxiv.org/abs/1203.3335}{{\tt
  1203.3335}}].

\bibitem{Nieves:2012yz}
J.~Nieves, F.~Sanchez, I.~Ruiz~Simo, and M.~J. Vicente~Vacas, {\it {Neutrino
  Energy Reconstruction and the Shape of the CCQE-like Total Cross Section}},
  {\em Phys. Rev.} {\bf D85} (2012) 113008,
  [\href{http://www.arxiv.org/abs/1204.5404}{{\tt 1204.5404}}].

\bibitem{Lalakulich:2012hs}
O.~Lalakulich, U.~Mosel, and K.~Gallmeister, {\it {Energy reconstruction in
  quasielastic scattering in the MiniBooNE and T2K experiments}},  {\em Phys.
  Rev.} {\bf C86} (2012) 054606,
  [\href{http://www.arxiv.org/abs/1208.3678}{{\tt 1208.3678}}].

\bibitem{Martini:2012uc}
M.~Martini, M.~Ericson, and G.~Chanfray, {\it {Energy reconstruction effects in
  neutrino oscillation experiments and implications for the analysis}},  {\em
  Phys. Rev. D} {\bf 87} (2013), no.~1 013009,
  [\href{http://www.arxiv.org/abs/1211.1523}{{\tt 1211.1523}}].

\bibitem{Aguilar-Arevalo:2013pmq}
{\bf MiniBooNE} {\bf Collaboration}, A.~Aguilar-Arevalo {\em et~al.}, {\it
  {Improved Search for $\bar \nu_\mu \rightarrow \bar \nu_e$ Oscillations in
  the MiniBooNE Experiment}},  {\em Phys. Rev. Lett.} {\bf 110} (2013) 161801,
  [\href{http://www.arxiv.org/abs/1303.2588}{{\tt 1303.2588}}].

\bibitem{Coloma:2013rqa}
P.~Coloma and P.~Huber, {\it {Impact of nuclear effects on the extraction of
  neutrino oscillation parameters}},  {\em Phys. Rev. Lett.} {\bf 111} (2013),
  no.~22 221802, [\href{http://www.arxiv.org/abs/1307.1243}{{\tt 1307.1243}}].

\bibitem{Mosel:2013fxa}
U.~Mosel, O.~Lalakulich, and K.~Gallmeister, {\it {Energy reconstruction in the
  Long-Baseline Neutrino Experiment}},  {\em Phys. Rev. Lett.} {\bf 112} (2014)
  151802, [\href{http://www.arxiv.org/abs/1311.7288}{{\tt 1311.7288}}].

\bibitem{Megias:2014qva}
G.~D. Megias {\em et~al.}, {\it {Meson-exchange currents and quasielastic
  predictions for charged-current neutrino-$^{12}C$ scattering in the
  superscaling approach}},  {\em Phys. Rev.} {\bf D91} (2015), no.~7 073004,
  [\href{http://www.arxiv.org/abs/1412.1822}{{\tt 1412.1822}}].

\bibitem{Ericson:2016yjn}
M.~Ericson, M.~V. Garzelli, C.~Giunti, and M.~Martini, {\it {Assessing the role
  of nuclear effects in the interpretation of the MiniBooNE low-energy
  anomaly}},  {\em Phys. Rev.} {\bf D93} (2016), no.~7 073008,
  [\href{http://www.arxiv.org/abs/1602.01390}{{\tt 1602.01390}}].

\bibitem{Aguilar-Arevalo:2018gpe}
{\bf MiniBooNE} {\bf Collaboration}, A.~Aguilar-Arevalo {\em et~al.}, {\it
  {Significant Excess of ElectronLike Events in the MiniBooNE Short-Baseline
  Neutrino Experiment}},  {\em Phys. Rev. Lett.} {\bf 121} (2018), no.~22
  221801, [\href{http://www.arxiv.org/abs/1805.12028}{{\tt 1805.12028}}].

\bibitem{Stowell:2016jfr}
P.~Stowell {\em et~al.}, {\it {NUISANCE: a neutrino cross-section generator
  tuning and comparison framework}},  {\em JINST} {\bf 12} (2017), no.~01
  P01016, [\href{http://www.arxiv.org/abs/1612.07393}{{\tt 1612.07393}}].

\bibitem{Hill:2009ek}
R.~J. Hill, {\it {Low energy analysis of nu N ---\ensuremath{>} nu N gamma in
  the Standard Model}},  {\em Phys. Rev. D} {\bf 81} (2010) 013008,
  [\href{http://www.arxiv.org/abs/0905.0291}{{\tt 0905.0291}}].

\bibitem{Zhang:2012xn}
X.~Zhang and B.~D. Serot, {\it {Can neutrino-induced photon production explain
  the low energy excess in MiniBooNE?}},  {\em Phys. Lett. B} {\bf 719} (2013)
  409--414, [\href{http://www.arxiv.org/abs/1210.3610}{{\tt 1210.3610}}].

\bibitem{Wang:2013wva}
E.~Wang, L.~Alvarez-Ruso, and J.~Nieves, {\it {Photon emission in neutral
  current interactions at intermediate energies}},  {\em Phys. Rev. C} {\bf 89}
  (2014), no.~1 015503, [\href{http://www.arxiv.org/abs/1311.2151}{{\tt
  1311.2151}}].

\bibitem{Wang:2014nat}
E.~Wang, L.~Alvarez-Ruso, and J.~Nieves, {\it {Single photon events from
  neutral current interactions at MiniBooNE}},  {\em Phys. Lett. B} {\bf 740}
  (2015) 16--22, [\href{http://www.arxiv.org/abs/1407.6060}{{\tt 1407.6060}}].

\bibitem{Ioannisian:2019kse}
A.~Ioannisian, {\it {A Standard Model explanation for the excess of
  electron-like events in MiniBooNE}},
  \href{http://www.arxiv.org/abs/1909.08571}{{\tt 1909.08571}}.

\bibitem{Giunti:2019sag}
C.~Giunti, A.~Ioannisian, and G.~Ranucci, {\it {A new analysis of the MiniBooNE
  low-energy excess}},  \href{http://www.arxiv.org/abs/1912.01524}{{\tt
  1912.01524}}.

\bibitem{Louis:privcomm}
W.~C. Louis, 2019.
\newblock private communication.

\bibitem{Andreopoulos:2015wxa}
C.~Andreopoulos, C.~Barry, S.~Dytman, H.~Gallagher, T.~Golan, R.~Hatcher,
  G.~Perdue, and J.~Yarba, {\it {The GENIE Neutrino Monte Carlo Generator:
  Physics and User Manual}},  \href{http://www.arxiv.org/abs/1510.05494}{{\tt
  1510.05494}}.

\bibitem{Casper:2002sd}
D.~Casper, {\it {The Nuance neutrino physics simulation, and the future}},
  {\em Nucl. Phys. B Proc. Suppl.} {\bf 112} (2002) 161--170,
  [\href{http://www.arxiv.org/abs/hep-ph/0208030}{{\tt hep-ph/0208030}}].

\bibitem{Golan:2012wx}
T.~Golan, C.~Juszczak, and J.~T. Sobczyk, {\it {Final State Interactions
  Effects in Neutrino-Nucleus Interactions}},  {\em Phys. Rev. C} {\bf 86}
  (2012) 015505, [\href{http://www.arxiv.org/abs/1202.4197}{{\tt 1202.4197}}].

\bibitem{Leitner:2008ue}
T.~Leitner, O.~Buss, L.~Alvarez-Ruso, and U.~Mosel, {\it {Electron- and
  neutrino-nucleus scattering from the quasielastic to the resonance region}},
  {\em Phys. Rev. C} {\bf 79} (2009) 034601,
  [\href{http://www.arxiv.org/abs/0812.0587}{{\tt 0812.0587}}].

\bibitem{AguilarArevalo:2010zc}
{\bf MiniBooNE} {\bf Collaboration}, A.~Aguilar-Arevalo {\em et~al.}, {\it
  {First Measurement of the Muon Neutrino Charged Current Quasielastic Double
  Differential Cross Section}},  {\em Phys. Rev. D} {\bf 81} (2010) 092005,
  [\href{http://www.arxiv.org/abs/1002.2680}{{\tt 1002.2680}}].

\bibitem{Ankowski:2005wi}
A.~M. Ankowski and J.~T. Sobczyk, {\it {Argon spectral function and neutrino
  interactions}},  {\em Phys. Rev. C} {\bf 74} (2006) 054316,
  [\href{http://www.arxiv.org/abs/nucl-th/0512004}{{\tt nucl-th/0512004}}].

\bibitem{BENHAR1994493}
O.~Benhar, A.~Fabrocini, S.~Fantoni, and I.~Sick, {\it Spectral function of
  finite nuclei and scattering of gev electrons},  {\em Nuclear Physics A} {\bf
  579} (1994), no.~3 493--517.

\bibitem{tune-list}
``Genie comprehensive model configurations and tunes.''
  \url{http://tunes.genie-mc.org/}.

\bibitem{REIN198179}
D.~Rein and L.~M. Sehgal, {\it Neutrino-excitation of baryon resonances and
  single pion production},  {\em Annals of Physics} {\bf 133} (1981), no.~1
  79--153.

\bibitem{Avanzini:2021qlx}
M.~B. Avanzini {\em et~al.}, {\it {Comparisons and challenges of modern
  neutrino-scattering experiments (TENSIONS 2019 report)}},
  \href{http://www.arxiv.org/abs/2112.09194}{{\tt 2112.09194}}.

\bibitem{Zyla:2020zbs}
{\bf Particle Data Group} {\bf Collaboration}, P.~A. Zyla {\em et~al.}, {\it
  {Review of Particle Physics}},  {\em PTEP} {\bf 2020} (2020), no.~8 083C01.

\bibitem{Betancourt:2018bpu}
M.~Betancourt {\em et~al.}, {\it {Comparisons and Challenges of Modern Neutrino
  Scattering Experiments (TENSIONS2016 Report)}},  {\em Phys. Rept.} {\bf
  773-774} (2018) 1--28, [\href{http://www.arxiv.org/abs/1805.07378}{{\tt
  1805.07378}}].

\bibitem{MiniBooNE:2008hfu}
{\bf MiniBooNE} {\bf Collaboration}, A.~A. Aguilar-Arevalo {\em et~al.}, {\it
  {The Neutrino Flux prediction at MiniBooNE}},  {\em Phys. Rev. D} {\bf 79}
  (2009) 072002, [\href{http://www.arxiv.org/abs/0806.1449}{{\tt 0806.1449}}].

\bibitem{Katori:2008zz}
T.~Katori, {\em {A Measurement of the muon neutrino charged current
  quasielastic interaction and a test of Lorentz violation with the MiniBooNE
  experiment}}.
\newblock PhD thesis, Indiana U., 2008.

\bibitem{PhysRevLett.26.445}
E.~J. Moniz, I.~Sick, R.~R. Whitney, J.~R. Ficenec, R.~D. Kephart, and W.~P.
  Trower, {\it Nuclear fermi momenta from quasielastic electron scattering},
  {\em Phys. Rev. Lett.} {\bf 26} (Feb, 1971) 445--448.

\bibitem{Martini:2012fa}
M.~Martini, M.~Ericson, and G.~Chanfray, {\it {Neutrino energy reconstruction
  problems and neutrino oscillations}},  {\em Phys. Rev. D} {\bf 85} (2012)
  093012, [\href{http://www.arxiv.org/abs/1202.4745}{{\tt 1202.4745}}].

\bibitem{Leitner:2008wx}
T.~Leitner, O.~Buss, U.~Mosel, and L.~Alvarez-Ruso, {\it {Neutrino induced pion
  production at MiniBooNE and K2K}},  {\em Phys. Rev. C} {\bf 79} (2009)
  038501, [\href{http://www.arxiv.org/abs/0812.1787}{{\tt 0812.1787}}].

\bibitem{Lalakulich:2012cj}
O.~Lalakulich and U.~Mosel, {\it {Pion production in the MiniBooNE
  experiment}},  {\em Phys. Rev. C} {\bf 87} (2013), no.~1 014602,
  [\href{http://www.arxiv.org/abs/1210.4717}{{\tt 1210.4717}}].

\bibitem{Mosel:2016cwa}
U.~Mosel, {\it {Neutrino Interactions with Nucleons and Nuclei: Importance for
  Long-Baseline Experiments}},  {\em Ann. Rev. Nucl. Part. Sci.} {\bf 66}
  (2016) 171--195, [\href{http://www.arxiv.org/abs/1602.00696}{{\tt
  1602.00696}}].

\bibitem{MBtalk}
``Miniboone collaboration: talk on april 11, 2007, slide 24.''
  \url{http://www-boone.fnal.gov/publicpages/First_Results.pdf}.

\bibitem{pdg}
\url{http://pdg.lbl.gov/2019/reviews/rpp2019-rev-passage-particles-matter.pdf}.

\bibitem{AguilarArevalo:2008qa}
{\bf MiniBooNE} {\bf Collaboration}, A.~A. Aguilar-Arevalo {\em et~al.}, {\it
  {The MiniBooNE Detector}},  {\em Nucl. Instrum. Meth.} {\bf A599} (2009)
  28--46, [\href{http://www.arxiv.org/abs/0806.4201}{{\tt 0806.4201}}].

\bibitem{Aguilar-Arevalo:2012fmn}
{\bf MiniBooNE} {\bf Collaboration}, A.~A. Aguilar-Arevalo {\em et~al.}, {\it
  {A Combined $\nu_\mu \rightarrow \nu_e$ and $\bar \nu_\mu \rightarrow \bar
  \nu_e$ Oscillation Analysis of the MiniBooNE Excesses}},  7, 2012.
\newblock \href{http://www.arxiv.org/abs/1207.4809}{{\tt 1207.4809}}.
\newblock accompanying data release at
  \url{https://www-boone.fnal.gov/for_physicists/data_release/nue_nuebar_2012/}.

\bibitem{Chanfray:2021wie}
G.~Chanfray and M.~Ericson, {\it {Gamma production in neutrino interaction}},
  \href{http://www.arxiv.org/abs/2105.02505}{{\tt 2105.02505}}.

\bibitem{Aguilar-Arevalo:2021odc}
A.~A. Aguilar-Arevalo {\em et~al.}, {\it {MiniBooNE Data Releases}},
  \href{http://www.arxiv.org/abs/2110.15055}{{\tt 2110.15055}}.

\bibitem{AguilarArevalo:2009ww}
{\bf MiniBooNE} {\bf Collaboration}, A.~A. Aguilar-Arevalo {\em et~al.}, {\it
  {Measurement of $\nu_\mu$ and $\bar{\nu}_\mu$ induced neutral current single
  $\pi^0$ production cross sections on mineral oil at $E_\nu \sim {\cal O}(1
  {\rm GeV})$}},  {\em Phys. Rev. D} {\bf 81} (2010) 013005,
  [\href{http://www.arxiv.org/abs/0911.2063}{{\tt 0911.2063}}].

\bibitem{Ronchen:2015vfa}
D.~R\"onchen, M.~D\"oring, H.~Haberzettl, J.~Haidenbauer, U.~G. Mei\ss{}ner,
  and K.~Nakayama, {\it {Eta photoproduction in a combined analysis of pion-
  and photon-induced reactions}},  {\em Eur. Phys. J. A} {\bf 51} (2015), no.~6
  70, [\href{http://www.arxiv.org/abs/1504.01643}{{\tt 1504.01643}}].

\bibitem{Kopp:2011qd}
J.~Kopp, M.~Maltoni, and T.~Schwetz, {\it {Are there sterile neutrinos at the
  eV scale?}},  {\em Phys.Rev.Lett.} {\bf 107} (2011) 091801,
  [\href{http://www.arxiv.org/abs/1103.4570}{{\tt 1103.4570}}].

\bibitem{Conrad:2012qt}
J.~Conrad, C.~Ignarra, G.~Karagiorgi, M.~Shaevitz, and J.~Spitz, {\it {Sterile
  Neutrino Fits to Short Baseline Neutrino Oscillation Measurements}},  {\em
  Adv.High Energy Phys.} {\bf 2013} (2013) 163897,
  [\href{http://www.arxiv.org/abs/1207.4765}{{\tt 1207.4765}}].

\bibitem{Archidiacono:2013xxa}
M.~Archidiacono, N.~Fornengo, C.~Giunti, S.~Hannestad, and A.~Melchiorri, {\it
  {Sterile Neutrinos: Cosmology vs Short-Baseline Experiments}},
  \href{http://www.arxiv.org/abs/1302.6720}{{\tt 1302.6720}}.

\bibitem{Kopp:2013vaa}
J.~Kopp, P.~A.~N. Machado, M.~Maltoni, and T.~Schwetz, {\it {Sterile Neutrino
  Oscillations: The Global Picture}},  {\em JHEP} {\bf 1305} (2013) 050,
  [\href{http://www.arxiv.org/abs/1303.3011}{{\tt 1303.3011}}].

\bibitem{Mirizzi:2013kva}
A.~Mirizzi, G.~Mangano, N.~Saviano, E.~Borriello, C.~Giunti, {\em et~al.}, {\it
  {The strongest bounds on active-sterile neutrino mixing after Planck data}},
  \href{http://www.arxiv.org/abs/1303.5368}{{\tt 1303.5368}}.

\bibitem{Giunti:2013aea}
C.~Giunti, M.~Laveder, Y.~Li, and H.~Long, {\it {Pragmatic View of
  Short-Baseline Neutrino Oscillations}},  {\em Phys.Rev.} {\bf D88} (2013)
  073008, [\href{http://www.arxiv.org/abs/1308.5288}{{\tt 1308.5288}}].

\bibitem{Gariazzo:2013gua}
S.~Gariazzo, C.~Giunti, and M.~Laveder, {\it {Light Sterile Neutrinos in
  Cosmology and Short-Baseline Oscillation Experiments}},  {\em JHEP} {\bf
  1311} (2013) 211, [\href{http://www.arxiv.org/abs/1309.3192}{{\tt
  1309.3192}}].

\bibitem{Collin:2016rao}
G.~H. Collin, C.~A. Argüelles, J.~M. Conrad, and M.~H. Shaevitz, {\it {Sterile
  Neutrino Fits to Short Baseline Data}},
  \href{http://www.arxiv.org/abs/1602.00671}{{\tt 1602.00671}}.

\bibitem{Gariazzo:2017fdh}
S.~Gariazzo, C.~Giunti, M.~Laveder, and Y.~F. Li, {\it {Updated Global 3+1
  Analysis of Short-BaseLine Neutrino Oscillations}},
  \href{http://www.arxiv.org/abs/1703.00860}{{\tt 1703.00860}}.

\bibitem{Giunti:2017yid}
C.~Giunti, X.~P. Ji, M.~Laveder, Y.~F. Li, and B.~R. Littlejohn, {\it {Reactor
  Fuel Fraction Information on the Antineutrino Anomaly}},
  \href{http://www.arxiv.org/abs/1708.01133}{{\tt 1708.01133}}.

\bibitem{Dentler:2017tkw}
M.~Dentler, A.~Hern\'{a}ndez-Cabezudo, J.~Kopp, M.~Maltoni, and T.~Schwetz,
  {\it {Sterile Neutrinos or Flux Uncertainties? - Status of the Reactor
  Anti-Neutrino Anomaly}},  \href{http://www.arxiv.org/abs/1709.04294}{{\tt
  1709.04294}}.

\bibitem{Dentler:2018sju}
M.~Dentler, A.~Hern\'{a}ndez-Cabezudo, J.~Kopp, P.~A.~N. Machado, M.~Maltoni,
  I.~Martinez-Soler, and T.~Schwetz, {\it {Updated global analysis of neutrino
  oscillations in the presence of eV-scale sterile neutrinos}},
  \href{http://www.arxiv.org/abs/1803.10661}{{\tt 1803.10661}}.

\bibitem{Adamson:2017uda}
{\bf MINOS} {\bf Collaboration}, P.~Adamson {\em et~al.}, {\it {Search for
  sterile neutrinos in MINOS and MINOS+ using a two-detector fit}},  {\em
  Submitted to: Phys. Rev. Lett.} (2017)
  [\href{http://www.arxiv.org/abs/1710.06488}{{\tt 1710.06488}}].

\bibitem{TheIceCube:2016oqi}
{\bf IceCube} {\bf Collaboration}, M.~G. Aartsen {\em et~al.}, {\it {Searches
  for Sterile Neutrinos with the IceCube Detector}},  {\em Phys. Rev. Lett.}
  {\bf 117} (2016), no.~7 071801,
  [\href{http://www.arxiv.org/abs/1605.01990}{{\tt 1605.01990}}].

\bibitem{Jones:2015}
B.~J.~P. Jones, {\em Sterile neutrinos in cold climates}.
\newblock PhD thesis, Massachusetts Institute of Technology, 2015.
\newblock available from \url{http://hdl.handle.net/1721.1/101327}.

\bibitem{Arguelles:2015}
C.~A. {Arg\"{u}elles}, {\em New Physics with Atmospheric Neutrinos}.
\newblock PhD thesis, University of Wisconsin, Madison, 2015.
\newblock available from
  \url{https://docushare.icecube.wisc.edu/dsweb/Get/Document-75669/tesis.pdf}.

\bibitem{AguilarArevalo:2009yj}
{\bf MiniBooNE} {\bf Collaboration}, A.~A. Aguilar-Arevalo {\em et~al.}, {\it
  {A Search for muon neutrino and antineutrino disappearance in MiniBooNE}},
  {\em Phys.Rev.Lett.} {\bf 103} (2009) 061802,
  [\href{http://www.arxiv.org/abs/0903.2465}{{\tt 0903.2465}}].

\bibitem{Cheng:2012yy}
{\bf MiniBooNE Collaboration, SciBooNE Collaboration} {\bf Collaboration},
  G.~Cheng {\em et~al.}, {\it {Dual baseline search for muon antineutrino
  disappearance at $0.1 {\rm eV}^2 < {\Delta}m^2 < 100 {\rm eV}^2$}},  {\em
  Phys.Rev.} {\bf D86} (2012) 052009,
  [\href{http://www.arxiv.org/abs/1208.0322}{{\tt 1208.0322}}].

\bibitem{Cyburt:2015mya}
R.~H. Cyburt, B.~D. Fields, K.~A. Olive, and T.-H. Yeh, {\it {Big Bang
  Nucleosynthesis: 2015}},  {\em Rev. Mod. Phys.} {\bf 88} (2016) 015004,
  [\href{http://www.arxiv.org/abs/1505.01076}{{\tt 1505.01076}}].

\bibitem{Ade:2015xua}
{\bf Planck} {\bf Collaboration}, P.~A.~R. Ade {\em et~al.}, {\it {Planck 2015
  results. XIII. Cosmological parameters}},
  \href{http://www.arxiv.org/abs/1502.01589}{{\tt 1502.01589}}.

\bibitem{Dasgupta:2013zpn}
B.~Dasgupta and J.~Kopp, {\it {A m\'enage \`a trois of eV-scale sterile
  neutrinos, cosmology, and structure formation}},  {\em Phys.Rev.Lett.} {\bf
  112} (2014) 031803, [\href{http://www.arxiv.org/abs/1310.6337}{{\tt
  1310.6337}}].

\bibitem{Hannestad:2013ana}
S.~Hannestad, R.~S. Hansen, and T.~Tram, {\it {How secret interactions can
  reconcile sterile neutrinos with cosmology}},  {\em Phys.Rev.Lett.} {\bf 112}
  (2014) 031802, [\href{http://www.arxiv.org/abs/1310.5926}{{\tt 1310.5926}}].

\bibitem{Chu:2018gxk}
X.~Chu, B.~Dasgupta, M.~Dentler, J.~Kopp, and N.~Saviano, {\it {Sterile
  Neutrinos with Secret Interactions -- Cosmological Discord?}},
  \href{http://www.arxiv.org/abs/1806.10629}{{\tt 1806.10629}}.

\bibitem{Yaguna:2007wi}
C.~E. Yaguna, {\it {Sterile neutrino production in models with low reheating
  temperatures}},  {\em JHEP} {\bf 06} (2007) 002,
  [\href{http://www.arxiv.org/abs/0706.0178}{{\tt 0706.0178}}].

\bibitem{Saviano:2013ktj}
N.~Saviano, A.~Mirizzi, O.~Pisanti, P.~D. Serpico, G.~Mangano, {\em et~al.},
  {\it {Multi-momentum and multi-flavour active-sterile neutrino oscillations
  in the early universe: role of neutrino asymmetries and effects on
  nucleosynthesis}},  {\em Phys.Rev.} {\bf D87} (2013) 073006,
  [\href{http://www.arxiv.org/abs/1302.1200}{{\tt 1302.1200}}].

\bibitem{Giovannini:2002qw}
M.~Giovannini, H.~Kurki-Suonio, and E.~Sihvola, {\it {Big bang nucleosynthesis,
  matter antimatter regions, extra relativistic species, and relic
  gravitational waves}},  {\em Phys. Rev.} {\bf D66} (2002) 043504,
  [\href{http://www.arxiv.org/abs/astro-ph/0203430}{{\tt astro-ph/0203430}}].

\bibitem{Bezrukov:2017ike}
F.~Bezrukov, A.~Chudaykin, and D.~Gorbunov, {\it {Hiding an elephant: heavy
  sterile neutrino with large mixing angle does not contradict cosmology}},
  \href{http://www.arxiv.org/abs/1705.02184}{{\tt 1705.02184}}.

\bibitem{Farzan:2019yvo}
Y.~Farzan, {\it {Ultra-light scalar saving the 3+1 neutrino scheme from the
  cosmological bounds}},  \href{http://www.arxiv.org/abs/1907.04271}{{\tt
  1907.04271}}.

\bibitem{Cline:2019seo}
J.~M. Cline, {\it {Viable secret neutrino interactions with ultralight dark
  matter}},  \href{http://www.arxiv.org/abs/1908.02278}{{\tt 1908.02278}}.

\bibitem{Archidiacono:2020yey}
M.~Archidiacono, S.~Gariazzo, C.~Giunti, S.~Hannestad, and T.~Tram, {\it
  {Sterile neutrino self-interactions: $H_0$ tension and short-baseline
  anomalies}},  \href{http://www.arxiv.org/abs/2006.12885}{{\tt 2006.12885}}.

\bibitem{AguilarArevalo:2010wv}
{\bf MiniBooNE} {\bf Collaboration}, A.~Aguilar-Arevalo {\em et~al.}, {\it
  {Event Excess in the MiniBooNE Search for $\bar \nu_\mu \rightarrow \bar
  \nu_e$ Oscillations}},  {\em Phys.Rev.Lett.} {\bf 105} (2010) 181801,
  [\href{http://www.arxiv.org/abs/1007.1150}{{\tt 1007.1150}}].

\bibitem{MiniBooNE:2009ozf}
{\bf MiniBooNE} {\bf Collaboration}, A.~A. Aguilar-Arevalo {\em et~al.}, {\it
  {A Search for muon neutrino and antineutrino disappearance in MiniBooNE}},
  {\em Phys. Rev. Lett.} {\bf 103} (2009) 061802,
  [\href{http://www.arxiv.org/abs/0903.2465}{{\tt 0903.2465}}].

\bibitem{Aartsen:2014yll}
{\bf IceCube Collaboration} {\bf Collaboration}, M.~Aartsen {\em et~al.}, {\it
  {Determining neutrino oscillation parameters from atmospheric muon neutrino
  disappearance with three years of IceCube DeepCore data}},
  \href{http://www.arxiv.org/abs/1410.7227}{{\tt 1410.7227}}.

\bibitem{deepcore:2016}
{\bf IceCube} {\bf Collaboration}, J.~P. Ya{\~n}ez {\em et~al.}, ``{IceCube
  Oscillations: 3 years muon neutrino disappearance data}.''
  \href{http://icecube.wisc.edu/science/data/nu\_osc}{\tt http://icecube.
  wisc.edu/science/data/nu\_osc}.

\bibitem{Dydak:1983zq}
F.~Dydak, G.~Feldman, C.~Guyot, J.~Merlo, H.~Meyer, {\em et~al.}, {\it {A
  Search for Muon-neutrino Oscillations in the Delta m**2 Range 0.3-eV**2 to
  90-eV**2}},  {\em Phys.Lett.} {\bf B134} (1984) 281.

\bibitem{Wendell:2010md}
{\bf Super-Kamiokande Collaboration} {\bf Collaboration}, R.~Wendell {\em
  et~al.}, {\it {Atmospheric neutrino oscillation analysis with sub-leading
  effects in Super-Kamiokande I, II, and III}},  {\em Phys.Rev.} {\bf D81}
  (2010) 092004, [\href{http://www.arxiv.org/abs/1002.3471}{{\tt 1002.3471}}].
  Long author list - awaiting processing.

\bibitem{Wendell:2014dka}
{\bf Super-Kamiokande} {\bf Collaboration}, R.~Wendell, {\it {Atmospheric
  Results from Super-Kamiokande}},  {\em AIP Conf. Proc.} {\bf 1666} (2015),
  no.~1 100001, [\href{http://www.arxiv.org/abs/1412.5234}{{\tt 1412.5234}}].

\bibitem{Adamson:2017zcg}
{\bf NOvA} {\bf Collaboration}, P.~Adamson {\em et~al.}, {\it {Search for
  active-sterile neutrino mixing using neutral-current interactions in NOvA}},
  \href{http://www.arxiv.org/abs/1706.04592}{{\tt 1706.04592}}.

\bibitem{MicroBooNE:2021rmx}
{\bf MicroBooNE} {\bf Collaboration}, P.~Abratenko {\em et~al.}, {\it {Search
  for an Excess of Electron Neutrino Interactions in MicroBooNE Using Multiple
  Final State Topologies}},  \href{http://www.arxiv.org/abs/2110.14054}{{\tt
  2110.14054}}.

\bibitem{MicroBooNE:2021nxr}
{\bf MicroBooNE} {\bf Collaboration}, P.~Abratenko {\em et~al.}, {\it {Search
  for an anomalous excess of inclusive charged-current $\nu_e$ interactions in
  the MicroBooNE experiment using Wire-Cell reconstruction}},
  \href{http://www.arxiv.org/abs/2110.13978}{{\tt 2110.13978}}.

\bibitem{MicroBooNE:2021jwr}
{\bf MicroBooNE} {\bf Collaboration}, P.~Abratenko {\em et~al.}, {\it {Search
  for an anomalous excess of charged-current quasi-elastic $\nu_e$ interactions
  with the MicroBooNE experiment using Deep-Learning-based reconstruction}},
  \href{http://www.arxiv.org/abs/2110.14080}{{\tt 2110.14080}}.

\bibitem{MicroBooNE:2021sne}
{\bf MicroBooNE} {\bf Collaboration}, P.~Abratenko {\em et~al.}, {\it {Search
  for an anomalous excess of charged-current $\nu_e$ interactions without pions
  in the final state with the MicroBooNE experiment}},
  \href{http://www.arxiv.org/abs/2110.14065}{{\tt 2110.14065}}.

\bibitem{MicroBooNE:2021zai}
{\bf MicroBooNE} {\bf Collaboration}, P.~Abratenko {\em et~al.}, {\it {Search
  for Neutrino-Induced Neutral Current $\Delta$ Radiative Decay in MicroBooNE
  and a First Test of the MiniBooNE Low Energy Excess Under a Single-Photon
  Hypothesis}},  \href{http://www.arxiv.org/abs/2110.00409}{{\tt 2110.00409}}.

\end{thebibliography}\endgroup

\end{document}